\documentclass{emulateapj}

\def\cm{\textrm{cm}}

\def\kpc{\textrm{kpc}}
\def\pc{\textrm{pc}}
\def\Mpc{\textrm{Mpc}}
\def\Kelv{\textrm{K}}

\def\ergps{\textrm{ergs}~\textrm{s}^{-1}}

\def\kms{\textrm{km}~\textrm{s}^{-1}}
\def\gcm2{\textrm{g}~\textrm{cm}^{-2}}
\def\ergscm3{\textrm{erg}~\textrm{s}^{-1}~\textrm{cm}^{-3}}
\def\ergcm3{\textrm{erg}~\textrm{cm}^{-3}}
\def\gscm2{\textrm{g}~\textrm{s}^{-1}~\textrm{cm}^{-2}}
\def\ergcmK34{\textrm{erg}~\textrm{cm}^{-3}~\textrm{K}^{-4}}
\def\cms31{\textrm{cm}^{-3}~\textrm{s}^{-1}}
\def\cmg21{\textrm{cm}^{2}~\textrm{g}^{-1}}
\def\DiffUnits{\textrm{cm}^{2}~\textrm{s}^{-1}}
\def\phcm2s1{\textrm{photons}~\textrm{cm}^{-2}~\textrm{s}^{-1}}

\def\phFluxUnitsWOph{\textrm{cm}^{-2}~\textrm{s}^{-1}}
\def\cm3{\textrm{cm}^{-3}}

\def\MeV{\textrm{MeV}}
\def\GeV{\textrm{GeV}}

\def\MHz{\textrm{MHz}}
\def\GHz{\textrm{GHz}}

\def\yr{\textrm{yr}}
\def\Myr{\textrm{Myr}}
\def\kyr{\textrm{kyr}}
\def\muGauss{\mu\textrm{G}}
\def\mGauss{\textrm{mG}}
\def\Gauss{\textrm{G}}

\def\GeVs1cm3{\textrm{GeV}~\textrm{s}^{-1}~\textrm{cm}^{3}}
\def\log{\textrm{log}}
\def\2phn{\phn\phn}
\def\3phn{\phn\phn\phn}
\def\4phn{\phn\phn\phn\phn}
\def\12phn{\4phn\4phn\4phn}
\def\Msun{\textrm{M}_{\sun}}
\newcommand{\mean}[1]{\ensuremath{\langle #1 \rangle}}

\begin{document}

\title{The Physics of the Far-Infrared-Radio Correlation. I. Calorimetry, Conspiracy, and Implications}
\author{Brian C. Lacki\altaffilmark{1}, Todd A. Thompson\altaffilmark{1,2}, and Eliot Quataert\altaffilmark{3}}
\altaffiltext{1}{Department of Astronomy, The Ohio State University, 140 West 18th Avenue, Columbus, OH 43210, USA; lacki@astronomy.ohio-state.edu}
\altaffiltext{2}{Center for Cosmology and Astroparticle Physics, The Ohio State University, 191 West Woodruff Avenue, Columbus, OH 43210, USA}
\altaffiltext{3}{Astronomy Department and Theoretical Astrophysics Center, University of California, Berkeley, 601 Campbell Hall, Berkeley, CA 94720, USA}

\begin{abstract}
The far-infrared (FIR) and radio luminosities of star-forming galaxies are linearly correlated over a very wide range in star formation rate, from normal spirals like the Milky Way to the most intense starbursts.  Using one-zone models of cosmic ray (CR) injection, cooling, and escape in star-forming galaxies, we attempt to reproduce the observed FIR-radio correlation (FRC) over its entire span.  The normalization and linearity of the FRC, together with constraints on the CR population in the Milky Way, have strong implications for the CR and magnetic energy densities in star-forming galaxies.  We show that for consistency with the FRC, $\sim$2\% of the kinetic energy from supernova explosions must go into high energy primary CR electrons and that $\sim$10\% - 20\% must go into high energy primary CR protons.  Secondary electrons and positrons are likely comparable to or dominate primary electrons in dense starburst galaxies.  We discuss the implications of our models for the magnetic field strengths of starbursts, the detectability of starbursts by \emph{Fermi}, and CR feedback.  Overall, our models indicate that both CR protons and electrons escape from low surface density galaxies, but lose most of their energy before escaping dense starbursts.  The FRC is caused by a combination of the efficient cooling of CR electrons (calorimetry) in starbursts and a conspiracy of several factors.  For lower surface density galaxies, the decreasing radio emission caused by CR escape is balanced by the decreasing FIR emission caused by the low effective UV dust opacity.  In starbursts, bremsstrahlung, ionization, and Inverse Compton cooling decrease the radio emission, but they are countered by secondary electrons/positrons and the dependence of synchrotron frequency on energy, which both increase the radio emission.  Our conclusions hold for a broad range of variations on our fiducial model, such as those including winds, different magnetic field strengths, and different diffusive escape times.
\end{abstract}


\section{Introduction}

The far-infrared (FIR) and radio luminosities of star-forming galaxies lie on a tight empirical relation, the ``FIR-radio correlation'' \citep[FRC;][]{vanDerKruit71,vanDerKruit73,deJong85,Helou85,Condon92,Yun01}.  The FRC spans over three decades in luminosity, remaining roughly linear across the range $10^9 L_{\sun} \la L \la 10^{12.5} L_{\sun}$, from dwarf galaxies to local ultra-luminous infrared galaxies (ULIRGs) like Arp 220 \citep{Yun01}.  At low luminosities ($L \la 10^9 L_{\sun}$), the correlation shows evidence of non-linearity \citep{Yun01,Bell03,Beswick08}.  The galaxies that make up the FRC span a large dynamic range, not just in bolometric luminosity, but also in gas surface density\footnote{$1~\gcm2 = 4800 \Msun \pc^{-2}$.} ($0.001~\gcm2 \la \Sigma_g \la 10~\gcm2$), photon energy density, and presumably magnetic field strength.  From the observed Schmidt law of star formation \citep{Schmidt59,Kennicutt98}, the range in gas surface density corresponds to a range of at least $4 \times 10^5$ in photon energy density.  Not only does the FRC hold on galactic scales, but it exists for regions within star-forming galaxies down to a few hundred parsecs \citep[e.g.,][]{Beck88,Bicay90,Murphy06a,Paladino06,Murphy06b,Murphy08}.

Star formation drives the FRC.  Young massive stars produce ultraviolet (UV) light, which is easily absorbed by dust grains.  The dust reradiates in the FIR, producing a linear correlation between star formation rate and the FIR luminosity, if the dust is optically thick to the UV light.  The non-thermal GHz radio continuum emission observed from star-forming galaxies is synchrotron radiation from cosmic ray (CR) electrons and positrons, believed to be accelerated in supernova (SN) remnants.  Since SNe mainly occur in young stellar populations, this means that star formation is directly linked to normal (non-active galactic nucleus) radio emission (reviewed in \citealt{Condon92}).

In this paper, we model the FRC, over its range in physical parameters from normal star-forming galaxies to the densest and most luminous starbursts.  Our motivation is that the normalization and linearity of the FRC has strong implications for the physical properties of star-forming galaxies and the CRs they contain.  For example, we can use the radio emission to estimate the energy injection rate and equilibrium energy density of both CR electrons and protons.  This is important because the CR pressure is known to be dynamically important in the Milky Way \citep[e.g.,][]{Boulares90}, and possibly starburst galaxies \citep{Socrates08}.  Furthermore, we can use the inferred CR proton energy density to calculate the flux of gamma-rays from pion production in the galaxies' host interstellar medium \citep[ISM; e.g.,][]{Torres04,Thompson07}.  Finally, the radio emission also constrains the magnetic field strength in galaxies on the FRC \citep[][]{Thompson06}.

Finding the causes of the linearity and span of the FRC is the other main purpose of this paper.  The FRC is affected by the density of CRs and the environment they propagate through.  For the Milky Way, the propagation of CRs has been well studied, both observationally and theoretically \citep[e.g.,][]{Strong98}.  However, given the vast range of environments in star-forming galaxies, it is not clear that our knowledge of CR propagation in the Galaxy can be extrapolated across the entire FRC.  Therefore, one aspect of our task in explaining the FRC is determining the extent to which the properties of CR injection, such as the initial spectral slope and proton-to-electron ratio, and CR propagation, such as the rate of escape by diffusion, can apply to all star-forming galaxies.

The diversity of star-forming galaxies on the FRC and the tightness of the correlation may imply a deeper, simpler principle at work.  In the calorimeter theory first proposed in \citet{Volk89}, the CR electrons lose all of their energy before escaping galaxies, with most of the energy radiated as synchrotron radio emission.  Thus, galaxies are \emph{electron calorimeters}, with the energy in CR electrons being converted into an observable form.  Calorimetry also requires that galaxies on the FRC are optically thick to UV light from young stars, which is reradiated in the FIR.  These galaxies would therefore also have to be \emph{UV calorimeters}.  If both electron calorimetry and UV calorimetry hold, and if synchrotron is the main energy loss mechanism, then the ratio of FIR to radio emission is simply the ratio of total starlight produced to the total energy supplied to CR electrons, which is naively expected to be a constant fraction of the energy from SNe, accounting for the FRC.  

Calorimeter theory has been questioned, however, both in its assumptions and its implications.  For example, the assumption that all normal galaxies are optically thick to UV light is probably false: the observed UV luminosity of normal star-forming galaxies is comparable to the observed FIR luminosity at low overall luminosities \citep[e.g.,][]{Xu95,Bell03,Buat05,Martin05,Popescu05}.  Nor is electron calorimetry believed to hold in the Milky Way (and presumably similar galaxies), since the inferred diffusive escape time is shorter than the typical estimated synchrotron cooling time (see equations~\ref{eqn:StandardCRLife} and~\ref{eqn:StandardB} later in this paper; or, e.g., \citealt{Lisenfeld96a}).

Even in cases when calorimetry holds, the implications of standard calorimeter theory may conflict with observations.  A long-standing problem with the predictions of calorimetry has been the radio spectral indices of star-forming galaxies.  If electron calorimetry holds, then the synchrotron cooling timescale is much less than the escape timescale.  The electron population will then be strongly cooled with a steep spectrum.  For an initial \emph{injection} spectrum of $N \propto E^{-p}$ where $p \approx 2 - 2.5$ and a final synchrotron-cooled \emph{steady-state} spectrum $N \propto E^{-{\cal P}}$, this would imply a synchrotron spectrum of $F_{\nu} \propto \nu^{-\alpha}$ with a spectral index of $\alpha = ({\cal P} - 1) / 2 = p / 2 \approx 1.0 - 1.2$.  The observed spectral indices are $0.7 - 0.8$ for normal galaxies, suggesting that, contrary to calorimeter theory, electrons escape before losing their energy.  \cite{Lisenfeld96a} consider a modified calorimeter model for normal galaxies that includes escape comparable to cooling losses, and \citet{Lisenfeld00} suggest that SN remnants in the galaxies can flatten the observed radio spectrum.  

More drastically, several non-calorimeter theories have been proposed \citep[e.g.,][]{Helou93,Niklas97a}, often involving a ``conspiracy'' to maintain the tightness of the FRC.  A potential pitfall of non-calorimeter models stems from the enormous dynamic range in physical properties for galaxies on the FRC.  For example, inverse Compton cooling alone is very quick in starbursts, implying that electrons cannot escape from these galaxies before losing most of their energy \citep{Condon91,Thompson06}.

Typical explanations of the FRC leave out two underappreciated but important effects: proton losses and non-synchrotron cooling.  Models of individual starbursts, which have gas densities $10^3 - 10^4$ times higher than the Milky Way, predict that CR protons lose most of their energy to pion creation as they interact with the ISM.  When a CR proton collides with a proton in the ISM, it produces a pion, either charged ($\pi^+$ or $\pi^-$), or uncharged ($\pi^0$).  Neutral pions decay into gamma rays, so that pion losses should act as a source of gamma-ray luminosity in starbursts.  Charged pions ultimately decay into neutrinos (which may eventually be observed with neutrino telescopes), as well as secondary electrons and positrons.  Therefore, dense starburst galaxies are expected to be \emph{proton calorimeters}: essentially all the injected energy in CR protons ends up converted to gamma rays, neutrinos, and secondary electrons and positrons.  Proton calorimetry would also imply that, unlike the Milky Way, secondary electrons and positrons may dominate over primary electrons and positrons, depending on the ratio of injected protons to electrons (\citealt{Rengarajan05}; secondary electrons and positrons are found to be more abundant than primary electrons in starbursts by \citealt{Torres04}, \citealt{Domingo05}, and \citealt{deCeaDelPozo09}).  Because secondary electrons and positrons radiate synchrotron, their presence poses a problem for any explanation of the FRC that requires the CR electron density to be directly proportional to the star formation rate, including both standard calorimeter theory and the theory of \citet{Niklas97a}.  

On the other hand, bremsstrahlung, ionization, and IC may all be more important in starburst galaxies.  \citet{Thompson06} point out that cooling by bremsstrahlung and ionization tends to flatten the radio spectra, thus saving calorimeter theory from the spectral index argument, at least for starbursts.  However, the energy CR electrons lose to bremsstrahlung, ionization, and IC cannot go into synchrotron radio emission, an obstacle for any theory that assumes radio emission is directly proportional to the injected power of primary CR electrons.  Therefore, even electron calorimetry and UV calorimetry are not enough to guarantee a linear FRC.

We address these issues with one-zone numerical models of CRs in star-forming galaxies.  These models include CR escape as well as the main cooling processes and secondary production, a combination that has not to our knowledge been done over the entire span of the FRC.  CRs in individual galaxies have been studied with similar one-zone models that fit the emission across the electromagnetic spectrum (e.g., Arp 220 in \citealt{Torres04}; M82 in \citealt{deCeaDelPozo09}), but in this paper our focus is on the FRC itself and not any individual galaxy.  Our one-zone approach allows us to efficiently parameterize unknown quantities like the magnetic field strength, and to try a large number of scenarios.  The primary independent variable in our calculations is the gas surface density $\Sigma_g$, which controls both the photon energy density through the observed Schmidt law and the average gas density.  These simplifying parameterizations allow us to qualitatively understand the FRC over the range of star-forming galaxies, although it ignores deviations and complications that may be important for individual galaxies.

We first describe the calculations necessary to find the CR spectra and observables for each galaxy (Section~\ref{sec:Procedure}).  We review the effects of each parameter on the observables (Section~\ref{sec:EffectsReview}), before presenting our results (Section~\ref{sec:Results}).  We discuss the implications of our work for CR physics (Section~\ref{sec:Discussion}), including whether calorimetry is correct (Section~\ref{sec:Calorimeter}), what causes the FRC (Section~\ref{sec:FRCCauses}), predictions for the FRC at other frequencies (Section~\ref{sec:OtherNuFRC}), the spectral slope problem (Section~\ref{sec:AlphaDiscussion}), the gamma-ray luminosities of galaxies (Section~\ref{sec:GammaRays}), and whether CR pressure and magnetic pressure are important as feedback mechanisms in galaxies (Section~\ref{sec:CRPressures}).  We finally summarize our results (Section~\ref{sec:Summary}).  In Appendix~\ref{sec:Variants}, we present results from a suite of variants on our standard model, and show that those consistent with the FRC have similar parameters to our standard model.  For the reader's convenience, we list the symbols we use in our calculations and discussions in Table~\ref{table:SymbolList}.

\section{Procedure}
\label{sec:Procedure}

We construct one-zone leaky box models of galaxies across the dynamic range of the observed FRC.  We treat star-forming galaxies as homogeneous disks of gas, characterized by a column density $\Sigma_g$, a star formation rate surface density $\Sigma_{\rm SFR}$, and a scale height $h$.  We solve the steady-state diffusion-loss equation for the equilibrium CR spectra of primary and secondary electrons and positrons, as well as primary CR protons.\footnote{In the steady-state approximation, the $\partial N / \partial t$ term in the diffusion-loss equation is assumed to be small.  For the Milky Way (and presumably other normal spirals), the CR flux is known to be constant within a factor of $\sim 2$ for the last billion years from solar system studies \citep{Arnold61,Schaeffer75}.  The tightness of the FRC combined with the long timescales for galactic evolution also imply the CR population in normal galaxies is steady state.  Additionally, in extreme starbursts, the IC cooling time alone for GHz-emitting CR electrons ($\la 10^4~\yr$) is much shorter than the characteristic timescale for the system to evolve.  Therefore, we expect that the steady-state assumption is valid.  However, we note that in weaker starbursts, where the cooling and escape times for CRs are several Myr, evolution may be important and the steady-state approximation may fail \citep[see][]{Lisenfeld96b}.}

Under these simplifying assumptions, the diffusion-loss equation for CRs becomes
\begin{equation}
\label{eqn:DiffusionLoss}
\frac{N(E)}{t_{\rm life}(E)} - \frac{d}{dE}[b(E)N(E)] - Q(E) = 0,
\end{equation}
where $E$ is the total energy, $N(E)$ is the CR spectrum, $t_{\rm life}(E)$ is the energy-dependent lifetime to diffusive or advective escape from the system, $Q(E)$ is the CR source term, and $b(E) = -(dE / dt)$ is the rate of energy loss for each particle.  The equilibrium CR spectrum is a competition at every energy between injection, cooling, and escape losses.  If the injected CRs initially have a spectrum of the form $Q(E) \propto E^{-p}$ and if escape is insignificant ($t_{\rm life}(E) \rightarrow \infty$), the final spectrum will have the form $N(E) \propto E^{1 - p} / b(E)$.  If instead cooling is insignificant compared to escape ($b(E) \rightarrow 0$), the final spectrum will have the form $N(E) \propto E^{-p} t_{\rm life}(E)$.

We solve the general form of equation (\ref{eqn:DiffusionLoss}) numerically using a Green's function for CR protons, electrons, and positrons \citep[see][]{Torres04}.  We include synchrotron, IC, bremsstrahlung, and ionization losses \citep[e.g.,][]{Rybicki79,Longair04}.  For CR protons, we also include pion losses in $t_{\rm life}(E)$ due to inelastic proton-proton collisions using the formalism of \citet{Torres04}.\footnote{Including pion losses in the cooling term $b(E)$ instead is formally incorrect, because the losses are catastrophic instead of continuous.  For $p = 2.0$, the resulting proton spectra are nearly identical, but for $p = 2.6$, including pion losses in $b(E)$ decreases the proton spectrum $N(E)$ by $\sim 40\%$ when pion losses are strong.}  The publicly available GALPROP code\footnote{Specifically, 
we use the ``PP\_MESON" subroutine, which calculates the cross sections for electron and photon production through pion production.  GALPROP is available at http://galprop.stanford.edu.} \citep{Strong98,Strong00,Moskalenko02} is used to calculate the differential cross section for electron or positron production from proton-proton collisions, as well as for calculating the spectrum of $\gamma$-rays produced by the decay of secondary $\pi^0$ mesons.  We also include knock-off electrons from CR proton collisions with atoms in the ISM \citep[see][]{Torres04}.

\subsection{Primary CR Injection Rates}
\label{sec:PrimaryInjection}

We assume that both the primary CR electrons and protons are injected into galaxies with power law spectra $Q(E) = C E^{-p}$ with $1 \le \gamma \le 10^6$ (where $\gamma = E / (mc^2)$ is the Lorentz factor), and we consider initial spectral slopes $p$ in the range $2.0 \le p \le 2.6$.  Integrating the injection spectrum times the kinetic energy per particle $K$ gives the total power injected per unit volume for each primary species, $\epsilon_{\rm CR} = \int K Q(E) dE$, to set the normalization.  Energetic and escape losses produce the final, steady-state spectrum as determined by the solution to equation (\ref{eqn:DiffusionLoss}).

In order to normalize the CR injection spectra, we assume that a constant fraction $\xi$ and $\eta$ of the kinetic energy of SN explosions ($E_{51}=E_{\rm SN}/10^{51}$\,erg) goes into accelerating primary CR electrons and protons, respectively.  The CR electron and proton emissivities $\epsilon_{\rm CR}$ are then proportional to the emissivity in starlight photons, $\epsilon_{\rm ph}$, produced by star formation, when averaged over the star formation episode.\footnote{Though we assume that the SN rate is proportional to the starlight, in reality, there will be a lag between the first massive stars and the first SNe when there will be very few CRs \citep[e.g.,][]{Roussel03}, which we do not account for.}  Following the discussion in \citet{Thompson07}, we calculate the starlight emissivity (here, in units of $\ergscm3$) as
\begin{equation}
\label{eqn:StarEmiss}
\epsilon_{\rm ph} = \varepsilon \Sigma_{\rm SFR} c^2 (2h)^{-1},
\end{equation}
where $\varepsilon = 3.8 \times 10^{-4}$ is a dimensionless initial mass function (IMF) dependent constant that relates the luminosity in young stars to the instantaneous star formation rate, $c$ is the speed of light, $h$ is the CR scale height (in cm), and $\Sigma_{\rm SFR}$ is the surface density of star formation in cgs\footnote{Note that $1\ \gscm2 = 1.5 \times 10^{11}\ \Msun\ \yr^{-1}\ \kpc^{-2}$} units of $\gscm2$ \citep{Kennicutt98}.  Equation~\ref{eqn:StarEmiss} essentially says that some proportion of the mass that forms stars is converted into starlight.  We take the surface density of star formation $\Sigma_{\rm SFR}$ from the observed Schmidt law \citep{Kennicutt98}.  

The emissivity in CR electrons can then be written as
\begin{equation}
\label{eqn:CReNorm}
\epsilon_{\rm CR,\,e} = 9.2 \times 10^{-5} E_{51} \psi_{17} 
 \epsilon_{\rm ph} \left(\frac{\xi}{0.01}\right),
\end{equation}
where $\psi_{17} = (\Gamma_{\rm SN}/\varepsilon)/(17 \Msun^{-1})$ and $\Gamma_{\rm SN}$ is the SN rate per unit star formation.  Similarly, the total emissivity of the primary CR protons can be written as 
\begin{eqnarray}
\label{eqn:CRpNorm}
\epsilon_{\rm CR,\,p} & = & \delta \,\epsilon_{\rm CR,\,e} \nonumber \\
& = & 9.2 \times 10^{-4}  \,\,E_{51} \psi_{17} \epsilon_{\rm ph} \left(\frac{\delta}{10}\right)\left(\frac{\xi}{0.01} \right),
\end{eqnarray}
where $\delta\equiv\eta/\xi$ is the ratio of the total energy injected in CR protons to that in CR electrons per SN.  Although we have normalized $\xi=0.01$ and $\delta=10$ in the above expressions for reference, one purpose of this paper is to show explicitly that numbers in this range are in fact compatible with observations of radio emission from star-forming galaxies. 

\subsection{Environmental Conditions}
\label{sec:Environment}

\subsubsection{Escape}
\label{sec:Escape}
The CR lifetime ($t_{\rm life}$) for both primary electrons and protons in equation (\ref{eqn:DiffusionLoss}) is uncertain, and probably varies from normal spirals like our own, where losses are mainly diffusive \citep[e.g.,][]{Longair04}, to dense starbursts like M82 and Arp 220, where CRs are likely advected in a large-scale galactic wind \citep[e.g.,][]{Seaquist91}.

We use a prescription for diffusive losses motivated by observations of beryllium isotope ratios (``CR clocks'') at the Solar Circle, which suggest that  the confinement timescale for CR protons with $E \ga 3\ \GeV$ is \citep[e.g.,][]{GarciaMunoz77,Engelmann90,Connell98,Webber03}
\begin{equation}
\label{eqn:StandardCRLife}
t_{\rm diff}(E) = 26~\Myr \left(\frac{E}{3~\GeV}\right)^{-1/2}.
\end{equation}
Identifying  $t_{\rm life}$ with a diffusion timescale on a physical scale of $\sim \kpc$ implies that CR protons diffuse in the ISM of the  Galaxy with a scattering mean free path of order $\sim \pc$.  Although the behavior of $t_{\rm life}$ at lower energies is uncertain
because of the effects of solar modulation (compare, e.g., Engelmann et al.~1990 \& Webber et al.~2003), we use equation (\ref{eqn:StandardCRLife}) for all fiducial models employing diffusive losses.  Variations to the diffusion constant are considered in Appendices~\ref{sec:FastEscape}-\ref{sec:MultipleEffects}.

There is ample evidence for large-scale mass-loaded winds in starburst galaxies with $\Sigma_g \gtrsim 0.05~\gcm2$ \citep{Heckman00,Heckman03}.  These winds can advect CRs out of their host galaxies on a short timescale with respect to equation (\ref{eqn:StandardCRLife}), thus affecting both the predicted emission and the overall equilibrium energy density of CRs. For this reason, we consider in Section\ \ref{sec:Winds} models of starburst galaxies with
\begin{equation}
\label{eqn:WindLife}
t_{\rm wind} = \frac{h}{v_{\rm wind}} \approx 300~\kyr~(\Sigma_g > 0.05~\gcm2),
\end{equation}
where we have taken $h = 100~\pc$, a wind speed of $v_{\rm wind} = 300~\kms$, and a cutoff between starburst and non-starburst galaxies of 
$\Sigma_g > 0.05~\gcm2$ as reference values.  The combined escape time in equation (\ref{eqn:DiffusionLoss}) is then given by $t_{\rm life}^{-1} = t_{\rm diff}^{-1} + t_{\rm wind}^{-1}$.

\subsubsection{Scale Height}
\label{sec:Height}

In our one-zone models, the galaxy scale height represents the volume in which the CRs are confined for $t_{\rm life}$ (eq.~\ref{eqn:DiffusionLoss}).  Because we specify the properties of galaxies by their surface density, $h$ is also important in determining the average gas density seen by CRs, which, in turn, is important for bremsstrahlung, ionization, and pion losses (see Section\ \ref{section:density}).
 
We adopt $h=1$\,kpc for normal galaxies ($\Sigma_g < 0.05~\gcm2$), although we consider several other scale heights in Appendix~\ref{sec:ScaleHeight}.  There are several relevant scale heights which are not identical, and we must choose one for our one-zone model.  CRs are injected in the gas disk ($h \approx 100\ \pc$), but we do not use this scale height since the CRs diffuse and emit synchrotron outside the gas disk.  The observed beryllium isotope ratios, as interpreted by CR diffusion models imply that the CRs have a scale height of 2 - 5 kpc in normal galaxies \citep{Lukasiak94,Webber98,Strong00}.  The magnetic field scale heights of normal galaxies are also several kpc \citep{Han94,Beck09}.  Finally, radio emission in most normal galaxies come from \emph{two} disks: a thin disk with $h \approx 0.3\ \kpc$ and a thick disk with $h \approx 2\ \kpc$ \citep[e.g.,][]{Beuermann85,Dumke98,Heesen09}.  On average, the thin and thick disks emit the same radio power at $\sim 1.4\ \GHz$, and one component fits of the radio emission of normal galaxies generally find a one-component scale height of $\sim 1 - 1.5\ \kpc$ \citep{Dumke95,Dumke00,Krause06}.  

Since we are most interested in the \emph{radio emission} of normal galaxies to explain the FRC, we use the value of $h = 1\ \kpc$ from the one component fits.  However, this one-zone approach does not capture all the relevant physics of the CRs.  In particular, the escape time in Equation~(\ref{eqn:StandardCRLife}) applies to the entire CR halo.  Escape from the radio-emitting regions is likely quicker and is probably underestimated in our models.  Conversely, a variant on our fiducial model with large $h$ in Appendix~\ref{sec:ScaleHeight} probably overestimates the synchrotron losses in normal galaxies, since it does not account for the lower magnetic field strengths far from the midplane.  Note that $h=1$\,kpc implies a vertical diffusion constant of $D_z \approx 7 \times 10^{27} \DiffUnits (E / \GeV)^{1/2}$ (for 1.4 GHz emitting electrons in the Milky Way, $D_z \approx 1.3 \times 10^{28}~\DiffUnits$; compare with the values in \citealt{Dahlem95} and \citealt{Ptuskin98}).  

Starburst galaxies ($\Sigma_g \ge 0.05~\gcm2$) are considerably more compact, both in terms of their star forming regions and in terms of their CR confinement zone, and for them we adopt $h=100$\,pc \citep[e.g.,][]{Downes98}.  

\subsubsection{ISM Density}
\label{section:density}
\label{sec:Density}

Estimates from beryllium isotopes imply that the average density experienced by CRs is about one fifth to one tenth that of the Galactic disk.  This may be because the ISM is clumpy and the CRs avoid the clumps, or because the CRs spend significant time in the low-density Galactic halo.  

CRs do not necessarily travel through gas with the mean ISM density.  The actual average density CRs experience depends upon the injection and propagation of the CRs, which depends on the small-scale ISM structure in galaxies and starbursts.  For example, most of the volume of the ISM in galaxies is low density material, and we can imagine the CRs are injected into this low density phase and lose their energy before encountering high density clumps.  Then the density experienced by CRs is lower than the average gas density.  Conversely, we can imagine that CRs are preferentially injected into high density clumps, and are confined there by magnetic fields in the clumps, in which case, the CRs experience a higher density than the average density of the galaxy or starburst.  

For this reason we include a parameter $f$, to account for these unknown propagation and injection effects, defined by 
\begin{equation}
\label{eqn:fDef}
n_{\rm eff} = f\mean{n},
\end{equation} that measures the effective density ``seen'' by CRs ($n_{\rm eff}$) with respect to the average density of the CR confinement volume, $\mean{n} = \Sigma_g/(2h)$.  For $f>1$ or $f<1$, the CRs traverse over- or under-dense material compared to $\mean{n}$, respectively.  Note that we are defining $f$ with respect to the CR confinement volume and \emph{not} the gas disk. 

The primary importance of the parameter $f$ is in determining the importance of bremsstrahlung and ionization losses for CR electrons and positrons, and of pion production from inelastic proton-proton collisions. 

Note that even though both the star formation rate and the magnetic fields of galaxies are taken to depend on the surface density (see Section~\ref{sec:MagneticField}), they are assumed to be independent of $f$.  This means that the radiation energy density and the magnetic field the CRs experience are assumed to be average, while the CRs are allowed to traverse through underdense or overdense material.  Although we allow ourselves this freedom in the modeling, it turns out that the models with $f \approx 1$ are most consistent with observations in the Milky Way.  For example, our adopted gas surface density at the Solar Circle, $\Sigma_g = 2.5 \times 10^{-3}~\gcm2$ \citep{Boulares90}, and scale height $h = 1~\kpc$ imply an average number density of $\mean{n} = 0.24~\cm3$.  Since the CRs are inferred to travel through material of density $n_{\rm eff} \approx 0.2 - 0.5 \,\cm3$ \citep[e.g.,][]{Connell98,Schlickeiser02}, this implies that $f \approx 1$.  

\subsubsection{Interstellar Radiation Field}
\label{sec:RadiationField}

The interstellar radiation field is important for determining the IC losses for CR electrons and positrons.   The primary contributions to the interstellar radiation field are starlight and the cosmic microwave background (CMB).  The latter is particularly important for low surface brightness galaxies where it dominates starlight.  Both sources of radiation are included in all models.

When the galaxy is optically-thin to the re-radiated FIR emission from young stars, then the energy density in starlight, which dictates the IC cooling timescale, is simply

\begin{eqnarray}
U_{\rm ph,\star} & = & F_\star / c = \varepsilon \Sigma_{\rm SFR} c\\
\label{eqn:UphThin}
& = & 3 \times 10^{-9} \left(\frac{\Sigma_g}{\gcm2}\right)^{1.4} \ergcm3,
\end{eqnarray}
where the surface density of star formation $\Sigma_{\rm SFR}$ is connected to the average gas surface density by the Schmidt law.  For large gas surface densities ($\Sigma_g\gtrsim0.1-1$\,g cm$^{-2}$) galaxies become optically thick to the reradiated FIR emission and 
\begin{eqnarray}
U_{\rm ph,\star} & = & (\tau_{\rm FIR} + 1) F_\star / c = (\tau_{\rm FIR} + 1) \varepsilon \Sigma_{\rm SFR} c\\
& = & 3 \times 10^{-9} (\tau_{\rm FIR} + 1) \left(\frac{\Sigma_g}{\gcm2}\right)^{1.4} \ergcm3,
\end{eqnarray}
where $\tau_{\rm FIR} = \kappa_{\rm FIR} \Sigma_g / 2$ is the vertical optical depth, and $\kappa_{\rm FIR}$ is the Rosseland mean dust opacity.  For parameters typical of starbursts and ULIRGs, $\kappa_{\rm FIR} \approx 1-10$\,cm$^2$ g$^{-1}$ for Galactic dust-to-gas ratio and solar metallicity \citep{Semenov03}.  For our standard models (Section~\ref{sec:StandardModel}), we assume that the CRs are always in optically thin regions, so that equation (\ref{eqn:UphThin}) holds.  However, we discuss models with $\tau_{\rm FIR} > 0$ in Section~\ref{sec:OpticalThickFIR}.

\subsubsection{Magnetic Fields}
\label{sec:MagneticField}

A primary motivation for this work is to determine how the average magnetic energy density of galaxies scales from normal galaxies like our own to dense ULIRGs like Arp 220.  Observations of Zeeman splitting in ULIRGs supports a relatively strong scaling of magnetic field strength with gas surface density \citep{Robinshaw08}.  To test a suite of models for consistency with observations, we parametrize the global average magnetic field of galaxies as
\begin{equation}
\label{eqn:StandardB}
B = 6\,\left(\frac{\Sigma_g}{0.0025~\gcm2}\right)^a\,\,\muGauss,
\label{bfield}
\end{equation}
where $a$ is determined from comparing with the FRC, and where the normalization has been chosen to match fiducial numbers at the Solar Circle \citep[as in][]{Boulares90,Strong00,Beck01}.  The magnetic field energy density is then just $U_B = B^2 / (8 \pi)$.  The $\Sigma_g^a$ dependence is motivated by the Parker instability: the magnetic energy density cannot exceed the gas disk midplane pressure $\pi G \Sigma_g^2$, or else the magnetic field will buoy up out of the disk and escape \citep{Parker66}.  A natural scaling for $B$ given the Parker limit would be $B \propto \Sigma_g$.  The $\Sigma_g^a$ scaling also arises if the magnetic field is in equipartition with the starlight, because the Schmidt law implies that $U_{\rm ph} \propto \Sigma_g^{1.4}$; in this case $a = 0.7$.  We consider $0.4 \le a \le 1.0$.    We assume that the magnetic field is constant within the confinement volume of scale height $h$, a reasonable assumption based on the observed radio halos of galaxies and Galactic pulsar rotation measures \citep{Han94}.  We also consider two other parameterizations of the magnetic field, $B \propto \rho^a$ and $U_B = U_{\rm ph}$, in Section~\ref{sec:Basn} and Section~\ref{sec:UBeqUph}, respectively.

\subsection{Observables and Constraints}
\label{sec:Constraints}
We use two broad conditions to select successful models.  We first ask if the model satisfies the FRC.  However, since the FRC alone does not necessarily demand CR protons at all, a second constraint is needed to fix the overall CR proton normalization.  We consider two sets of constraints on the protons, either using Earth-based measurements of CRs, or observations of the entire Milky Way.  

\begin{enumerate}
\item\emph{Reproduce the FIR-radio correlation.}
The nonthermal radio luminosity is calculated directly from our synchrotron spectrum as $\nu \epsilon_{\nu}$ at $\nu = 1.4~\GHz$.  We do not include the thermal free-free contribution to the radio luminosity; however, the thermal radio luminosity of most galaxies is typically small at GHz frequencies.  Nor do we consider the effects of free-free absorption.  The total infrared (TIR) luminosity\footnote{While some of the light absorbed by dust is emitted as far-infrared (40 - 120 $\mu$m), some is also emitted in near or mid infrared.  For simplicity, we assume that TIR light is directly proportional to the FIR emission.  For starburst galaxies, $L_{\rm TIR} \approx 1.75 L_{\rm FIR}$ \citep{Calzetti00}, which we apply to every galaxy for simplicity.  This correction has been applied to the reported $L_{\rm FIR}/L_{\rm radio}$ in \citet{Yun01} to get our quoted observed $L_{\rm TIR}/L_{\rm radio}$.  \citet{Bell03} reports a similar $L_{\rm TIR} / L_{\rm FIR} \approx 2$ for galaxies on the FRC.} is $\epsilon_{\rm ph} [1 - (1 - \exp(-\tau_{\rm UV})) / \tau_{\rm UV}]$, with the UV optical depth $\tau_{\rm UV}$ through the entire disk calculated as $\kappa_{\rm UV} \Sigma_g$.  We adopt a UV opacity of $500~\cmg21$, which is roughly appropriate at wavelengths of $\sim 1000~\textrm{\AA}$ ($T_{\rm eff} \approx 30,000~\Kelv$) and Galactic metallicity and dust-to-gas ratios (\citealt{Li01}; \citealt{Bell03} use $\kappa_{\rm UV} = 190~\gcm2$, using a smaller dust-to-gas ratio and $1500~\textrm{\AA}$).  Then, $L_{\rm TIR}/L_{\rm radio}$ is simply the ratio of these luminosities, and can easily be converted into $q_{\rm FIR}$, an observable quantity we calculate\footnote{Our version of this equation divides our calculated $L_{\rm TIR}$ by $L_{\rm TIR}/L_{\rm FIR} \approx 1.75$ \citep{Calzetti00} to get to the true, observed FIR emission.  See \citet{Helou85} for the usual definition of $q$.} as
\begin{equation}
q_{\rm FIR} = \log_{10}\left(\frac{L_{\rm TIR}}{L_{\rm radio}}\right) - 3.67
\end{equation}
and defined in \citet{Helou85}.  The normalization of the FRC is $L_{\rm TIR}/L_{\rm radio} = 9 \times 10^5$ \citep{Yun01}, which we match by adjusting $\xi$ appropriately, therefore fixing the primary CR electron injection rate in galaxies (Section~\ref{sec:PrimaryInjection}).

Once we have the ratio $L_{\rm TIR}/L_{\rm radio}$, our primary constraint is that we require a linear FRC to exist.  

\emph{We require that \begin{equation}\frac{{\rm max}(L_{\rm TIR}/L_{\rm radio})}{{\rm min}(L_{\rm TIR}/L_{\rm radio})} \le 2. \label{eqn:FRCExists}\end{equation}}

\item \emph{Fix the proton normalization.}
We then use two sets of constraints to fix the proton normalization, the local constraints and the integrated constraints.  Each is considered independently for each model.  For simplicity, the proton normalization is assumed to be constant across the entire range of star-forming galaxies.

The ``local'' set of constraints is based on in-situ measurements of CRs at Earth.  These are: 
\begin{enumerate}
\item For each electron energy, we calculate the ratio of the CR positron number density to the total number density of positrons and electrons at \GeV~energies.  Below \GeV~energies, solar modulation of CRs can affect the observed CR spectrum.  Above $\sim 1$ \GeV, the observed positron flux exceeds the predicted flux even in detailed models (see \citealt{Moskalenko98,Beatty04,Adriani08}; but see \citealt{Delahaye08}).  The observed value of $e^+ / (e^+ + e^-)$ is 0.1 at \GeV~energies \citep[e.g.,][]{Schlickeiser02,Adriani08}.  

\emph{We require as a local constraint that~$0.05 \le e^+ / (e^+ + e^-) \le 0.2$ at 1~\GeV~when $\Sigma_g = 0.0025~{\rm \gcm2}$.}

\item We also compute the ratio of the proton number flux and electron number flux at \GeV~energies.  The ratio is observed to be $p/e \approx 100$ at Earth at energies of a few GeV \citep[e.g.,][]{Ginzburg76,Schlickeiser02}.  This value is also inferred from SN remnants, which are believed to accelerate CRs \citep{Warren05}.  

\emph{We require as a local constraint that~$50 \le p/e \le 200$ at 10~\GeV~when $\Sigma_g = 0.0025~{\rm \gcm2}$.}

\end{enumerate}

As an alternative way to find the CR proton normalization, we considered a separate ``integrated'' constraint for the entire Milky Way galaxy using an average $\Sigma_g$ inferred from the Galactic scale radius and star formation rate.  Our purpose was to assess the possibility that the Earth is not in a representative location of the Galaxy; for example, it sits in the Local Bubble.  

\begin{enumerate}
\item We calculate the gamma-ray luminosity of the Galaxy from $\pi^0$ decay.  We approximate the Milky Way as a uniform disk with $R = 4~\kpc$, and a surface density of $\Sigma_g = 0.01~\gcm2$ derived from the Schmidt law (Section~\ref{sec:Environment}) and the Milky Way luminosity, $L_{\star} \approx 2 \times 10^{10} L_{\sun}$ (\citealt{Freudenreich98}; similar results are obtained by using the starlight radiation field in \citealt{Strong00}, or the SN rate in \citealt{Ferriere01}).  \citet{Strong00} calculate the total $\pi^0$ gamma-ray luminosity to be $L_{\pi_0} \approx 2 \times 10^{39} \ergps$. 

\emph{We require as the integrated constraint that $1 \times 10^{39} {\rm \ergps} \le L_{\pi_0} \le 4 \times 10^{39} {\rm \ergps}$ when $\Sigma_g = 0.01~{\rm \gcm2}$.}

\end{enumerate}
\end{enumerate}

\emph{Additional checks:} As an added check, we have the observed CR spectrum at Earth.  At high energies ($\gamma \gg 1$), the observed CR electrons have \citep[e.g.,][]{Longair04}
\begin{equation}
\frac{dI_e}{dE} = \frac{c N_e(E)}{4\pi} = 0.07 \left(\frac{E}{\GeV}\right)^{-3.3}~\textrm{cm}^{-2}~\textrm{s}^{-1}~\textrm{sr}^{-1}~\GeV^{-1}
\end{equation}
and the observed CR protons have \citep[e.g.,][]{Mori97,Menn00,AMS02}
\begin{equation}
\frac{dI_p}{dE} = \frac{c N_p(E)}{4\pi} \approx 1.5 \left(\frac{E}{\GeV}\right)^{-2.7}~\textrm{cm}^{-2}~\textrm{s}^{-1}~\textrm{sr}^{-1}~\GeV^{-1}.
\end{equation}
The predicted CR spectrum does not determine whether a model was considered formally ``successful'', but it was used to select among the adequate models for the best standard set of parameters.

Although we did not use it directly, we also calculate the spectral slopes at 1.4 GHz from the radio synchrotron spectrum as a sanity check.  These include the instantaneous spectral slope $\alpha_{1.4}$ as well as the spectral slopes to 4.8 GHz ($\alpha_{1.4}^{4.8}$) and 8.4 GHz ($\alpha_{1.4}^{8.4})$.  Unless otherwise stated, $\alpha$ refers to $\alpha_{1.4}^{4.8}$, the spectral slope from 1.4 GHz to 4.8 GHz.  Typical values of $\alpha$ are $0.7 - 0.8$.  As a constraint, $\alpha$ can be very sensitive to minor details in the model; we note that a difference of $0.2$ in $\alpha$ results in only a $60 \%$ difference in the specific flux after one decade in frequency, and we are mainly concerned with factor of $2$ accuracy in our models.  \citet{Lisenfeld00} have argued that $\alpha$ is decreased by $\sim 0.1$ by SN remnants within galaxies, so our value of $\alpha$ is uncertain at that level.  We also do not include free-free emission, which can flatten the spectral slope, especially in low surface density galaxies.  In ULIRGs like Arp 220, the observed $\alpha$ is typically $\sim 0.5$ \citep{Clemens08}, but radio emission in these galaxies may suffer free-free absorption which flattens the spectrum; the unabsorbed synchrotron spectrum $\alpha$ may be as high as $0.7$ \citep{Condon91}.  To some extent, a small to moderate difference in $\alpha$ from its observed value can be adjusted by altering $p$, since decreasing $p$ by $0.1$ generally decreases $\alpha$ by $0.05$, and $p$ often is not well constrained in the considered range $2 - 2.6$.  Given these uncertainties, caveats, and sensitivities in $\alpha$, and given the vast range of galaxies and starbursts we are considering, and the simplified parameterizations we are using, we do not impose any direct constraint on $\alpha$.  Of course, models of individual galaxies should and do account for $\alpha$ when they model the radio emission.

Throughout this work, we assume that the local values of the proton normalization and propagation -- in particular, $\delta$, $\eta$, and $f$ -- are the same for both normal galaxies and starbursts.  We use this assumption for simplicity, and to keep the number of free parameters reasonable.  In practice, the CR acceleration efficiency and the proton-to-electron ratio may change somewhat from normal galaxies and starbursts, but we do not consider small variations necessary for a basic understanding of the FRC.  More detailed models of individual systems can and do take these changes into account, and we refer readers to these models if they wish to understand starburst galaxies in detail.  It is also conceivable that $f$ changes dramatically from normal galaxies to starbursts.  Again, we do not consider this possibility in this paper, although we will explore the consequences of very low $f$ applying to only starbursts in a future paper.

\section{Review of Physical Effects of Parameters}
\label{sec:EffectsReview}

To search for models that satisfy the observational constraints listed in Section\ \ref{sec:Constraints}, our grid of models spanned values of $a$ (eq.~\ref{eqn:StandardB}), $f$ (eq.~\ref{eqn:fDef}), $\xi$ and $\delta$ (eqs.~\ref{eqn:CReNorm} and \ref{eqn:CRpNorm}), and $p$ (Section \ref{sec:PrimaryInjection}).  For a listing of these parameters of the model, see Table~\ref{table:SymbolList}.  As background for interpreting our results in \S\ref{sec:Results}, we briefly review the effects of these quantities on observables.

\subsection{Injection Parameters: $\xi$, $\eta$, $\delta$, and $p$}
\label{sec:xiEffects}

The parameter $\xi$ is the normalization of the injected primary CR electron spectrum, and with $\delta=\eta/\xi$, the injected CR proton spectrum normalization (eqs.~\ref{eqn:CReNorm} \& \ref{eqn:CRpNorm}).  Changes in $\xi$ do not affect the shape of any of the equilibrium CR spectra.  For fixed $\delta$, larger $\xi$ linearly increases the CR luminosity and energy density within galaxies, and thus --- for fixed galaxy parameters --- the luminosity of CRs in all wavebands, including the radio, neutrino, and gamma-ray luminosities.  

An increase in $\delta$ at fixed $\xi$ raises the number of secondaries from protons, the ratios $e^+ / (e^+ + e^-)$ and $e^-_{\rm sec} / e^-$, and the luminosity from pion decay, $L_{\pi}$.  

Of note is the ratio of injected protons to electrons at relativistic energies.  Suppose the electrons are injected with a spectrum $Q_e (E) = C_e E^{-p}$ and the protons are injected with a spectrum $Q_p (E) = C_p E^{-p}$.  Then, given our normalization conditions $\epsilon_{\rm CR,\,e} = \int_{m_e c^2}^{\gamma_{\rm max} m_e c^2} C_e K E^{-p} dE$ and $\epsilon_{\rm CR,\,p} = \int_{m_p c^2}^{\gamma_{\rm max} m_p c^2} C_p K E^{-p} dE$, where $K$ is the kinetic energy (see Section~\ref{sec:PrimaryInjection}), it can be shown that 
\begin{equation}
\label{eqn:DeltaTilde}
\tilde\delta \equiv \frac{C_p}{C_e} = \delta \left(\frac{m_p}{m_e}\right)^{p - 2}.
\end{equation}
The quantity $\tilde\delta$ represents the proton to electron ratio at high energies ($m_p c^2 \ll E \ll \gamma_{\rm max} m_e c^2$) if there were no escape, energy losses, or secondary production.  Note that it is not generally equal to $\delta$, since $\delta$ is largely dependent on the shape of the spectrum at low energies.  Our injection spectra go as $E^{-p}$, where $E$ is the total energy: the electron spectra stretch down to $m_e c^2$ while the proton spectra only extend down to $m_p c^2$.  For steep spectra ($p > 2$), the low-energy particles receive most of the energy, so that electrons with $E < m_p c^2$ act as a hidden reservoir of energy.\footnote{Conversely, the proton spectrum extends to a maximum energy of $\gamma_{\rm max} m_p c^2$, much greater than the maximum energy of the electrons; for shallow spectra ($p < 2$), the reservoir of energy in these high energy protons would lower $C_p / C_e$ at $E < \gamma_{\rm max} m_e c^2$.}  This reservoir is unconstrained because the observables we use do not constrain the shape of the CR spectra at low energies (see Section~\ref{sec:Constraints}).  This follows from the fact that the FRC is observed at GHz frequencies, implying electron energies of order 100 MeV to 10 GeV (eq.~\ref{eqn:nuCSynch}).  Thus, the actual quantity we constrain is $\tilde\delta$.  Note that the relationship between $C_p / C_e$ and $\delta$ would be different for another spectrum, such as $K^{-p}$ or $\gamma^{-p}$.

\label{sec:SpectralSlopeEffects}
The spectral slope $p$ of the injected CRs in part controls the final, propagated spectral slope ${\cal P} \equiv d\log N(E) / d\log E$.  The spectral slope, in turn, determines how much the secondary particles are diluted.  Protons at energy $E$ produce secondary electrons and positrons of energy $E' < E$; a steeper primary spectrum increases the number of primaries at these lower energies compared to the proton energy $E$.  Therefore, a larger $p$ (and thus a bigger ${\cal P}$ for primary electrons) implies a smaller $e^+ / (e^+ + e^-)$ and $e^-_{\rm sec} /e^-$. This dilution implies that even in the limit of full proton calorimetry, primary electrons may be more important than secondaries.  Similarly, the secondary fraction is not a good measure of proton calorimetry in itself.  For our standard model, though, we find that in proton calorimeters, secondary electrons and positrons outnumber the primary electrons $\sim 4$-1.

\subsection{Magnetic Field}
\label{sec:BEffects}

The magnetic field strength affects the CR spectra in several ways:

1.~It determines the importance of synchrotron cooling relative to other radiative and escape losses.  The synchrotron cooling timescale for CR electrons and positrons emitting at frequency $\nu_{\rm GHz}=\nu/{\rm GHz}$ is
\begin{equation}
t_{\rm synch}\approx 4.5\times10^7\,B_{10}^{-3/2}\nu_{\rm GHz}^{-1/2}\,\,{\rm yr},
\label{eqn:tSynch}
\end{equation}
where $B_{10}=B/10$\,$\mu$G. For normal galaxies, $t_{\rm synch}$ is comparable to, but somewhat longer than, the inferred diffusive escape timescale for the CR electrons producing GHz emission in normal galaxies (eq.~\ref{eqn:StandardCRLife}).  For the $\sim$\,mG (and larger) fields thought to exist in the densest starbursts, $t_{\rm synch}$ is shorter than even the advection timescale (eq.~\ref{eqn:WindLife}).
  
2.~The relative importance of synchrotron also affects the propagated equilibrium spectral slope ${\cal P}$ of electrons and positrons; stronger magnetic fields imply steeper final spectra (see Section~\ref{sec:SpectralSlopeEffects}).  In the limit that cooling dominates escape, and that synchrotron is the main form of cooling, the equilibrium spectral slope is ${\cal P}=1+p$.

3.~The magnetic field strength determines the critical synchrotron frequency ($\nu_C$) for electrons and positrons:
\begin{equation}
\label{eqn:nuCSynch}
\nu_C \approx 3.3 \left(\frac{\gamma}{10^4}\right)^2 B_{10}~\GHz.
\end{equation}
At a fixed observed frequency (such as 1.4\,GHz), a stronger magnetic field implies that we see lower energy electrons and positrons.

4.~When synchrotron cooling dominates over other cooling and escape losses, a stronger magnetic field lowers the equilibrium energy density of CR electrons and positrons, because of increased losses.  However, in this calorimeter limit, each electron and positron has a higher luminosity.  Therefore, $L_{\rm radio}$ approaches a maximum set by $\xi$, and is not affected by further increases in the magnetic field strength.  This effect is the essence of the original calorimeter theory.

All else being equal in our models of non-calorimetric galaxies, larger magnetic fields imply that a larger share of injected CR electron power is lost to synchrotron, because of the faster synchrotron cooling time.  In cases when synchrotron does not already dominate, increasing the magnetic field strength thus increases $L_{\rm radio}$ .  

Note that the magnetic field in our models is normalized to the local Solar Circle gas surface density (eq.~\ref{bfield}), so that changing $a$ has no effect on local Milky Way constraints discussed in Section\ \ref{sec:Constraints}.

\subsection{Effective Density}
\label{sec:DensityEffects}

The ISM density encountered by CR protons controls the production rate of secondary electrons and positrons, as well as gamma rays and high-energy neutrinos, from inelastic proton-proton collisions.  The proton lifetime to pion losses is 
\begin{equation}
t_{\pi} \approx 5\times10^7\,\,{\rm yr} \left(\frac{f\langle n\rangle}{\cm3}\right)^{-1},
\label{eqn:tPion}
\end{equation}
from \citet{Mannheim94} \citep[see also][]{Torres04}.  Higher $n_{\rm eff}=f\langle n\rangle$ (eq.~\ref{eqn:fDef}) means more secondaries, higher $L_{\rm radio}$, and higher gamma-ray and neutrino luminosities.  The secondary electrons and positrons raise the ratios $e^+ / (e^+ + e^-)$ and $e^-_{\rm sec} / e^-$, and they lower the equilibrium ratio $p/e$.  Additionally, if the ratio of primaries to secondaries changes with energy, then the combined spectral slope ${\cal P}$ for electrons and positrons can be altered, which affects the observed radio spectral slope (see Section~\ref{sec:SpectralSlopeEffects}).

The effective ISM density also determines the efficiency of bremsstrahlung and ionization losses for CR electrons and positrons, with higher densities making these processes more efficient.  The bremsstrahlung and ionization energy loss timescales are  
\begin{equation}
\label{eqn:tBrems}
t_{\rm brems}\approx 3.7\times10^7\,\,{\rm yr} \left(\frac{f\langle n\rangle}{\cm3}\right)^{-1},
\end{equation}
and
\begin{equation}
\label{eqn:tIon}
t_{\rm ion}\approx 2.1 \times 10^8\,B_{10}^{-1/2}\nu_{\rm GHz}^{1/2}\left(\frac{f\langle n\rangle}{\cm3}\right)^{-1}\,\,{\rm yr},
\end{equation}
respectively, where we have again scaled the energy dependence of $t_{\rm ion}$ for CR electrons and positrons emitting at GHz frequencies for comparison with $t_{\rm synch}$ (eq.~\ref{eqn:tSynch}).  Importantly, energy lost to bremsstrahlung and ionization is not radiated in the radio, so higher $f$ implies lower $L_{\rm radio}$ from these processes, all else being equal.  

The energy dependence of these cooling processes also flattens the propagated equilibrium electron and positron spectra (see the discussion after eq.~\ref{eqn:DiffusionLoss}; Section~\ref{sec:SpectralSlopeEffects}).  For example, when $t_{\rm synch} = t_{\rm brems}$ at some energy and all other losses are negligible, then ${\cal P} = p + 1/2$ and $\alpha = p/2 - 1/4$.  Similarly, when $t_{\rm synch} = t_{\rm ion}$ and there are no other losses, ${\cal P} = p$ and $\alpha = p/2 - 1/2$.

\subsection{The Schmidt Law and the Photon Energy Density}
\label{sec:UphEffects}
The energy density of photons, and thus the importance of IC losses for CR electrons and positrons in star-forming galaxies, is set by the slope and normalization of the Schmidt law.  The IC cooling timescale for CR electrons and positrons emitting radio synchrotron at frequency $\nu$ is 
\begin{equation}
\label{eqn:tIC}
t_{\rm IC}\approx 1.8\times10^8\,B_{10}^{1/2}\nu_{\rm GHz}^{-1/2}U_{\rm ph,\,-12}^{-1}\,{\rm yr}
\end{equation}
where $U_{\rm ph,\,-12} = U_{\rm ph}/10^{-12}$\,ergs cm$^{-3}$ is the photon energy density scaled to that for a typical star-forming galaxy.   For optically-thin galaxies obeying the Schmidt law of \citet{Kennicutt98} and ignoring the CMB, $F_{\star} / c = U_{\rm ph, \star} \propto \Sigma_{\rm SFR} \propto \Sigma_g^{1.4}$ (eq.~\ref{eqn:UphThin}).  Then, for fixed frequency, the IC lifetime therefore scales as $t_{\rm IC} \propto \Sigma_g^{a/2-1.4}$ if the magnetic field strength varies as $B \propto \Sigma_g^a$.  We do not consider variations on the Schmidt law \citep[e.g.,][]{Bouche07}, but their effects can be inferred from equation~\ref{eqn:tIC}: if $\Sigma_{\rm SFR}$ has a steeper increase with $\Sigma_g$, then $t_{\rm IC}$ will fall more rapidly with surface density.  In practice, the CMB will make IC losses more efficient in the lowest density galaxies, and any FIR opacity (Section~\ref{sec:RadiationField}) will make them more efficient in high-density starbursts.

Just as bremsstrahlung and ionization losses can reduce the share of energy left for synchrotron radiation, a greater photon energy density and IC power decreases $L_{\rm radio}$.  Unlike bremsstrahlung and ionization, though, IC losses produce a steep spectrum.  In the limit where they dominate other losses and escape, ${\cal P}=1+p$, as in the case of pure synchrotron cooling, and $\alpha = p / 2$.

\section{Results}
\label{sec:Results}

\subsection{Standard Model}
\label{sec:StandardModel}

\begin{figure*}
\includegraphics[width=18cm]{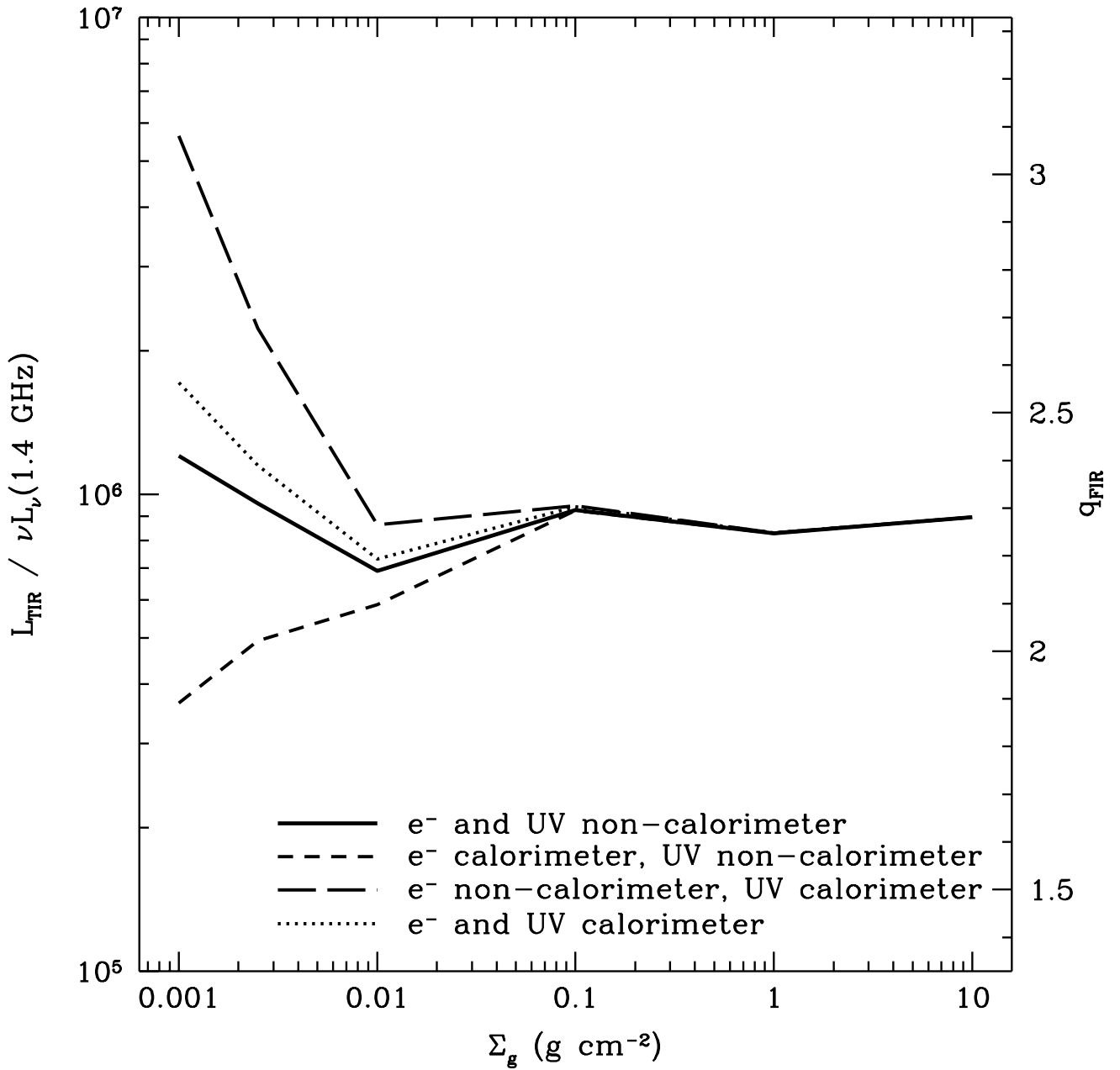}
\figcaption[figure]{The non-thermal FRC, as reproduced in our standard model ($p = 2.3$, $f = 1.5$, $a = 0.7$, $\delta = 5$, $\xi = 0.023$).  While low CR escape times and low UV optical depth on their own would break the correlation at low surface densities, the two effects cancel each other out, creating a largely linear FRC.\label{fig:Calorimetry}}
\end{figure*}

We adopt $p = 2.3$, $f = 1.5$, $a = 0.7$, $\tilde\delta = 48$ ($\delta = 5.0$), and $\xi = 0.023$ as our fiducial model.  This model reproduces the FRC, as seen in Figure~\ref{fig:Calorimetry} (solid line).  In this particular model, we require $\xi = 0.023$ to match the normalization of the FRC.  The ratio of FIR to $1.4~\GHz$ luminosities varies by only 1.7 over the entire range of $\Sigma_g$, and shows no obvious trend.  However, the scatter appears to be concentrated at the low-$\Sigma_g$ end of the FRC, with $L_{\rm TIR}/L_{\rm radio}$ varying by less than 12\% in the starbursts in this model.

The standard model also satisfies both local and integrated constraints on the proton normalization, as well as the observed CR spectrum. Our positron ratio at 1 GeV, $e^+ / (e^+ + e^-) = 0.10$, and proton-to-electron ratio at 10 GeV, $p/e = 82$, are good matches to the observed values.  The Milky Way $\gamma$-ray luminosity in this model, $2.0 \times 10^{39}~\ergps$, is also a good match to the value of \citet{Strong00}.  

\begin{figure}
\centerline{\includegraphics[width=8cm]{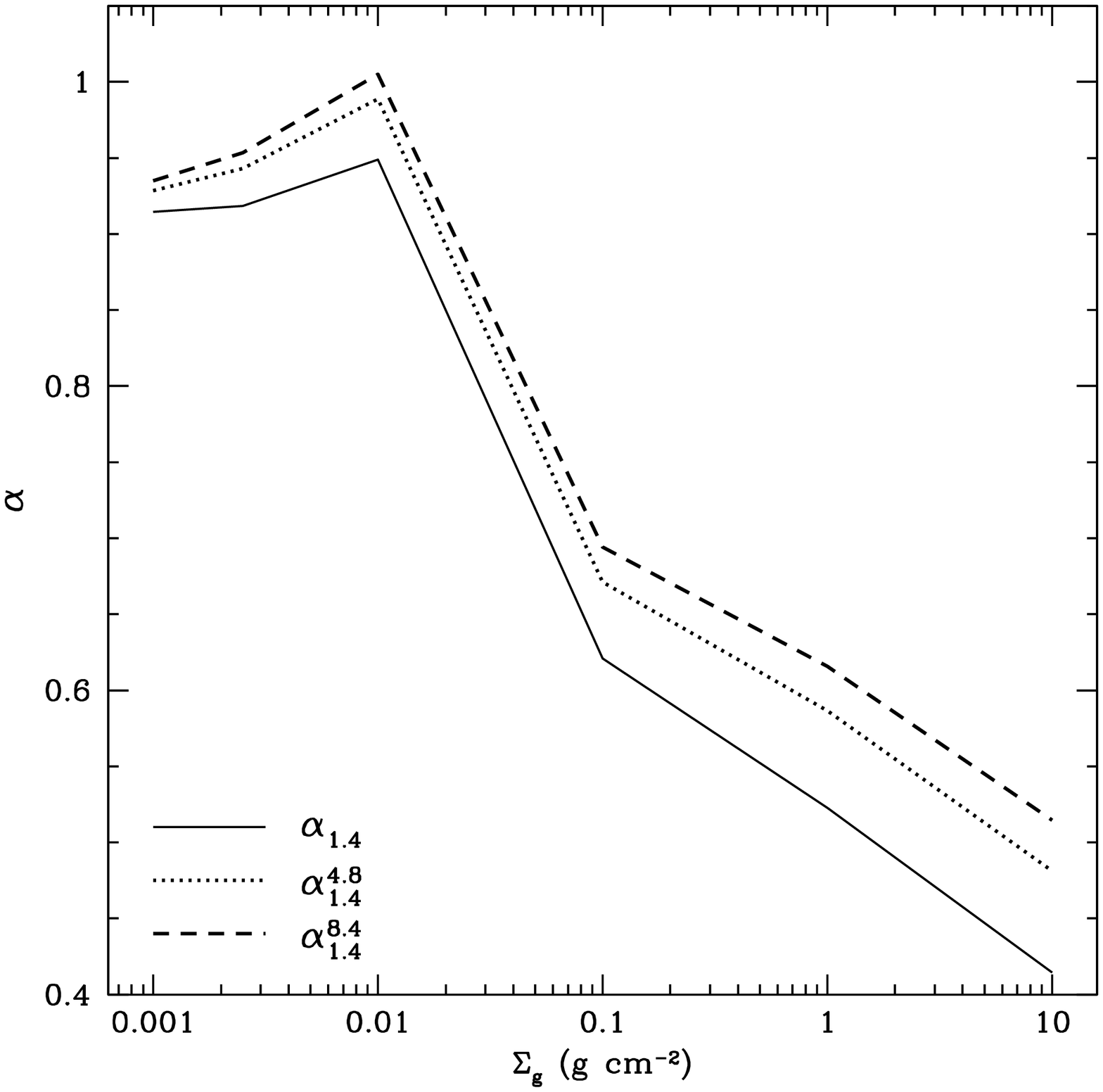}}
\figcaption[figure]{Spectral slope as a function of the gas surface density.  In this plot, $\alpha_{1.4}$ is the instantaneous spectral slope, $d\log~F_{\nu} / d\log~\nu$, at 1.4 GHz.  Elsewhere in the paper, $\alpha$ is the observable $\alpha_{1.4}^{4.8}$.  The parameters have their fiducial values.  For our standard $p = 2.3$, strong cooling by synchrotron and IC alone would imply that $\alpha = 1.15$.  Instead the spectral indices are significantly flatter, especially in the strong cooling calorimeter limit at high $\Sigma_g$, as a result of ionization and bremsstrahlung losses.\label{fig:Alpha}}
\end{figure}

The predicted proton spectrum at Earth in this model is $91\% - 118\%$ of its observed value at 1, 10, and 100 GeV, implying that $p$ is well matched to the Galactic CR spectrum.  Similarly, the CR electron flux at Earth is 122\% of its observed value at 10 GeV.  The least satisfactory aspect of this model is the predicted spectral slope $\alpha$ for Milky Way-type galaxies ($\alpha \approx 0.9 - 1.0$), which is somewhat too high.  Our results are, however, reasonable for starbursts ($\alpha \approx 0.5 - 0.7$; see Figure~\ref{fig:Alpha}). 

We emphasize that the parameters of our standard model are adequate for \emph{all} star-forming galaxies on the FRC.  We discuss the many competing effects that yield the FRC in Section~\ref{sec:FRCCauses}. 

\subsection{Degeneracy in the Standard Model}
Our local set of constraints (Section~\ref{sec:Constraints}) narrow down the allowed parameter space considerably.  The models that survive have $\tilde\delta \approx 34 - 100$ and $a = 0.6 - 0.7$.  To get the correct normalization of the FRC, we must set $0.019 \le \xi \le 0.027$ when $p = 2.3$, so that $0.097 \le \eta \le 0.22$.  Flatter injection spectra generally have lower $\xi$ (down to $0.006$ for $p = 2.0$) and higher $\eta$ (reaching $0.28$ for $p = 2.1$), while steeper injection spectra generally have higher $\xi$ (up to $0.18$ for $p = 2.6$) and lower $\eta$ (as low as $0.09$, which occurs when $p = 2.4$).  However, $\alpha_{1.4}^{4.8}$ is somewhat high when $p \ga 2.2$ ($0.9 - 1$ predicted compared to $0.7 - 0.9$ observed) for normal galaxies, but is close to observed values for $p \la 2.2$ ($0.8 - 0.9$ predicted).  The spectral slope is sufficiently low ($\alpha_{1.4}^{4.8} \approx 0.5 - 0.7$ predicted) for starbursts. 

The integrated Milky Way $\gamma$-ray luminosity from $\pi^0$ decay provides similar, but somewhat weaker constraints, favoring lower $\tilde\delta$.  At low $p = 2.0$, models with $10 \la \delta \la 50$ are selected by the $\gamma$-ray luminosity.  Higher $p$ models continue to work so long as $\delta$ decreases, because the normalization depends on the spectrum at very low energies as discussed in Section~\ref{sec:xiEffects}.  We can take the normalization into account by comparing $\tilde\delta$ (eq. \ref{eqn:DeltaTilde}), and we find that the allowed $\tilde\delta$ ($10 \la \tilde\delta \la 100$) slowly increases with $p$ (from roughly $25$ at $p = 2.1$ to $91$ at $p = 2.6$ when $f = 1.0$) and decreases with $f$ (from roughly 60 at $p = 2.2$, $f = 1.0$ to 50 at $p = 2.2$, $f = 2.0$).  Even $p = 2.6$ models predict an FRC and the correct $\pi^0$ luminosity of the Milky Way; we would need to take into account either the observed $\alpha$ in normal galaxies or the local observed CR spectral slope to further constrain $p$ in our standard model.  For example, $p = 2.6$ works when $\tilde\delta \approx 90$ ($\delta \approx 1$), and $\xi \approx 0.17$.  As discussed in Section~\ref{sec:SpectralSlopeEffects}, a higher $p$ dilutes secondaries and lowers the fraction of electrons that are secondaries.  Therefore, the secondaries contribute a smaller fraction of the radio luminosity and are less likely to break the FRC as galaxies become proton calorimeters at high density.  The FRC can then tolerate a higher secondary production rate (and ultimately higher $\tilde\delta$) for high $p$.  A higher $\xi$ is needed, though, since more of the energy goes into unobserved low energy electrons.  These high $p$ solutions also produce steep synchrotron spectra with $\alpha \approx 1.1 - 1.2$ in normal galaxies, and can be constrained by the spectral slopes.

More broadly, we also consider variations on our usual parametrization, such as lifetimes including advection, different scale heights, and FIR optical depths.  Many of these variations are inconsistent with the constraints in Section~\ref{sec:Constraints}.  However, those that do satisfy the constraints had similar values for $\xi$, $\tilde\delta$, $p$, and $a$ as the fiducial model.  We describe these variants in detail in Appendix~\ref{sec:Variants}.

\subsection{General Features of the Particle Spectra}
\label{sec:ParticleFeatures}
\begin{figure*}
\centerline{\includegraphics[width=9cm]{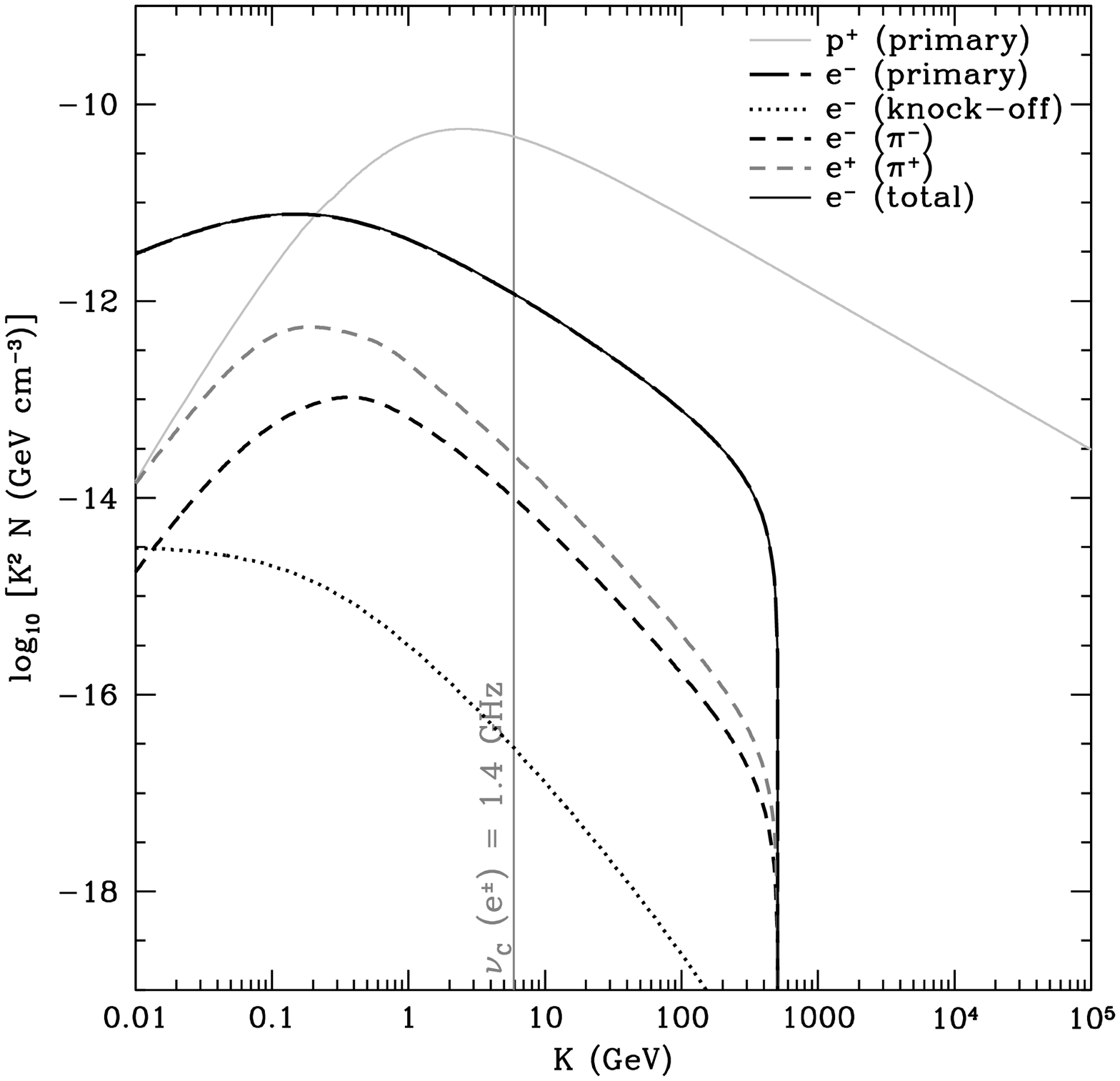}\includegraphics[width=9cm]{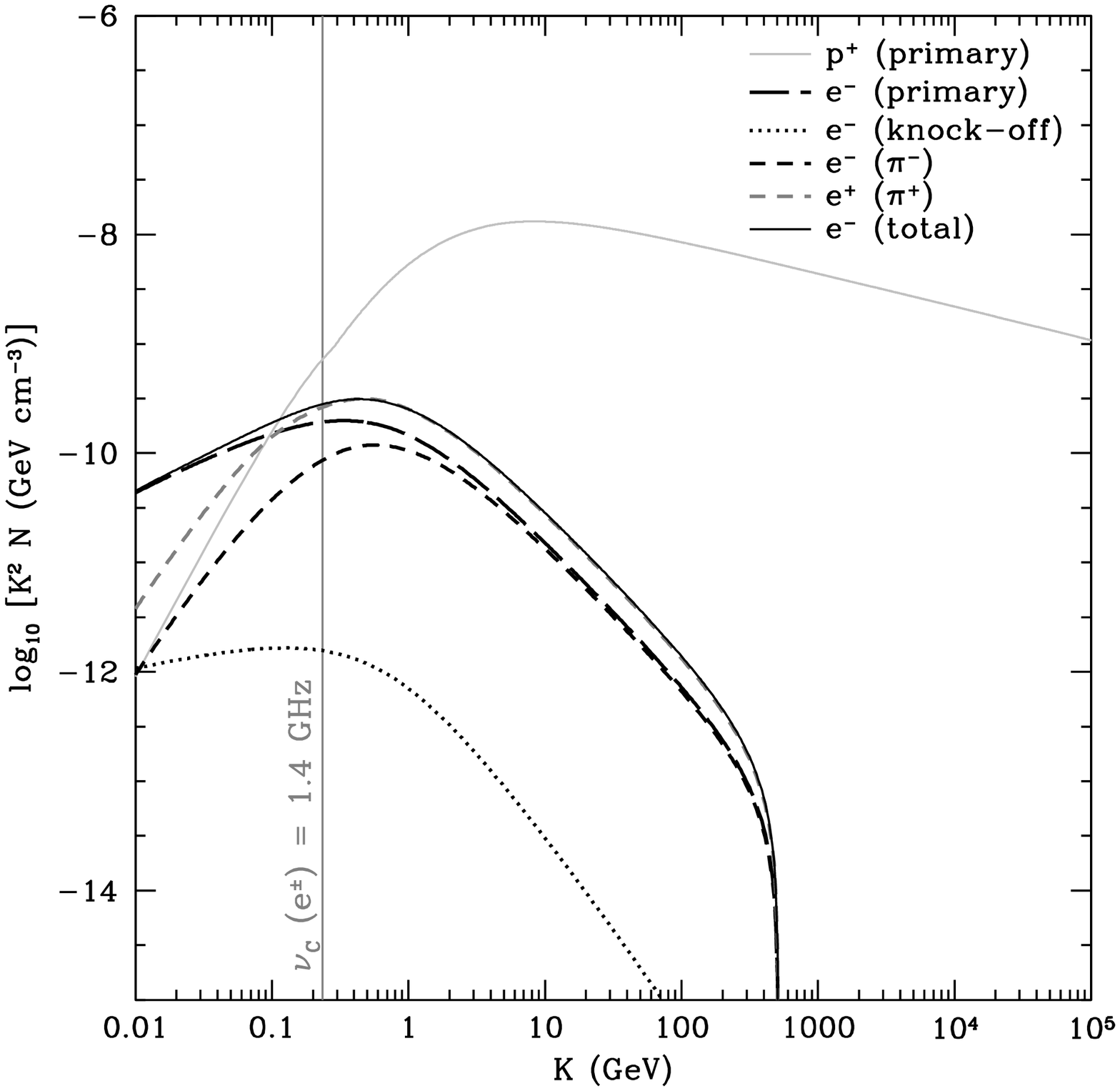}}
\figcaption[simple]{The predicted kinetic energy spectra of cosmic rays in a low density galaxy (\emph{left}, $\Sigma_g = 0.001~\gcm2$) and a high density starburst (\emph{right}, $\Sigma_g = 10~\gcm2$) for our standard model ($p = 2.3$, $f = 1.5$, $a = 0.7$, $\tilde\delta = 48$, $\xi = 0.023$).  We mark the kinetic energy where electrons and positrons emit synchrotron radiation at $\nu \approx 1.4~\GHz$.  The cutoff in the lepton spectra at $\sim 500~\GeV$ is caused by our use of $\gamma_{\rm max} = 10^6$; see Section~\ref{sec:ParticleFeatures} for further discussion.\label{fig:CRSpectraLow}\label{fig:CRSpectraHigh}}
\end{figure*}

We show typical predicted CR spectra in Figure~\ref{fig:CRSpectraLow} for $\Sigma_g = 0.001~\gcm2$ and $10~\gcm2$, the lowest and highest surface densities we consider.

In low surface density galaxies, the protons with $\gamma \gg 1$ have a power-law spectrum with ${\cal P}_p$ about 0.5 greater than the injected spectral index $p$.  The increased steepness comes from faster diffusive escape at higher energies (eq.~\ref{eqn:StandardCRLife}; \citealt{Ginzburg76}).  At lower energies, the CR proton spectrum flattens due to ionization losses, which are constant with energy \citep{Schlickeiser02,Torres04}.  High surface density galaxies have harder proton spectra with ${\cal P}_p = p$ at high energies, since pion losses overwhelm escape and are roughly energy independent.

\begin{figure*}
\centerline{\includegraphics[width=9cm]{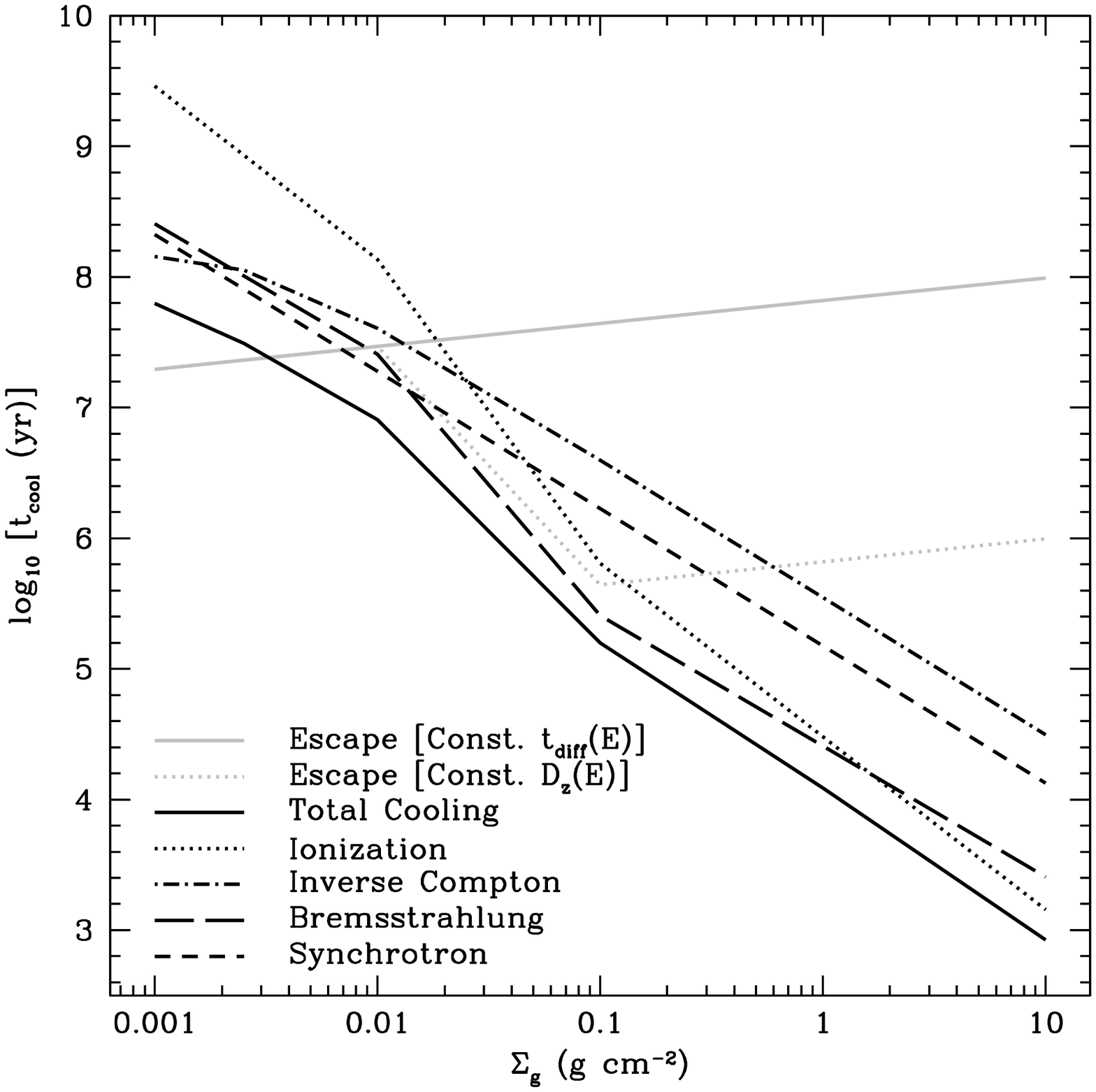}\includegraphics[width=9cm]{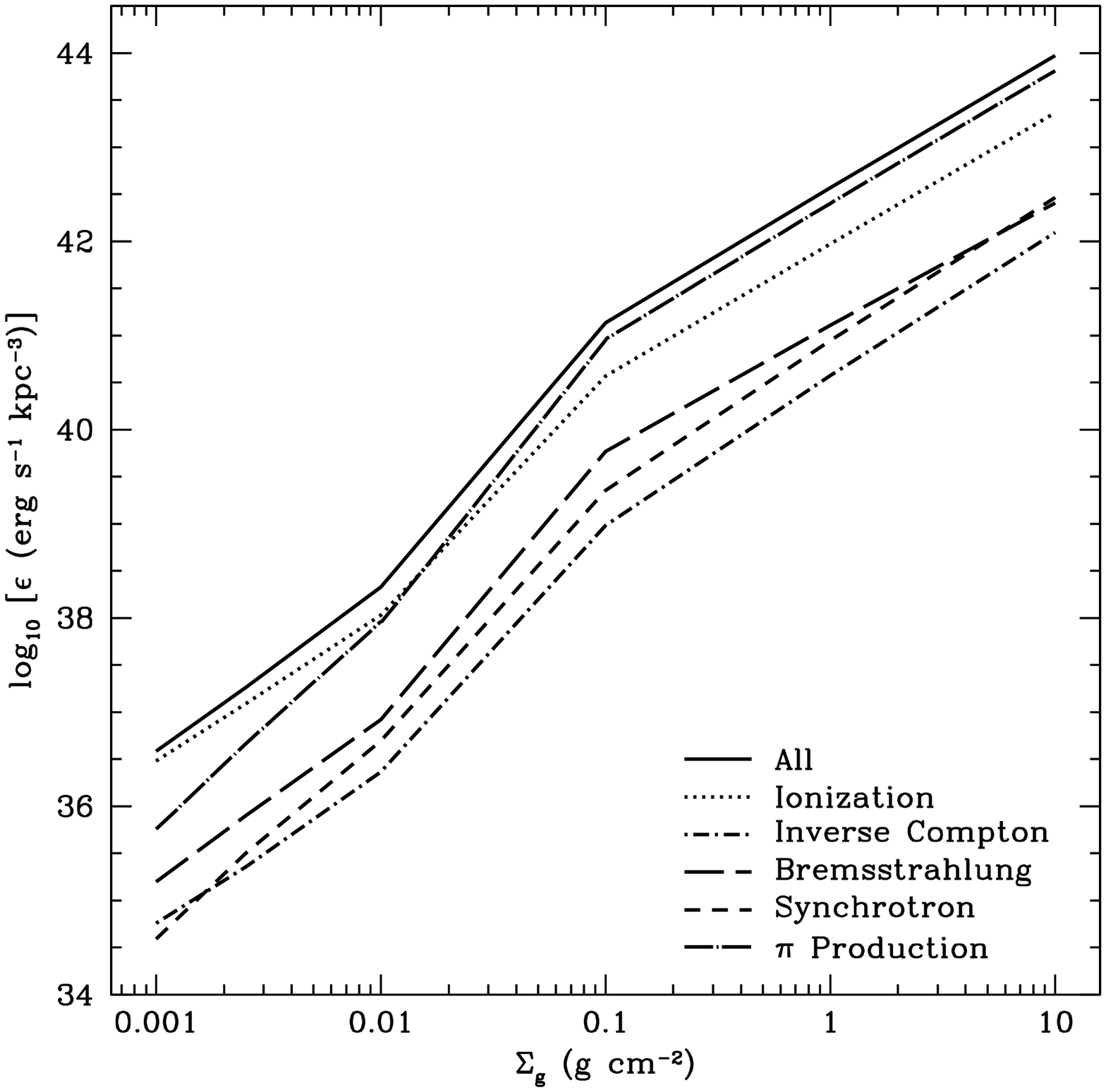}}
\figcaption[simple]{{\it Left:} cooling times \emph{(for electrons and positrons)} when $\nu_C = 1.4~\GHz$.
{\it Right:} emissivity (energy lost per volume per unit time) for each process from protons, electrons, and positrons, integrated over energy.  Pion losses include all of the energy going into secondary production as well as $\gamma$-rays and neutrinos.\label{fig:tCool}\label{fig:Emiss}}
\end{figure*}

In low surface density galaxies, the primary electrons behave similarly to protons.  For most low energies, they have a power-law spectrum with ${\cal P}_e \approx 0.5 + p$, caused by diffusive escape losses \citep[e.g.,][]{Ginzburg76}.  However, synchrotron and IC losses steepen the spectrum at high energies.  Bremsstrahlung and ionization flatten the spectrum at lower energies ($E \la 1~\GeV$) (compare Figure~\ref{fig:Emiss}; see also \citealt{Thompson06}; \citealt{Condon92}).  In high surface density galaxies, diffusive losses are negligible compared to synchrotron and IC losses, thus forcing ${\cal P}_e \approx 1 + p$ at higher energies \citep[e.g.,][]{Ginzburg76}.

\begin{figure*}
\centerline{\includegraphics[width=9cm]{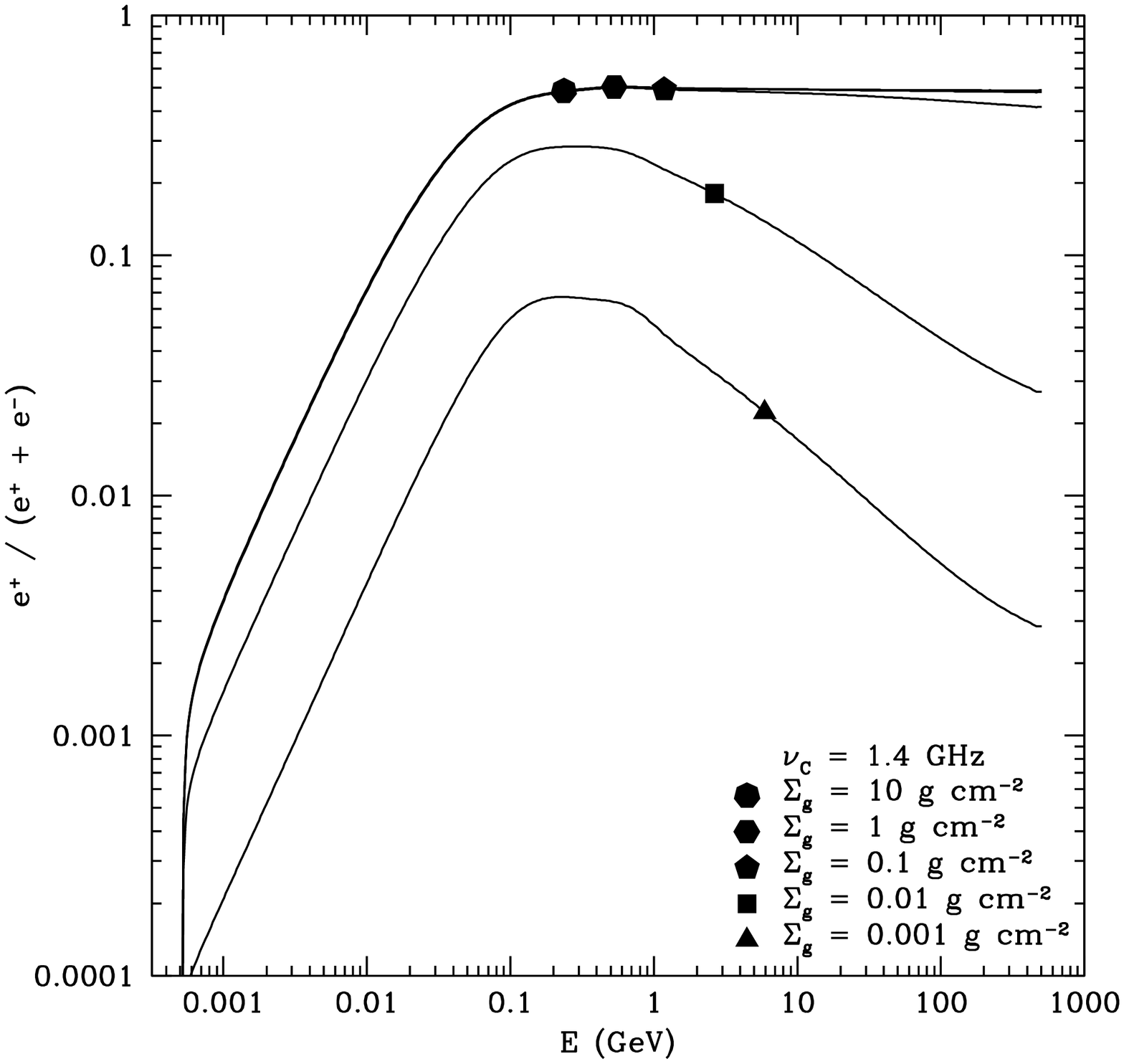}\includegraphics[width=9cm]{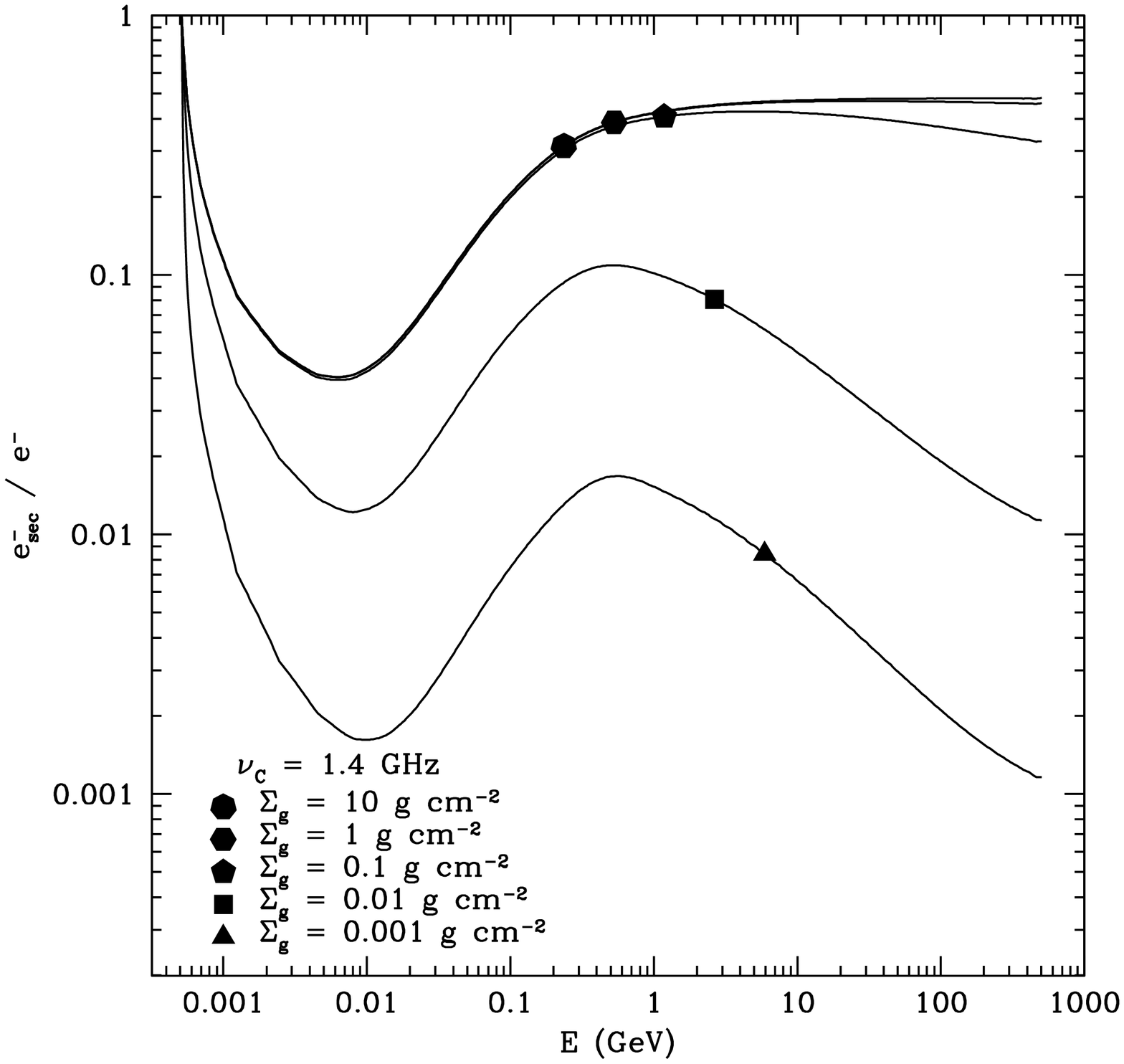}}
\figcaption[simple]{{\it Left:} the abundance of positrons as a function of energy.  {\it Right:} the abundance of secondary electrons compared to all electrons, as a function of energy.  The saturation of the positron and secondary electron ratios at high $\Sigma_g$, and the lack of energy dependence for $E \ga 1~\GeV$, is a sign of proton calorimetry (see Section~\ref{sec:ParticleFeatures} and Section~\ref{sec:Calorimeter}). We mark with filled symbols the energies where the critical synchrotron frequency $\nu_C$ is 1.4 GHz.\label{fig:ePlusRatio}\label{fig:eMinusSecRatio}}
\end{figure*}

The spectra of secondary electrons and positrons show additional features with respect to the primary electron spectrum.  The secondary (pion-produced) spectrum is flatter than the primary spectrum at low energies, because the production cross sections decreases near the pion production threshold (see Figure~\ref{fig:eMinusSecRatio}; \citealt{Strong98,Torres04}).  At high energies, the pion electrons and positrons are \emph{injected} with a spectrum proportional to the \emph{steady-state} proton spectrum.  This means that in low surface density galaxies, the pion electrons and positrons have a steeper spectrum than the primary electrons at high energies \citep{Ginzburg76}, while in high surface density galaxies, the secondary and primary spectral slopes are the same at high energies (as seen in Figure~\ref{fig:eMinusSecRatio}).  Furthermore, there are always more secondary pion positrons than secondary pion electrons (ultimately due to charge conservation).  Knock-off electrons become increasingly important at very low energies and dominate as $\gamma$ approaches one \citep[see][]{Torres04}.

\begin{figure*}
\centerline{\includegraphics[width=9cm]{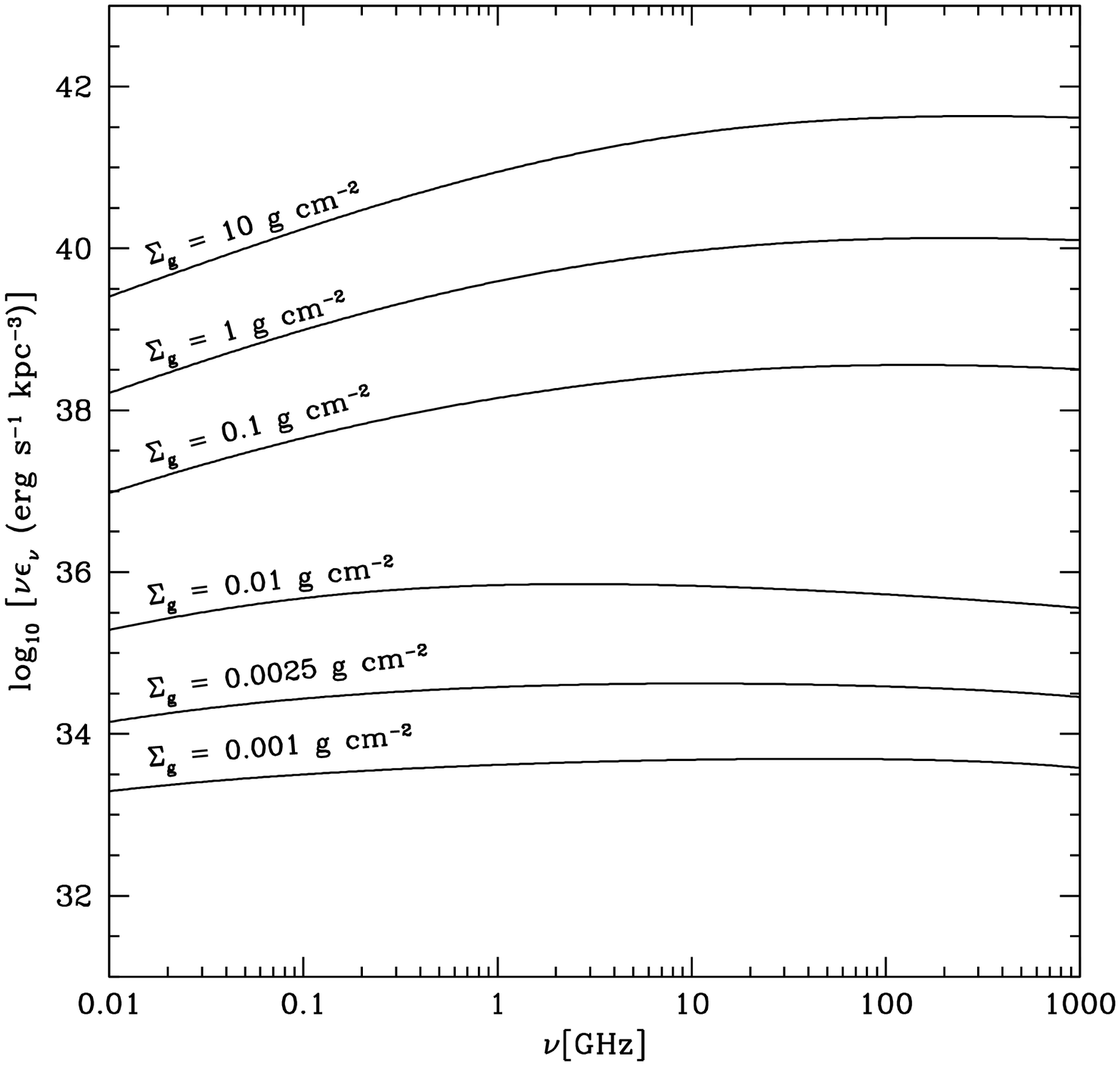}\includegraphics[width=9cm]{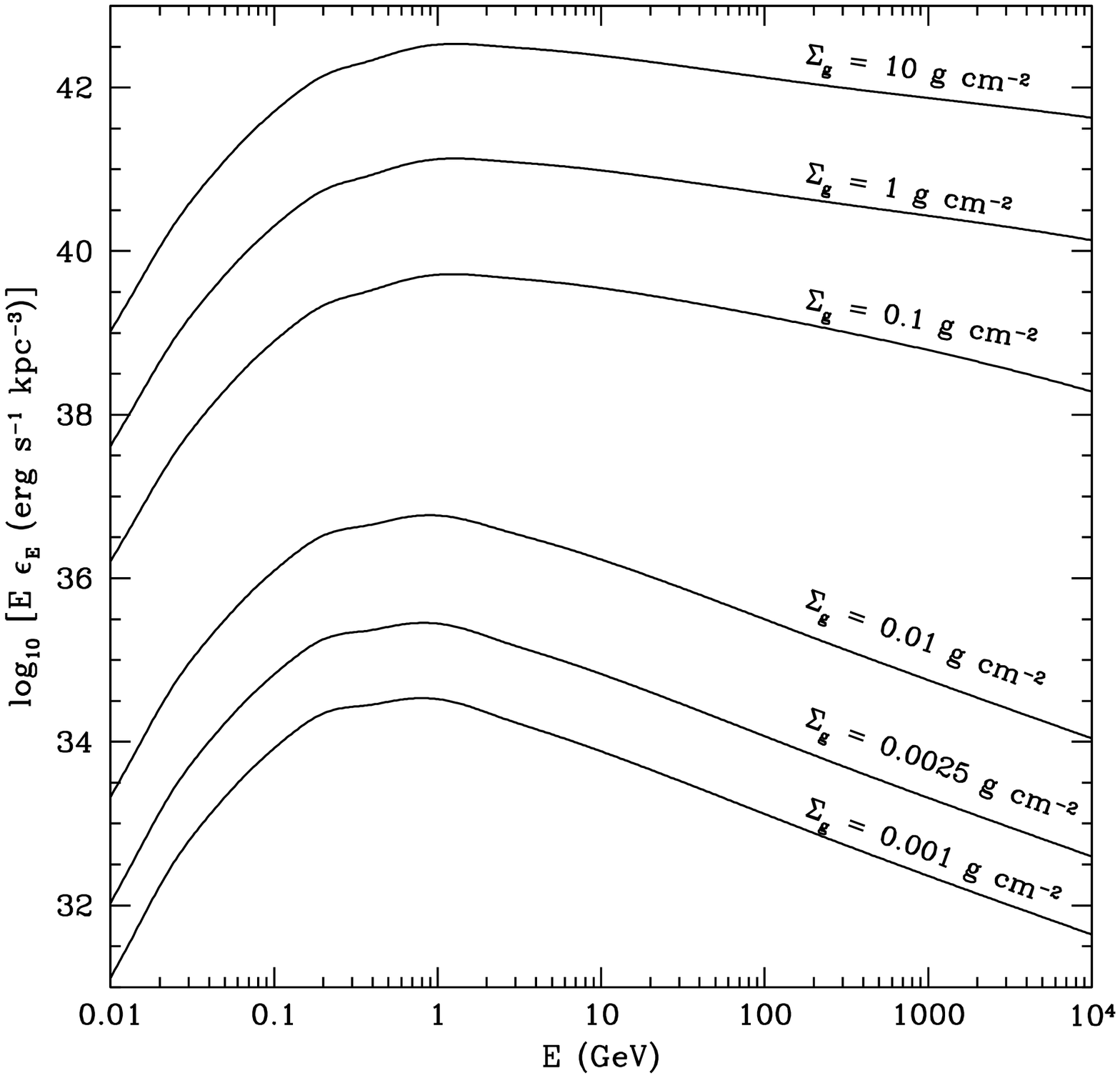}}
\figcaption[simple]{The spectra of radio synchrotron (\emph{left}) and $\pi^0$ $\gamma$-rays (\emph{right}) predicted by our standard model ($p = 2.3$, $f = 1.5$, $a = 0.7$, $\tilde\delta = 48$, $\xi = 0.023$).  Note that the synchrotron spectra steepen somewhat with frequency (also see Figures~\ref{fig:Alpha} and \ref{fig:AlphaNu}).  Proton calorimetry flattens the $\pi^0$ $\gamma$-ray spectrum in the starbursts.\label{fig:SynchSpectra}\label{fig:Pi0Spectra}}
\end{figure*}

\begin{figure*}
\centerline{\includegraphics[width=9cm]{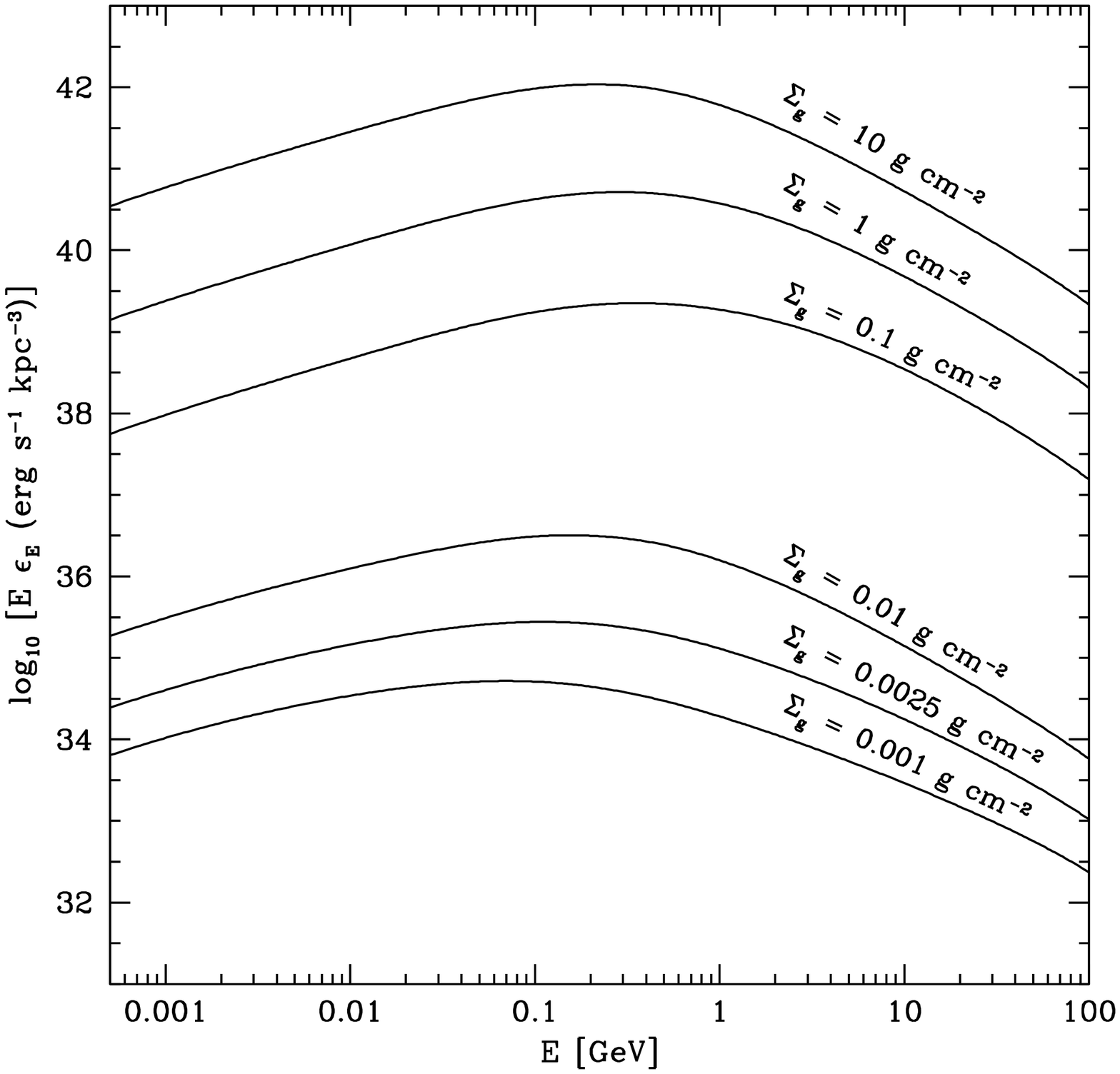}\includegraphics[width=9cm]{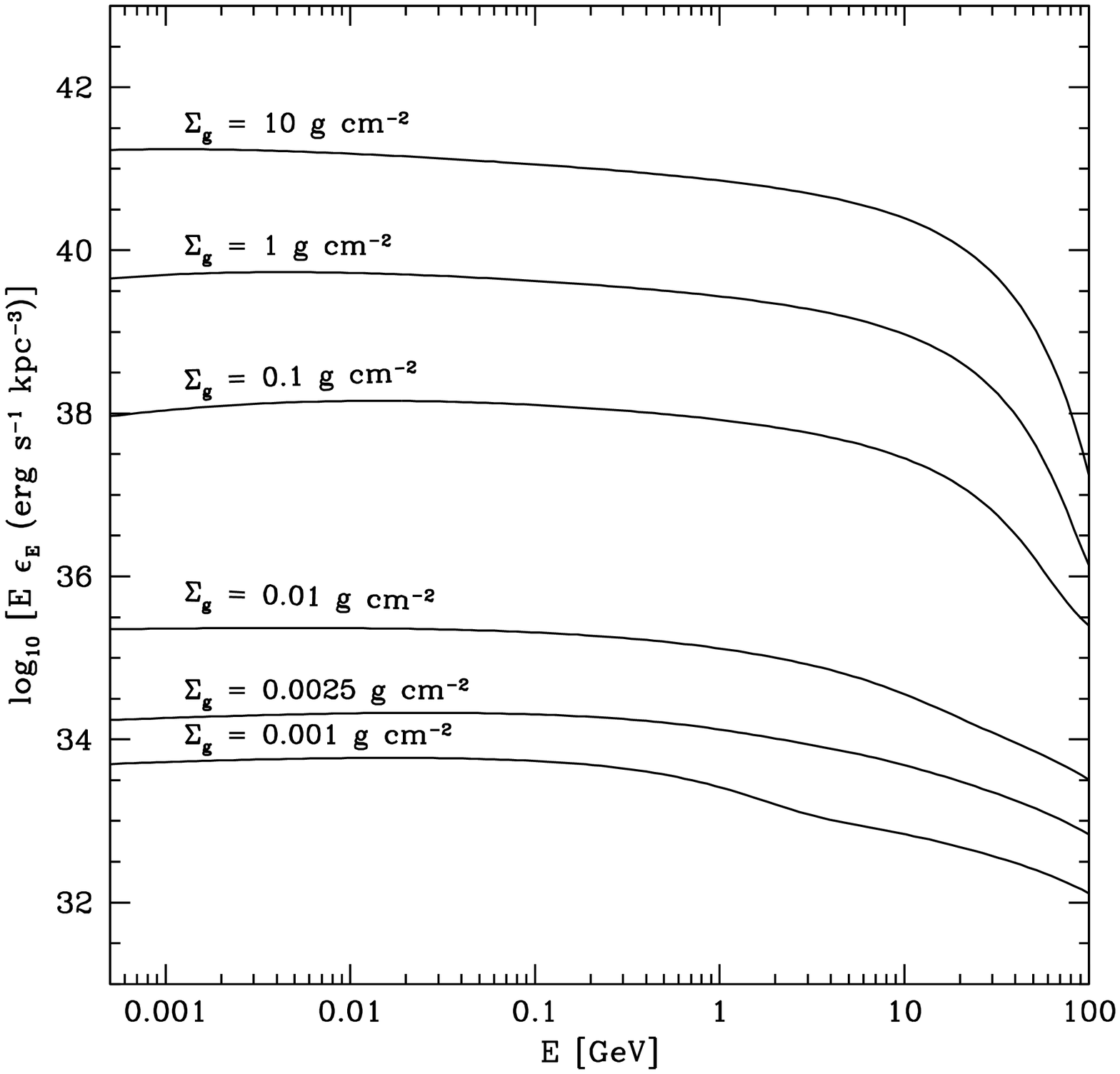}}
\figcaption[simple]{The spectra of bremsstrahlung $\gamma$-rays (\emph{left}) and estimated IC (\emph{right}) predicted by our standard model ($p = 2.3$, $f = 1.5$, $a = 0.7$, $\tilde\delta = 48$, $\xi = 0.023$).  We discuss our assumptions for the IC emission in Section~\ref{sec:GammaRays}. \label{fig:BremsSpectra}\label{fig:ICSpectra}}
\end{figure*}

We also show radiation spectra in Figures~\ref{fig:SynchSpectra} and~\ref{fig:BremsSpectra} for synchrotron, $\pi^0$ $\gamma$-rays, relativistic bremsstrahlung, and IC emission (see Section~\ref{sec:GammaRays} for the assumptions used to estimate the IC emission).  The radio, bremsstrahlung, and IC emission generally steepen with increasing frequency (compare with Figure~\ref{fig:Alpha}; see also \citealt{Lisenfeld96a,Thompson06}), whereas pion $\gamma$-rays peak at a few hundred MeV.  Although we do not calculate it here, the overall high-energy neutrino emission is comparable to the $\gamma$-ray emission from $\pi^0$ decay \citep{Stecker79}.

\section{Discussion}
\label{sec:Discussion}
\subsection{Is Calorimetry Correct?}
\label{sec:Calorimeter}

We show the effects of forcing electron and UV calorimetry to hold in Figure~\ref{fig:Calorimetry} for our standard model (cf. Section~\ref{sec:StandardModel}).  It is clear that most of the energy in 1.4 GHz electrons is lost radiatively in galaxies with $\Sigma_g \ga 0.01~\gcm2$: calorimetry holds in high- but not low-density galaxies (this behavior was first described in \citealt{Chi90} and was also predicted by \citealt{Lisenfeld96a}).  At lower surface densities, electron calorimetry begins to fail (decreasing the radio luminosity), but the effect of this on the FRC is largely mitigated by the decreasing optical thickness to UV photons (decreasing the FIR luminosity).  This conspiracy saves the FRC, as discussed by \citet{Bell03}, but only applies in our standard model over one decade in $\Sigma_g$.  In our standard model, electron escape at low $\Sigma_g$ eventually becomes the stronger effect, so that low-density galaxies would be radio dim with respect to the FRC.  A high value of $q$ is in fact observed as a nonlinearity in the FRC at low luminosities (\citealt{Yun01}, though \citealt{Beswick08} find the opposite).  Unfortunately, studies of the low luminosity FRC are complicated by the presence of thermal radio emission \citep[e.g.,][for the Large Magellanic Cloud]{Hughes06}, which also correlates with FIR light and overwhelms the nonthermal synchrotron emission considered here in the lowest density galaxies.

While the standard models predict that electron calorimetry holds in the inner Milky Way, there are physically motivated variants (Appendix~\ref{sec:Variants}) which predict that electron calorimetry fails for normal galaxies and the weakest starbursts.  In particular, the ``strong wind'' variants predict a non-calorimetric inner Milky Way, because of the wind inferred by \citet{Everett08} (see Appendix~\ref{sec:Winds}).

Although the transition to calorimetry is model dependent, it seems unavoidable that extreme starbursts like Arp 220 are electron calorimeters.  We can derive the speed $v_{\rm esc}$ at which CRs would have to stream out of galaxies for electron calorimetry to fail, according to the cooling rates in our standard model.  We can also compare these numbers to standard CR confinement theory, where CRs are limited to propagate at the Alfv\'en speed ($v_A = B / \sqrt{4 \pi \rho}$) by a streaming instability in the ionized ISM \citep{Kulsrud69}.  We can also invert the problem and determine the magnetic field with a high enough Alfv\'en speed\footnote{This estimate assumes that CRs are streaming through material with the mean ISM density.  In a lower density phase, $v_A$ will be larger and $B_{\rm esc}$ will be smaller.} for CRs to stream out of the galaxy in one cooling time ($B_{\rm esc}^2 = 4 \pi \rho v_{\rm esc}^2$), as well the diffusion constant ($D_{\rm esc} = h^2 / t_{\rm cool}$) needed to diffuse out of the system in one cooling time.  The environmental conditions needed to allow CRs to escape before cooling significantly are reasonable for weak $\Sigma_g \le 0.1~\gcm2$ starbursts.  We find that $v_{\rm esc} = 620~\kms$, and winds of several hundred kilometers per second are in fact observed in starbursts.  Similarly, we calculate $D_{\rm esc} = 1.9 \times 10^{28}~\DiffUnits$,  and diffusion constants of order $10^{28}~\DiffUnits$ are inferred for starburst galaxies \citep[e.g.,][]{Dahlem95}.  However, if the CRs stream through mean density ISM, then $B_{\rm esc} \approx 3~\mGauss$ is higher than the equipartition magnetic field strength $B_{\rm eq} = \sqrt{8 \pi^2 G \Sigma_g^2} \approx 300~\muGauss$ for $\Sigma_g = 0.1~\gcm2$, so that CR escape would have to be super-Alfv\'enic.  For higher $\Sigma_g$, though, escape would require extreme wind speeds ($8000~\kms$ when $\Sigma_g = 1~\gcm2$ and $120,000~\kms \approx 0.4 c$ when $\Sigma_g = 10~\gcm2$), extremely high diffusion rates ($2.5 \times 10^{29}~\gcm2$ for $\Sigma_g = 1~\gcm2$ and $3.6 \times 10^{30}~\gcm2$ for $\Sigma_g = 10~\gcm2$), or extremely strong magnetic fields ($B_{\rm esc} = 0.11~\Gauss$ in the $\Sigma_g = 1~\gcm2$ case and $B_{\rm esc} = 5~\Gauss$ in the $\Sigma_g = 10~\gcm2$ case) that are unreasonable.  \emph{We therefore conclude that electron calorimetry must hold in dense starbursts.}

\begin{figure}
\centerline{\includegraphics[width=8cm]{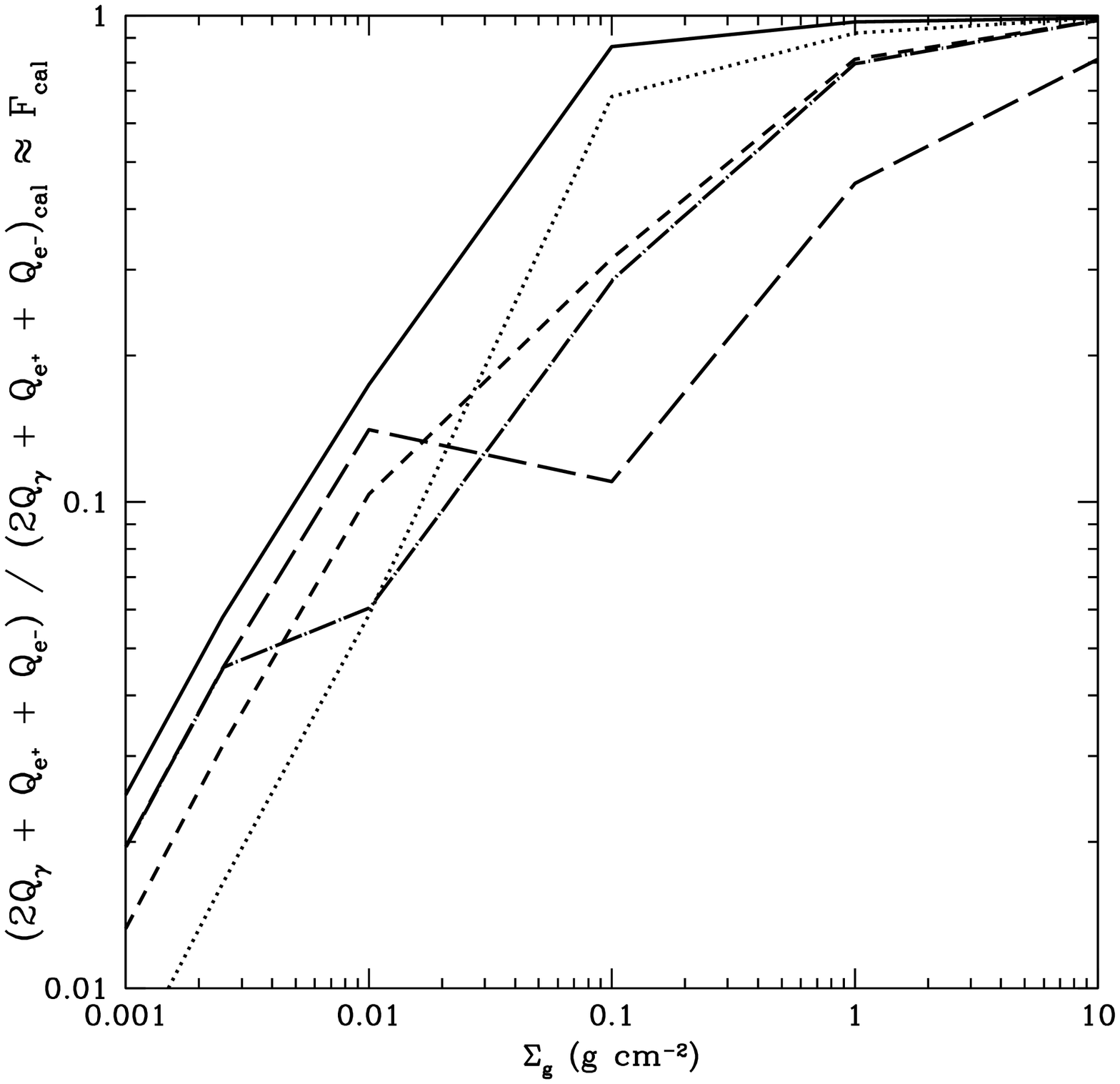}}
\figcaption[figure]{The estimated proton calorimetry fraction $F_{\rm cal}$ in our models for several variants.  The values are normalized so that models with no escape have $F_{\rm cal} = 1$. Normal galaxies are not proton calorimeters, while ULIRGs with $\Sigma_g \approx 10~\gcm2$ are in all of our variants (note the convergence of all models to $\sim 1$ at high $\Sigma_g$).  In all variants, proton calorimetry holds for starbursts with $\Sigma_g \ga 1~\gcm2$.  Variants shown are our standard model (solid; $p = 2.3$, $f = 1.5$, $\tilde\delta = 48$, $a = 0.7$); $B \propto \rho^a$ with 300~$\kms$ wind in starbursts (dash; $p = 2.2$, $f = 1.0$, $\tilde\delta = 67$, $a = 0.5$); strong winds of 175~$\kms$ in $\Sigma_g = 0.01~\gcm2$, 600~$\kms$ for starbursts (long dash dot; $p = 2.2$, $f = 2.0$, $\tilde\delta = 45$, $a = 0.5$); constant $D_z$, $B \propto \rho^a$, and winds of 300~$\kms$ in starbursts (long dash; $p = 2.2$, $f = 1.5$, $\tilde\delta = 34$, $a = 0.6$); and fast diffusive escape with $B \propto \Sigma_g^a$ and no winds (dotted; $p = 2.2$, $f = 2.0$, $\tilde\delta = 45$, $a = 0.6$).\label{fig:pCalorimetry}}
\end{figure}

We can similarly ask whether galaxies are proton calorimeters.  The low pion luminosity of the Galaxy and the secondary positron fraction at Earth imply that normal galaxies like the Milky Way are not proton calorimeters.  We have estimated the proton calorimetry fraction $F_{\rm cal}$ in our models by adding the emissivity in pion products\footnote{Since we do not calculate the neutrino spectrum, we simply assume that $Q_{\nu} = Q_{\gamma}$, which is a reasonable approximation at high energies.} to the emissivity in CR protons with energy greater than 1.22 GeV, the pion production threshold energy.  However, when we do this we find that even \emph{explictly} proton calorimetric models with \emph{no} diffusive or advective escape have $F_{\rm cal} \approx 0.5$.  This appears to be caused by an inconsistency between the pionic lifetime we use (eqn~\ref{eqn:tPion}) from \citet{Mannheim94} and the GALPROP cross sections: if we add up all of the energy in all of the pionic products of a CR proton of energy $\sim \GeV$, the effective energy loss rate is several times smaller than implied by \cite{Mannheim94}\footnote{As far as we are aware, this discrepancy has not been discussed in the literature.}.  We also note that the \cite{Mannheim94} pionic lifetime is twice as short at $\sim \GeV$ energies as the pionic lifetime in \citet{Schlickeiser02}. Finally, this approach ignores ionization losses, which do not create secondaries but will prevent lower energy protons from escaping.  

To account for this discrepancy in the energetics, we normalize our estimate of $F_{\rm cal}$ so that an explicitly proton calorimetric model of the same CR injection rate, $\Sigma_g$, $p$, and $f$ has $F_{\rm cal} = 1$.  We then see in Figure~\ref{fig:pCalorimetry} that dense starbursts with $\Sigma_g \approx 10~\gcm2$ all are proton calorimeters with $F_{\rm cal} \approx 1$ for several variants (Appendix~\ref{sec:Variants}).   As with electron calorimetry, proton calorimetry sometimes breaks down for the $\Sigma_g = 0.1~\gcm2$ weak starbursts, because the time to cross the 100 pc starburst scale height is short.  However, when $\Sigma_g \ga 1~\gcm2$, proton calorimetry holds in our models; a model with winds and strong diffusive losses has proton calorimetry breaking down at $\Sigma_g = 1~\gcm2$ ($F_{\rm cal} = 0.45$).  

As with the electrons, we can derive the speed that the CRs would need to stream out of a starburst for proton calorimetry to fail, which is $v_{\rm esc} = h / t_{\pi} = 1900\ \kms (\Sigma_g / \gcm2) f$.  While $v_{\rm esc}$ is easily attained by winds in starbursts with $\Sigma_g = 0.1~\gcm2$, only the fastest winds are capable of breaking proton calorimetry in $\Sigma_g = 1~\gcm2$ starbursts.  Diffusive escape limited to the mean Alfv\'en speed of the starburst would require strong magnetic fields ($B_{\rm esc} = 27\ \mGauss (h / 100 \pc)^{-1/2} (\Sigma_g / \gcm2)^{3/2} f$) with energy densities greater than the midplane gas pressure in starbursts to break proton calorimetry.  We therefore conclude that proton calorimetry is difficult to avoid in $f \approx 1$ starbursts with $\Sigma_g \ga 1~\gcm2$.

\subsection{What Causes the FIR-Radio Correlation?}
\label{sec:FRCCauses}
\subsubsection{Calorimetry and the $\nu_C$ Effect}
\label{sec:nuCEffect}
Calorimetry provides a simple way to explain the FRC.  We find that both electron and UV calorimetry hold for starbursts, and possibly the inner regions of normal galaxies, depending on the variant on our underlying model (Appendix~\ref{sec:Variants}).  Calorimetry therefore serves as the foundation of our explanation for the FRC.  Other effects alter the radio luminosity, both at low density and high density, but by a factor of $\sim 10$, compared to the dynamic range of $10^4$ in $\Sigma_g$.  At the order-of-magnitude level, calorimetry can be said to cause the FRC, and other effects are relatively moderate corrections.  

However, in more detail, we find that $L_{\rm TIR}/L_{\rm radio}$ is not in fact flat even in the simple calorimeter model, with no escape, non-synchrotron cooling, or secondaries (the light dotted line in Figure~\ref{fig:HighSGConspiracy}).  Instead, $L_{\rm TIR}/L_{\rm radio}$ decreases by a factor of 2.6 as $\Sigma_g$ increases, because 1.4 GHz observations probe lower CR electron energies as the magnetic field strength increases.  We call this decrease in $L_{\rm TIR}/L_{\rm radio}$ with $\Sigma_g$ the ``$\nu_C$ effect''.  In general, the effect becomes more significant as $p$ increases past 2.0, because the electron spectrum becomes steeper. It can be shown that in this simplest calorimeter limit, $L_{\rm TIR}/L_{\rm radio} \propto B^{p/2 - 1}$.

\subsubsection{High-$\Sigma_g$ Conspiracy}
\label{sec:HighSGConspiracy}
\begin{figure}
\centerline{\includegraphics[width=8cm]{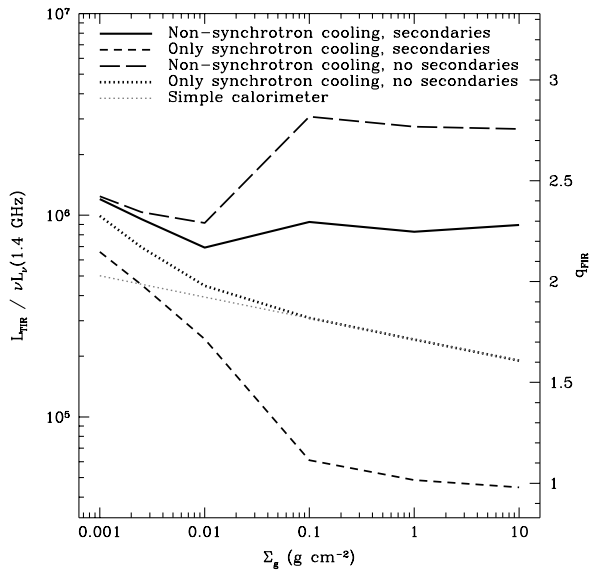}}
\figcaption[figure]{The high-$\Sigma_g$ conspiracy in our standard model ($p = 2.3$, $f = 1.5$, $a = 0.7$, $\tilde\delta = 48$, $\xi = 0.023$).  The simple calorimeter model has perfect UV calorimetry and electron calorimetry, with only synchrotron cooling and no secondaries.  Non-synchrotron cooling and secondaries alone each create a broken FIR-radio correlation, but conspire to make it linear at high density.\label{fig:HighSGConspiracy}}
\end{figure}

The radio luminosity in high-density galaxies is altered from the calorimetric luminosity mainly by two mechanisms, non-synchrotron cooling and the appearance of secondary electrons and positrons.  We illustrate these effects in Figure~\ref{fig:HighSGConspiracy}.

In normal galaxies, synchrotron cooling dominates the energy losses, though bremsstrahlung and IC off the CMB can be competitive within a factor of a few or less.  However, in starbursts, energy loss is mainly by bremsstrahlung and ionization.  This decreases the proportion of energy lost that goes into radio.  The energy diverted to bremsstrahlung and ionization therefore increases $L_{\rm TIR}/L_{\rm radio}$ by a factor of up to $\sim 20$ in starbursts compared to normal galaxies (compare the dotted and long-dashed lines in Figure~\ref{fig:HighSGConspiracy}).

Secondary electrons and positrons themselves radiate in the radio.  In the starbursts, which are proton calorimeters, there are several times more secondaries than primary electrons, while in normal galaxies, the secondary contribution is small.  Secondaries increase the radio emission by a factor of $\sim 4$ in starbursts compared to normal galaxies (compare the dotted and short-dashed lines in Figure~\ref{fig:HighSGConspiracy}).  

These effects each on their own alter the calorimetric radio luminosity by up to an order of magnitude.  Since both are density dependent, they both become important in starbursts.  However, combined with the $\nu_C$ effect (Section~\ref{sec:nuCEffect}) in the simple calorimeter model, they largely cancel each other out to maintain a linear FRC.  The exact magnitudes of these effects are model dependent, but they are always important and the direction each works in is the same in every case.  It is possible that relaxing the assumptions of our approach, such as including time dependence or spatial variation, could avoid the severe non-synchrotron losses and secondary electrons and positrons giving rise to this particular high-$\Sigma_g$ conspiracy.  However, any new effects would have to be tuned to avoid the processes we already include while still reproducing the FRC, trading one conspiracy for another.

There are two other effects that appear in our variants (Appendix~\ref{sec:Variants}), but not our standard model, which can change the FRC.  First, if the magnetic field is assumed to depend on density instead of surface density (Section~\ref{sec:Basn}), the magnetic fields will be much stronger in the starbursts for the $B \propto \rho^a$ case, since the starbursts are more compact.  This will make synchrotron cooling dominant again, upsetting the high-$\Sigma_g$ conspiracy.  This effect can be compensated by winds and a weak magnetic field dependence on $\rho$ (low $a$).  Second, if the FIR optical depth is significant (Section~\ref{sec:OpticalThickFIR}), the photon energy density inside the galaxy is greater by a factor of $\sim\tau_{\rm FIR}$ than inferred from the photon flux alone.  While typical FIR opacities are small, the optical depth is appreciable in dense starbursts ($1~\cmg21 \la \kappa_{\rm FIR} \la 10~\cmg21$).  This increases IC losses dramatically at the high densities, decreasing the radio luminosity.  Models with large FIR optical depths have trouble reproducing the FRC (Section~\ref{sec:OpticalThickFIR}).

\subsubsection{Low-$\Sigma_g$ Conspiracy}
The radio luminosity in low density normal galaxies is modified by a different pair of opposing mechanisms, the failure of electron calorimetry and the failure of UV calorimetry.  This conspiracy is illustrated in Figure~\ref{fig:Calorimetry} for our standard model.

Normal galaxies are not generally electron calorimeters -- both diffusive and advective escape can operate faster than cooling.  In weak starbursts ($\Sigma_g \la 0.1~\gcm2$), escape can be competitive with cooling processes, but not in stronger starbursts.  Escape therefore decreases the radio emission in normal galaxies compared to the calorimetric expectation.  

However, normal galaxies are generally not UV calorimeters either; a substantial fraction of the UV light emitted by star formation can escape without being reprocessed into FIR light \citep[e.g.,][]{Xu95,Bell03,Buat05,Martin05,Popescu05}.  Therefore, normal galaxies also have a lower FIR luminosity compared to the calorimetric expectation.

As can be seen in Figure~\ref{fig:Calorimetry}, each of these effects alters the FRC by a factor of $\sim 4$ for $\Sigma_g = 0.001~\gcm2$ in our standard model.  Since they work in opposite directions, the resulting $L_{\rm TIR}/L_{\rm radio}$ nonetheless remains the same as the calorimetric prediction \citep[as suggested by][]{Bell03}.

\subsubsection{The Intermediate Case}
The boundary between these two conspiracies occurs when $\Sigma_g = 0.1~\gcm2$.  In some variants (Appendix~\ref{sec:Variants}), factors from both surface density regimes must be tuned to maintain the FRC at this surface density: escape time, secondaries, non-synchrotron cooling, and magnetic field strength all have an effect on the radio luminosity.  However, these starbursts are unavoidably opaque to UV light, so the full low-$\Sigma_g$ conspiracy cannot work for these galaxies.  This becomes a problem when CR escape is quick, such as when strong winds are present (see Section~\ref{sec:Winds}), causing these galaxies to be radio-dim.  Since the conspiracies begin to break down for the weakest starbursts, the transition from normal galaxies to starbursts may prove important in testing models of the FRC. 

\subsubsection{Summary}
The many factors described above conspire to produce the FRC, both in low-density non-calorimetric galaxies and high-density calorimetric starbursts.  The traditional distinction between calorimeter and conspiracy explanations of the FRC is not clear cut in our models.  We find that the FRC requires both calorimetry \emph{and} conspiracy.

\subsection{The FIR-Radio Correlation at Other Frequencies}
\label{sec:OtherNuFRC}

\begin{figure}
\centerline{\includegraphics[width=8cm]{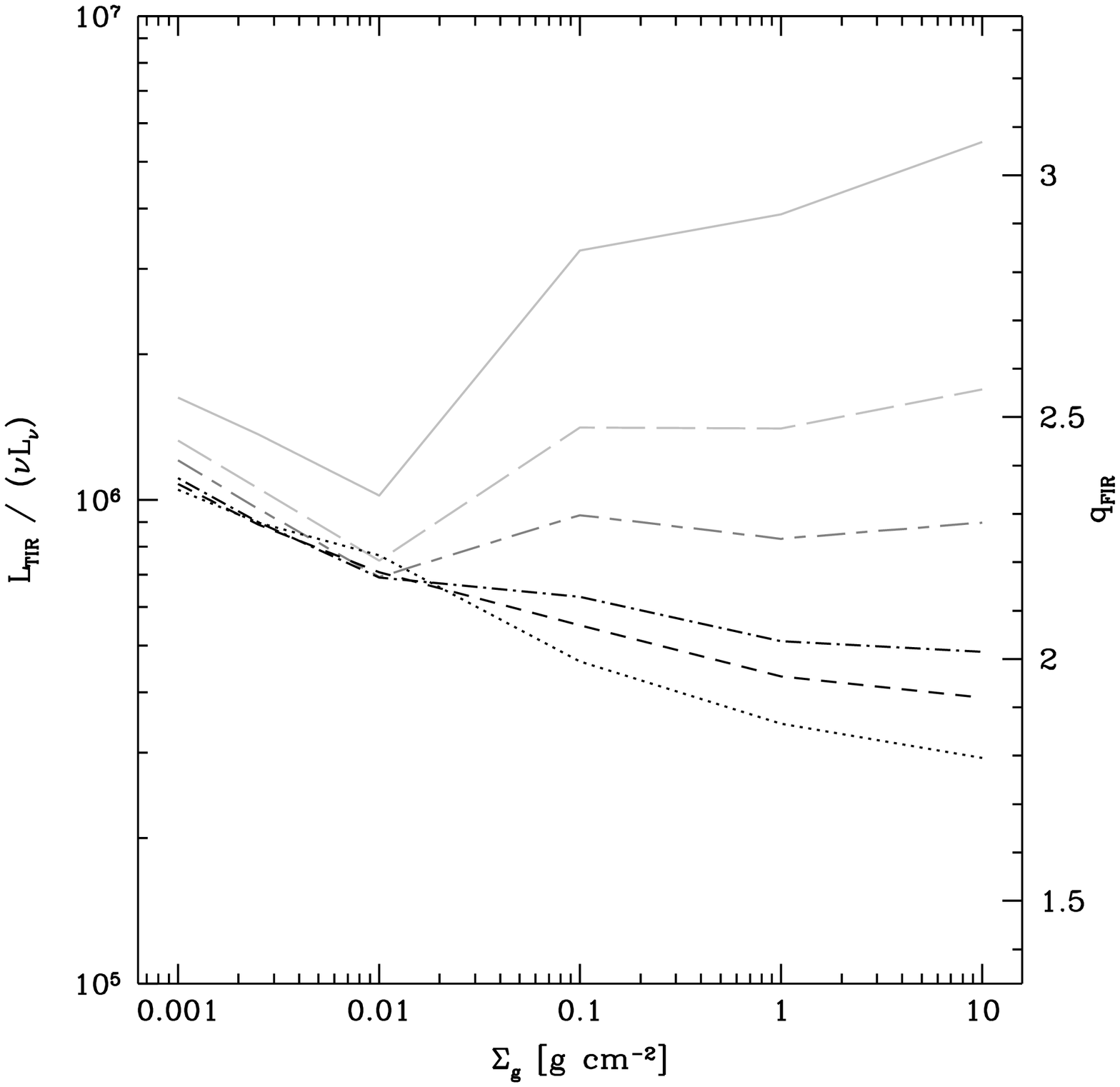}}
\figcaption[figure]{The FIR-radio correlation at other frequencies.  Galaxies toward the top of the plot are radio-dim for their FIR emission, while galaxies toward the bottom of the plot are radio-bright.  The shown frequencies are 100 MHz (light gray, solid), 500 MHz (light gray, long-dashed), 1.4 GHz (medium gray, long dashed/short dashed), 4.8 GHz (black, dash dot), 8.4 GHz (black, short dash), and 22.5 GHz (black, dotted).  All lines are for our standard model ($p = 2.3$, $f = 1.5$, $a = 0.7$, $\tilde\delta = 48$, $\xi = 0.023$).\label{fig:OtherNuLFIRRadio}}
\end{figure}

We have mainly considered the well-studied FRC at 1.4 GHz.  However, the FRC is also known to exist at 150 MHz \citep{Cox88}, 4.8 GHz \citep{deJong85,Wunderlich87}, and 10.55 GHz \citep{Niklas97b}.  The correlation holds for both normal galaxies and starbursts at these frequencies, and remains even after thermal radio emission is subtracted.  We show the predicted ratios of FIR to synchrotron radio fluxes in Figure~\ref{fig:OtherNuLFIRRadio}.

Our standard model predicts increased nonlinearity at other frequencies for a set of galaxies that span from normal galaxies to starbursts.  While $L_{\rm TIR}/L_{\rm radio}$ varies by only 1.7 at 1.4 GHz over the full range in $\Sigma_g$, it varies by 2.3 at 500 MHz, and a factor of 5.4 at 100 MHz.  At higher frequencies, the situation is similar, though the linear FRC is somewhat better preserved: $L_{\rm TIR}/L_{\rm radio}$ varies by a factor of 2.3 at 4.8 GHz, 2.8 at 8.4 GHz, and 3.6 at 22.5 GHz.   As can be seen in Figure~\ref{fig:OtherNuLFIRRadio}, at low frequencies the FRC is predicted to tilt to the FIR with increasing $\Sigma_g$, while at high frequencies the correlation is predicted to tilt to the radio in starbursts.  Our fiducial model with winds and $B \propto \rho^a$ (Section~\ref{sec:Winds}) predicts a similar increase in scatter at other frequencies ($L_{\rm TIR}/L_{\rm radio}$ varies by 2.1 at 500 MHz; $L_{\rm TIR}/L_{\rm radio}$ varies by 2.4 at 4.8 GHz).

Our models also predict that the normalization of the FRC should change with the observed frequency.  In general, $L_{\rm TIR}/L_{\rm radio}$ decreases with increasing frequency.  This effect is stronger for the starburst galaxies, where the nonlinearities in the predicted FRCs appear.  The radio-brightness at high frequencies is a direct consequence of the strong bremsstrahlung and ionization cooling in our models: synchrotron losses are more efficient relative to bremsstrahlung and ionization at higher energies, so that more energy goes into radio emission.  Only when $\alpha_{\nu}$ reaches $1$ does the radio emission begin to decrease with frequency.\footnote{This is also a generic prediction if there are loss processes that dominate synchrotron at low energies.  For example, galaxies are radio dim at low frequencies if they have strong diffusive losses ($t_{\rm diff} \propto E^{-1/2}$) or winds ($t_{\rm wind}$ constant with $E$).  A large $L_{\rm TIR}/L_{\rm radio}$ also arises if there is radio absorption at low frequencies.}

Direct comparison between our models and observations can be difficult, because the FRC is usually considered in terms of luminosity rather than $\Sigma_g$ and because the FRC is often fit as a nonlinear function.  We can nonetheless make some qualitative comparisons between observations and our models.  The observed 151 MHz correlation appears to be nonlinear, with luminous galaxies being brighter in the radio than would be predicted from the FIR (\citealt{Cox88} find $L_{\rm FIR} \propto L_{\nu}^{0.87 \pm 0.04}$).  Our models predict the opposite effect if $L_{\rm TIR}$ increases monotonically with $\Sigma_g$, with $L_{\rm TIR}/L_{\rm radio}$ increasing with $\Sigma_g$.  \citet{Fitt88} attribute the observed non-linearity at these frequencies to the FIR emission of old stars, and infer a linear FRC when they remove this effect.  It is also worth noting that Arp 220 is radio dim at 151 MHz, though this may be due to free-free absorption \citep{Sopp89,Condon91}.  At 4.8 GHz, the FRC is known to be tight (0.2 dex dispersion) and approximately linear \citep{deJong85, Wunderlich87}, though our models predict that starbursts should be radio bright compared to their FIR fluxes at these frequencies.  At 10.5 GHz, most of the radio emission is thermal and not from synchrotron.  \citet{Niklas97b} estimates the contribution from synchrotron alone and finds a nonlinear dependence $L_{\nu} \propto L_{\rm FIR}^{1.25 \pm 0.09}$, so that the FRC tilts towards stronger radio emission at higher luminosities.  Assuming that $L_{\rm TIR}$ increases with $\Sigma_g$, we find a qualitatively similar behavior in our models.  However, \citet{Niklas97b} also finds a non-linear FRC at 1.4 GHz, with no dependence on frequency for the slope of the FRC, in contrast to \citet{Yun01} who find a linear correlation (except at low luminosities) but only consider the FRC at 1.4 GHz.

At least two effects we do not include would complicate our predictions.  At low frequencies, free-free absorption may significantly lower the radio flux beyond what we predict in starbursts.  \citet{Condon91} argue that free-free absorption is important even at GHz frequencies in starbursts like Arp 220, and it becomes more effective at low frequency.  This effect would make the low frequency FRC even more nonlinear than we predict.  Thermal emission becomes significant at high frequencies (\citealt{Niklas97b} estimates that $\sim 30 \%$ of the radio emission is thermal at 10.5 GHz).  While the thermal contribution can be estimated and subtracted off, at very high frequencies it may so overwhelm the synchrotron radiation that studying the correlation between FIR and nonthermal radio becomes impossible.\footnote{A linear correlation between \emph{thermal} radio emission and radio emission is predicted and observed \citep[e.g.,][]{Condon92,Niklas97b}, though it provides no information on the cosmic rays or magnetic fields in a galaxy and is beyond the scope of this paper.}

\subsection{The Spectral Slope $\alpha$}
\label{sec:AlphaDiscussion}

\begin{figure}
\centerline{\includegraphics[width=8cm]{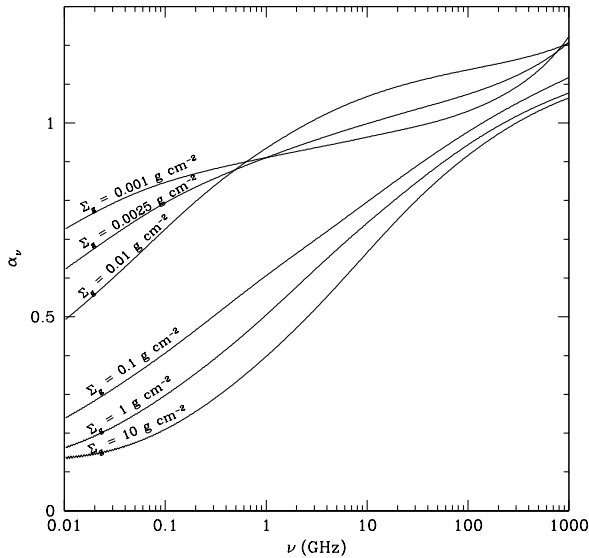}}
\figcaption[figure]{The instantaneous spectral slope of the (nonthermal) synchrotron emission as a function of frequency.  In this plot, $\alpha_{\nu}$ is the instantaneous spectral slope, $d\log~F_{\nu} / d\log~\nu$.  Elsewhere in the paper, $\alpha$ is the observable $\alpha_{1.4}^{4.8}$.\label{fig:AlphaNu}}
\end{figure}

In the Milky Way, the observed spectral slope $\alpha$ increases (the spectrum steepens) with frequency.  At low frequencies ($\la 100~\MHz$), the spectral slope is only $\approx 0.4 - 0.5$ \citep[e.g.,][]{Andrew66,Rogers08} but $\alpha$ reaches $\approx 0.75 - 0.8$ at GHz frequencies and reaches $\approx 0.8 - 0.9$ at several GHz \citep{Webster74,Platania98,Platania03} before free-free emission flattens the spectrum \citep[e.g.,][]{Kogut09}, though there are variations with direction and Galactic latitude \citep[for example,][]{Reich88}.  In fact, our models predict a steepening with frequency, though in our standard model $\alpha_{\nu}$ is higher than observed at all frequencies for $\Sigma_g = 0.0025~\gcm2$: it is $0.79$ at 100 MHz, $0.91$ at 1 GHz, and $1.00$ at 10 GHz (Figure~\ref{fig:AlphaNu}; see also Figures~\ref{fig:SynchSpectra} and~\ref{fig:Alpha}).  We note that $\alpha$ can be decreased by adjusting $p$; a value of $p = 2.1$ can decrease $\alpha$ by 0.1.  We also note that we used an escape time that increased as energy decreased; if the escape time is constant or even decreasing at low energies \citep[e.g.,][]{Engelmann90,Webber08}, then our low frequency $\alpha$ will also decrease, and be more in line with observations.  The predicted $\alpha_{\nu}$ for $\Sigma_g = 0.01~\gcm2$ are somewhat better at low frequencies: $0.73$ at 100 MHz, $0.93$ at 1 GHz, and $1.07$ at 10 GHz.  Models with $p \approx 2.0 - 2.1$ do a better job of matching the observed spectral slopes of the Milky Way.

As can be seen by the cooling and escape times in Figure~\ref{fig:tCool}, our standard model implies that escape, synchrotron, and bremsstrahlung all can shape the spectrum in normal galaxies.  Escape dominates at low surface densities, while all three are comparable for the inner regions of galaxies ($\Sigma_g \approx 0.01~\gcm2$).  Our greatest problem with $\alpha$ for normal galaxies is that it is predicted to increase slightly with $\Sigma_g$ (Figure~\ref{fig:Alpha}).  This would imply that the inner regions of spirals would have steeper spectra than the outer regions, when in fact the opposite effect is observed \citep[e.g.,][]{Murgia05}.  

The reason for the steepening is that escape becomes less effective as the galaxies become denser, so that cooling prevails.  In normal galaxies, synchrotron dominates bremsstrahlung by a factor of a few (and ionization by an order of magnitude), and the ratio of the synchrotron to bremsstrahlung cooling times is only weakly dependent on density (compare the short-dashed synchrotron and the long-dashed bremsstrahlung lines in Figure~\ref{fig:tCool}).  For a constant scale height so that $B \propto \langle n \rangle^a$, we have from equations~\ref{eqn:tSynch} and~\ref{eqn:tBrems} that $t_{\rm synch} / t_{\rm brems} \propto \langle n \rangle B^{-3/2} \propto \langle n \rangle^{1 - 3a/2}$, which is essentially constant for $a = 0.7$ and slowly changing for $a = 0.5$.  In contrast, from equations~\ref{eqn:StandardCRLife}, \ref{eqn:tSynch}, and~\ref{eqn:nuCSynch}, we find that at fixed frequency $t_{\rm synch} / t_{\rm diff} \propto B^{-7/4} \propto \mean{n}^{-7a/4}$, roughly inversely proportional to $\mean{n}$.  This implies that synchrotron losses become much more effective than escape as density increases, but the bremsstrahlung losses remain a factor of a few less important than synchrotron losses.  Therefore, as normal galaxies become calorimetric, their radio spectra will become steep in our model, since bremsstrahlung is only important enough to flatten the spectrum from its pure synchrotron-cooled limit of $\alpha_{\nu} = p / 2 \approx 1.1$ to $\alpha_{\nu} \approx 0.9 - 0.95$.   

This problem remains for all of the variants (Appendix~\ref{sec:Variants}) that satisfy local or integrated constraints, except in the strong wind variant (Section~\ref{sec:Winds}) in which we include advective escape that would result from the wind inferred by \citet{Everett08}, and the fast diffusive escape variant (Section~\ref{sec:FastEscape}).  If normal galaxies all host similar winds from their inner regions, escape prevents electrons from fully cooling, and our strong wind variant would imply that $\alpha$ would decrease to $\sim 0.75 - 0.8$ at these densities.  In our fast diffusive escape model, the electrons are similarly prevented from fully cooling, and $\alpha$ is slightly reduced in normal galaxies to $\sim 0.85 - 0.90$.  However, the efficient escape in these models tends to break the FRC.  

Although calorimeter theory often is said to produce too high $\alpha$, we consistently find that $\alpha$ is relatively low for starbursts (Figures~\ref{fig:Alpha} and~\ref{fig:AlphaNu}).  The spectral slope at 1.4 GHz ranges from 0.7 for weak starbursts ($\Sigma_g = 0.1~\gcm2$) to 0.5 for extreme starbursts ($\Sigma_g = 10~\gcm2$) in our standard model.  In our models, the high densities in the starbursts (relative to the low-density radio disk of the Milky Way) cause the flat spectra.  CR electrons and positrons experience severe cooling by bremsstrahlung and ionization, lowering $\alpha$ \citep[cf.][]{Thompson06}.  Extreme starbursts are in fact observed to have flat spectra \citep{Condon91,Clemens08}, though \citet{Condon91} attribute the flat spectra to free-free absorption and argues that the intrinsic $\alpha$ is 0.7.  We also note that models that include the FIR optical depth in $U_{\rm ph}$ predict steeper spectra, since IC losses are more effective: our $\kappa_{\rm FIR} = 10~\cmg21$ model (Section~\ref{sec:OpticalThickFIR}) implies that $\alpha \approx 0.65$ in starbursts. 

\begin{figure}
\centerline{\includegraphics[width=8cm]{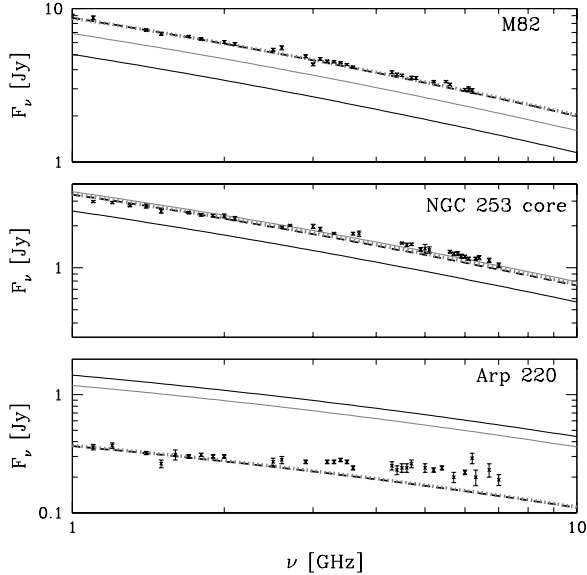}}
\figcaption[figure]{The predicted synchrotron radio spectra of the starbursts in M82, NGC 253, and Arp 220, compared with Allen Telescope Array observations from \citet{Williams10}.  Our fiducial model is in black (solid -- using Schmidt law and estimated starburst volume; dashed -- scaled to 1.4 GHz flux) and our fiducial model with winds and $B \propto \rho^a$ is in gray (solid -- using Schmidt law and estimated starburst volume; dotted -- scaled to 1.4 GHz flux).  We do not include free-free absorption or thermal emission. \label{fig:StarburstRadio}}
\end{figure}

As an example of the power and limitations of our approach, we show the predicted synchrotron radio spectra of the starbursts in M82, NGC 253, and Arp 220 of our fiducial model in Figure~\ref{fig:StarburstRadio}.  We calculate the radio emission using the Schmidt law, and assuming a disk geometry with the radius of the starburst from \citet{Thompson06} and scale height $h = 100\ \pc$.  Our fiducial model underpredicts the radio emission of M82 by a factor of $\sim 2$ and overpredicts the radio emission of Arp 220 by a factor of $\sim 4$.  This is caused by scatter in the Schmidt law and the FRC, which our models do not currently account for.  However, if we normalize the radio spectra to the observed 1.4 GHz (dashed line in Figure~\ref{fig:StarburstRadio}), we find that our models predict the radio spectra surprisingly well.  The spectra of NGC 253 and Arp 220 are slightly flatter than predicted, which is probably due to free-free absorption.  A variant with winds and $B \propto \rho^a$ similarly predicts the spectral shape, although not the normalization (gray lines in Figure~\ref{fig:StarburstRadio}).  While the fiducial model is no replacement for individual models of galaxies, which predict the correct normalization of the radio spectra and model the thermal emission and absorption, Figure~\ref{fig:StarburstRadio} demonstrates that the GHz radio spectra of starbursts in general can be understood well in terms of the high-$\Sigma_g$ conspiracy.

\subsection{The $\gamma$-Ray (and Neutrino) Luminosities of Starbursts}
\label{sec:GammaRays}

\begin{figure}
\centerline{\includegraphics[width=8cm]{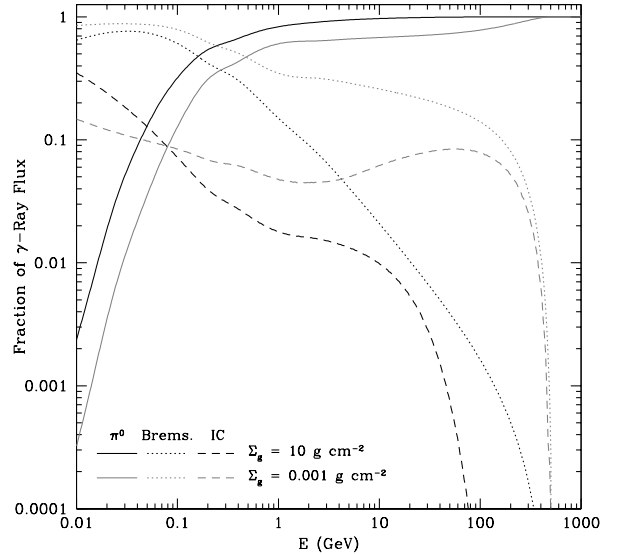}}
\figcaption[figure]{The fractional contributions of $\pi^0$ decay (solid), bremsstrahlung (dotted), and IC (dashed) to the $\gamma$-ray flux for $\Sigma_g = 0.001~\gcm2$ (gray) and $\Sigma_g = 10~\gcm2$ (black).  The drop in bremsstrahlung and IC past 100 GeV comes from the 500 GeV cutoff in our electron spectra. \label{fig:GammaRayRatios}}
\end{figure}

Starburst galaxies are predicted to be strong sources of $\gamma$-rays, observable with \emph{Fermi} and Very High Energy (VHE) telescopes.  Previous studies have considered NGC 253 \citep{Domingo05}, M82 \citep{Persic08,deCeaDelPozo09}, Arp 220 \citep{Torres04}, and the diffuse $\gamma$-ray background \citep{Thompson07}.  Several starburst galaxies have already been observed in VHE $\gamma$-rays to search for the emission.  Until recently, only upper limits were available on their $\gamma$-ray emission \citep[e.g.,][]{Aharonian05,Albert07}.  However, detections of NGC 253 and M82 have now been announced with VHE telescopes \citep{Acciari09,Acero09} and \emph{Fermi} \citep{Abdo10}.  

Pionic $\gamma$-rays come from CR protons in the ISM of the starbursts.  Since our explanation of the FRC requires that secondary electrons and positrons contribute to the radio emission, the $\gamma$-ray luminosities of starbursts are a useful test of the high-$\Sigma_g$ conspiracy.

We calculate the $\gamma$-ray flux\footnote{We do not include any optical depth to $\gamma$-rays in our calculations.  However, \citet{Torres04} found that Arp 220 was opaque to $\gamma$-rays only at energies above 1 TeV, and this should also be true for galaxies with a lower surface density.} from secondary $\pi^0$ decay for M31, NGC 253, M82, and Arp 220 in Table~\ref{table:GammaRayFluxes} as a check on our models.  We use the Schmidt law and the $\Sigma_g$ from \citet{Kennicutt98} and \citet{Thompson06} to calculate the emissivities of gamma rays for these systems, which we then multiply by volume (from the radii given in \citealt{Thompson06} and the scale heights in Section~\ref{sec:Height}) to get total luminosities to be converted to fluxes.  Since we are using approximate relations such as the Schmidt law, our models will be less accurate than more detailed models of individudal galaxies, and the predicted $\gamma$-ray luminosities are rough estimates only.  These models are \emph{not} meant to replace individual models of starburst galaxies.  The main advantage of our approach is only that we consider starbursts like M82 and NGC 253 in the broad context of all star-forming galaxies spanning the range between normal galaxies and ULIRGs; our models are necessarily more qualitative than more specific predictions.

Inelastic proton-proton collisions will also create neutrinos and antineutrinos.  The total neutrino ($\nu + \bar{\nu}$) flux is approximately equal to the $\pi^0$ $\gamma$-ray flux at energies $E \gg m_{\pi} c^2 \approx 140~\MeV$ \citep{Stecker79,Loeb06}.  Although we do not calculate the neutrino flux directly, we note that the values listed in Table~\ref{table:GammaRayFluxes} would also be good estimates for the neutrino fluxes, summed over all flavors and including both neutrinos and antineutrinos.

Bremsstrahlung and IC emission also are expected to contribute to the gamma-ray luminosities, especially at low energies.  We calculate the bremsstrahlung spectrum for M31, NGC 253, M82, and Arp 220 with our standard parameters.  Both our standard model and our fiducial wind model imply that in starbursts bremsstrahlung emission equals the total pion emission at 100 MeV and decreases at higher energies (see Figure~\ref{fig:GammaRayRatios}).  In less dense galaxies, bremsstrahlung grows in importance, but is still a minority contributor above 100 MeV.  The high energy fall-off for bremsstrahlung comes from the steepness of the electron and positron spectra relative to the proton spectra.  About half of the energy in the bremsstrahlung emission is below 100 MeV, because the electron spectrum steepens above 100 MeV.

The IC emission, when integrated over energy, is less than the bremsstrahlung or pion $\gamma$-ray emission (Figure~\ref{fig:Emiss}).  An IC gamma-ray spectrum would require an incident spectrum including CMB, dust, and stellar emission.  To get a feel for the IC emission, we model the background emission as three blackbodies: the CMB, a dust component (20 K in normal galaxies and 50 K in starbursts), and a direct stellar component (10000 K).  The dust component and the stellar component have a total energy density of $U_{\rm ph,\star}$ (eq. \ref{eqn:UphThin}) and are scaled according the UV optical depth $\tau_{\rm UV}$ (see Section~\ref{sec:Constraints}).  

\begin{figure}
\centerline{\includegraphics[width=8cm]{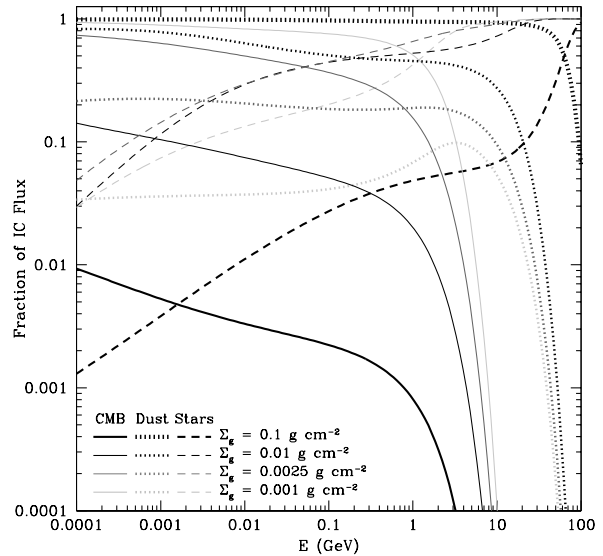}}
\figcaption[figure]{The fractional contributions of the CMB, dust emission, and starlight to the IC emission of galaxies.  For starbursts with $\Sigma_g > 0.1~\gcm2$, upscattered emission from dust dominates at all shown energies.  The drop in the contribution of the CMB near $\sim 1\ \GeV$ and dust past $\sim 10\ \GeV$ is an artifact of our 500 GeV cutoff in the electron and positron spectra. \label{fig:ICRatios}}
\end{figure}

We find that IC is smaller than bremsstrahlung for energies above about 1 MeV, and smaller than pion $\gamma$-rays above about 50-80 MeV.  At low energies, dust emission dominates the IC emission in galaxies with $\Sigma_g \ga 0.01~\gcm2$ (see Figure~\ref{fig:ICRatios}).  Upscattered UV emission from stars only dominates at high energies, and the CMB dominates the low energy IC emission in low surface density galaxies.  Our predicted spectra for starbursts have a precipitous fall-off in IC emission for the starbursts past 10 GeV, because there are no electrons above our 500 GeV cutoff to boost FIR photons to higher energies.  In low surface density normal galaxies, the CMB continues to provide the photons for most of the IC emission, so the drop is at $\sim 1\ \GeV$, with UV starlight providing higher energy photons (see also Figure~\ref{fig:ICSpectra}).  Most of the energy in the IC emission is at low energies, with more than half of the upscattered IC photons having less than 5 MeV.  This is because the electron spectrum steepens for $E > 100~\MeV$ ($\gamma \approx 200$) and incident photons are upscattered in energy by a factor of about $\gamma^2$; an incident 10 eV photon would typically only be boosted to a few hundred keV.

Considering the uncertainties and approximations in our approach, our models are in loose order-of-magnitude agreement with more sophisticated models (Table~\ref{table:GammaRayFluxes}).  The high-energy pionic $\gamma$-ray spectra ($\ga 1~\GeV$) are largely the same as previous models for M82 and Arp 220, although our M82 models are near the low range of the predictions of \citet{deCeaDelPozo09}.  We predict lower fluxes for NGC 253 than previous models by \citet{Domingo05} and \citet{Rephaeli09}, by a factor of $\sim 4 - 13$.  In particular, we predict an integrated flux of $F_{\gamma} (\ge 100\ \MeV) \approx 4 \times 10^{-9}$ for NGC 253.  This is substantially smaller than the predictions of $2.3 \times 10^{-8}$ by \citet{Domingo05} and $1.8^{+1.5}_{-0.8} \times 10^{-8}$ by \citet{Rephaeli09}.  

Interestingly, our predictions for M82 and NGC 253 are comparable to the \emph{Fermi} and VHE detections \citep{Acciari09,Acero09,Abdo10}.  However, this agreement is caused by a fortuitous cancellation of factors.  In Table~\ref{table:GammaRayFluxes}, we show that the predicted SN rates for M82 and NGC 253 using the Schmidt law are very small.  If we combine the IR luminosities of \citet{Sanders03} with equations~\ref{eqn:StarEmiss} and~\ref{eqn:CRpNorm}, we find SN rates of $0.065~\yr^{-1}$ for M82 and $0.039~\yr^{-1}$ for NGC 253.  Only half of the FIR luminosity of NGC 253 comes from its starburst core \citep{Melo02}, so its starburst has a supernova rate of $\sim 0.019~\yr^{-1}$.   By contrast, our model using the Schmidt law implies SN rates that are smaller by a factor of~$\sim 3$ for M82 and $\sim 2$ for NGC 253's starburst core.\footnote{If we scale to the 1.4 GHz radio luminosity from \citet{Williams10} instead of the TIR luminosity, we find that M82 must be scaled up by $\sim 1.8$ and NGC 253 by $\sim 1.3$ in our fiducial model.  The agreement is better still in our fiducial model with winds and $B \propto \rho^a$.}  If we scale the $\gamma$-ray fluxes to these SN rates, then the $\gamma$-ray fluxes are near or somewhat above the upper ranges of previous models.  Furthermore, our rescaled fluxes for M82 and NGC 253 are then about twice as high as observed.

Other differences with the models arise because we also use different distances to the starbursts (we use 3.5 Mpc for NGC 253 instead of 2.5 Mpc, as \citealt{Domingo05} and \citealt{Rephaeli09} did), but the other models fit the observed radio emission so a greater distance would be fit with a greater luminosity in these models.  We also use different low-energy energy spectra (we simply use $E^{-p}$ instead of $K^{-p}$ or $q^{-p}$ where $q$ is momentum), which will tend to underestimate the low energy CR proton spectrum, although the higher energy CR proton spectrum will be largely the same.  We again emphasize that our current generic models cannot replace existing models, but are meant as a demonstration of principle for the broad range of star-forming galaxies.

Our predicted fluxes do not change significantly if we consider models with winds and $B \propto \rho^a$ (see Section~\ref{sec:Winds}), though the TeV fluxes of starbursts are higher with this variant because we use $p = 2.2$ instead of $2.3$.  The $\gamma$-ray fluxes also provide good tests of the high-$\Sigma_g$ conspiracy in our models.  The $\pi^0$ $\gamma$-ray fluxes we predict for starbursts are mainly determined by proton calorimetry, the fraction of electron power lost to synchrotron, and the Milky Way CR proton normalization.  Note that our fluxes are several times greater than those predicted by \citet{Thompson07}, who assumed proton calorimetry but did not take into account non-synchrotron losses.  The significant bremsstrahlung, ionization, and IC losses in our model requires more (secondary) electrons and positrons to get the same radio emission, in turn requiring more protons.  

Data from \emph{Fermi} and VHE telescopes can distinguish these scenarios.  Proton calorimetry implies a hard $E^{-2.2}$ $\gamma$-ray spectrum instead of the Galactic $E^{-2.7}$ spectrum: \emph{proton calorimetry increases the high-energy $\gamma$-ray emission}.  The detections of starburst galaxies with  VHE telescopes support a hard spectrum and proton calorimetry.  Note that M31 is much fainter than the starbursts in VHE $\gamma$-rays, because of its steeper escape-dominated CR proton spectrum, though the flux of $E \la 1~\GeV$ $\gamma$-rays from M31 is similar to that from nearby starbursts.  Bremsstrahlung and IC $\gamma$-rays overwhelm $\pi^0$ $\gamma$-rays at $E \la 100~\MeV$ in our model, as seen in Figure~\ref{fig:GammaRayRatios}: \emph{non-synchrotron losses increase the low-energy $\gamma$-ray emission}.  \emph{Fermi} detection of this low energy emission would support the importance of non-synchrotron cooling.  Finally, a different proton normalization simply changes the amount of both CR protons and secondary electrons and positrons: \emph{high proton normalization increases $\gamma$-ray emission at all energies, without changing the spectrum}.  

The current \emph{Fermi} and VHE detections of M82 and NGC 253 are somewhat ambiguous, because these starbursts are relatively weak and there is no spectral information at 100~MeV yet \citep{Abdo10}.  The implied GeV-to-TeV spectral slopes are $\sim 2.2 - 2.3$, which is consistent with proton calorimetry \citep{Acciari09,Acero09,Abdo10}.  However, the fluxes are lower than our predicted fluxes scaled to the IR luminosities of these galaxies.  This can imply that either proton calorimetry is weaker, or the high-$\Sigma_g$ conspiracy is weaker, particularly in NGC 253.  We note that several groups have estimated $\Sigma_g \approx 0.1 - 0.2~\gcm2$ for these starbursts, so that the observed winds could be sufficient to break proton calorimetry.  More data and more sophisticated modeling are needed to fully understand the implications of these $\gamma$-ray observations.  Future detections of ULIRGs, which are more likely to be proton calorimeters, would be particularly helpful in understanding whether there is a high-$\Sigma_g$ conspiracy.  

\subsection{The Dynamical Importance of Cosmic Ray Pressure}
\label{sec:CRPressures}

\begin{figure}
\centerline{\includegraphics[width=8cm]{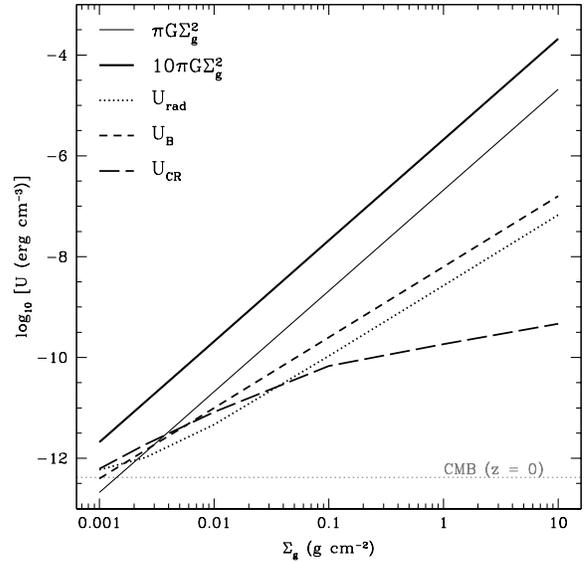}}
\figcaption[figure]{The importance of magnetic, radiation, and CR pressures compared to the hydrostatic pressure needed to support a galactic disk.  The hydrostatic pressure needed to support the gas alone is $\pi G \Sigma_g^2$.  In low-density galaxies, the mass of the stars implies that $P_{\rm hydro} = 10 \pi G \Sigma_g^2$ (see the discussion in Section~\ref{sec:CRPressures}).  The cosmic ray energy density does not increase as quickly as radiation and magnetic field energy densities in starburst galaxies.  None of the three components provides enough pressure to support starburst galaxies. \label{fig:Pressure}}
\end{figure}

Near the Solar Circle in the Milky Way, the CR energy density approximately equals magnetic field energy density, gas pressure, and radiation energy density.  Their pressure is also comparable to the pressure needed to support the Milky Way hydrostatically, $P_{\rm hydro} \approx \pi G \Sigma_g \Sigma_{\rm tot} \approx 10 \pi G \Sigma_g^2$, where $\Sigma_{\rm tot} \approx 10 \Sigma_g$ is the surface density of all matter in the Galactic disk.  Extrapolating from the Milky Way \citep{Chevalier84,Everett08}, \citet{Socrates08} hypothesize that CRs continue to provide significant pressure support and that they drive strong winds. \citet{Jubelgas08} have also explored the dynamical importance of CRs in galaxies, and conclude that they are not important for starbursts.

We show in Figure~\ref{fig:Pressure} the pressure from magnetic fields, radiation, and CRs in our standard model, compared to the hydrostatic pressure needed to support a galactic disk.  Magnetic fields and radiation (including FIR light) remain comparable as $\Sigma_g$ increases, as predicted by \citet{Thompson06}.  For the inner Milky Way, the predicted CR pressure is $2.7 \times 10^{-12}~\ergcm3$, within a factor of 2 of the derived best-fit CR pressure in \citet{Everett08}.  CR pressure remains in rough equipartition with magnetic field and radiation pressure until the weak starbursts, but then increases much more slowly.  As CR pressure is mainly provided by protons, the failure of CR pressure is caused by pion losses: starbursts are proton calorimeters, converting most of the CR proton energy into gamma rays and neutrinos that escape the system.  Our results are consistent with the low average CR pressure in starbursts found by \citet{Jubelgas08}, though the CR contribution may increase near the starburst edges \citep[Appendix C of][]{Socrates08}.

\section{Summary and Future Improvements}
\label{sec:Summary}

We model the FRC across the range $0.001~\gcm2 \le \Sigma_g \le 10~\gcm2$, from normal spirals to the densest starbursts.  The correlation holds in several scenarios described in Appendix~\ref{sec:Variants}.  We find that:

\begin{itemize}
\item We are able to reproduce a linear FRC (Figure~\ref{fig:Calorimetry}) consistent with both local and integrated Galactic constraints on the energy in CR protons.  We find that $\xi \approx 0.021$ of an SN's energy goes into CR electrons and $\eta \approx 0.1$ goes into CR protons when $p \approx 2.3$ and using an $E^{-p}$ spectrum.

\item Starburst galaxies ($\Sigma_g \ga 0.1~\gcm2$) are UV, electron, and proton calorimeters for most possible scenarios.  In our standard model, normal galaxies with $\Sigma_g = 0.01~\gcm2$ are UV and electron calorimeters, but they are not proton calorimeters with only $\sim 5\% - 15\%$ of CR proton energy going into pion losses.

\item The FRC is caused by calorimetry combined with two conspiracies operating in different density regimes.  At low $\Sigma_g$, decreasing electron calorimetry causes lower radio emission, but is balanced by decreasing UV opacity, which causes lower IR emission (Figure~\ref{fig:Calorimetry}).  At high $\Sigma_g$, bremsstrahlung, ionization, and IC losses decrease the synchrotron radio emission, while the appearance of secondaries and the effects of $B$ on $\nu_C$ increase the radio emission (Figure~\ref{fig:HighSGConspiracy}).

\item The magnetic field strength scales as $B \propto \Sigma_g^{0.6 - 0.7}$, implying $B \approx 1 - 2~\mGauss$ in extreme starbursts with $\Sigma_g = 10~\gcm2$.  Magnetic fields are significantly below equipartition with respect to gravity in starbursts.

\item The CR pressure remains in equipartition with radiation and magnetic field pressure for galaxies with $\Sigma_g \la 0.1~\gcm2$.  In starbursts, the CR pressure is significantly below equipartition, because of pion losses (long dashed line in Figure~\ref{fig:Pressure}).

\item Despite the short synchrotron and IC cooling timescales, our models reproduce the observed flattened radio spectra of starbursts (Figure~\ref{fig:Alpha}), because of the strong bremsstrahlung and ionization cooling in these galaxies.

\item Our models predict that FRCs exist at frequencies other than 1.4 GHz, though with increased non-linearity (Figure~\ref{fig:OtherNuLFIRRadio}).

\item Our predictions for the $\gamma$-ray emission from M82 and NGC 253 are within an order of magnitude of the \emph{Fermi}, VERITAS, and HESS detections.  However, these models assume the Schmidt law holds exactly for these starbursts.  If we normalize our models' IR emission to the observed IR emission, which should scale as star formation and CR injection power, we find that our $\gamma$-ray predictions are $\sim 1.5$ times higher than observations of M82 and NGC 253, possibly because of strong winds in these starbursts.  Our predictions Arp 220 are roughly in line with previous theoretical models, considering the approximations we make (Table~\ref{table:GammaRayFluxes}).  Full understanding of the $\gamma$-ray fluxes of these individual galaxies probably requires more refined models.

\end{itemize}

Our models still have several unresolved issues.  We have trouble matching the spectral slope $\alpha$ to observations of normal galaxies: we predict spectra that are too steep.  A possible solution may be the presence of a wind lowering the escape time.  However the addition of a wind, as observed by \citet{Everett08}, to our models of the Milky Way tends to break the FRC.  It is also possible that stronger diffusive escape is present in the radio halos of normal galaxies than we used in our models, because the normal galaxy radio scale height is typically less than the CR scale height (see Section~\ref{sec:ScaleHeight}).

Our one-zone models include CR cooling processes and escape through diffusion (winds are considered in Appendix~\ref{sec:Winds}), and can test a variety of parameterizations for the environment CRs travel through.  Not every issue was considered in this paper, though.  A natural question would be how robust the FRC is to scatter in the properties of the host galaxy environments.  For example, the Schmidt law has a scatter of $\ge 0.3$ dex \citep{Kennicutt98}, comparable to the FRC's own scatter of about 0.26 dex \citep[e.g.,][]{Yun01}.  It is also unlikely that the magnetic field exactly scales as $\rho^a$ or $\Sigma_g^a$, or that the overdensity $f$ of ISM gas that CRs travel through would be exactly the same from galaxy to galaxy.

We focused on star formation and the CRs it produces in our models of the FRC.  However both radio and FIR emission have other sources.  Star formation drives thermal radio emission, which is important at high frequencies \citep[e.g.,][]{Condon92}.  Thermal free-free emission probably also dominates the radio for very low density galaxies like the Large Magellanic Cloud, where the luminosity is low and electrons escape easily \citep{Hughes06}.  In normal galaxies, old stellar populations contribute a significant amount of FIR light without generating CRs.  To account for this, we might have to distinguish between a warm component of FIR directly related to star formation and a cool FIR component that includes old stars, and make predictions for the FIR colors of galaxies and starbursts \citep{Helou86}.  

An obvious modification would be to apply our models to higher redshifts.  Although the FRC has been mainly studied in the low-$z$ universe, there have been several recent studies of high-$z$ star-forming galaxies.  Recent observations by \citet{Vlahakis07} have found that starbursts become radio brighter at high redshift relative to the $z \sim 0$ FRC \citep[see also, e.g.,][]{Kovacs06,Murphy09,Michalowski09}.  In general, though, calorimeter theory predicts that $L_{\rm TIR}/L_{\rm radio}$ should not change in starburst galaxies.  Other studies have found that the FRC holds unchanged at high redshifts \citep[e.g.,][]{Appleton04}.  At high redshifts, the CMB will have a greater energy density, implying greater IC losses.  However, the CMB will not be important in dense galaxies, except at the greatest redshifts (c.f. the CMB line and the starlight radiation line in Figure~\ref{fig:Pressure}).  More important are the morphology changes.  Many starbursts at high redshifts are observed to be kiloparsecs in radius instead of $\sim 100~\pc$ \citep[e.g.,][]{Chapman04,Biggs08,Younger08}, usually with moderate $\Sigma_g$ but at least one with a surface density comparable to Arp 220 \citep{Walter09}.  In these starbursts, the high-$\Sigma_g$ conspiracy can be unbalanced, altering $L_{\rm TIR}/L_{\rm radio}$.  We will explore these effects in detail in \citet{Lacki09}.

While we assumed that galaxies and starbursts are homogeneous, future improvements can be made by using simple few-zone models of the ISM.  In normal galaxies, the CRs are injected from a gas-rich thin disk, but can diffuse within a thicker radio disk containing much lower density ionized gas.  This is reflected in the radio emission in the Milky Way, which has both a thin and a thick disk, the latter providing most of the luminosity \citep{Mills59,Beuermann85}.  Two-zone models can account for these density variations.  Even within the gas-rich disk, the density can fluctuate wildly between the high-density molecular clouds and the low-density coronal phase.  The magnetic field and density also change from spiral arms to interarm regions, as well as with distance from the centers of galaxies.  Similarly in starbursts, most of the gas is believed to be in a phase with high density, while most of the volume is relatively low density \citep{Greve09}.  If the CR populations in such phases are not well mixed, bremsstrahlung and ionization losses would be weak in a low-density phase, but strong in a high-density phase.  The low-density phase, with more volume, would contain most of the CRs but might have a steep spectrum, while the high density phase, with more mass, would have fewer CRs but with a harder spectrum.  However, including these different phases would add additional parameters to the models.  The structure of the ISM phases would have to remain generic, because detailed information is only available for the Milky Way, but the FRC spans a vast range in star formation rate and gas surface density.  These parameters would complicate the conspiracy even further, since there would be more parameters to tune.

Ultimately, abandoning the one-zone (or few-zone) approach would be necessary for a full understanding of the FRC.  Our approach only considers the global properties, but the FRC holds locally in galaxies to sub-kiloparsec scales.  We could address the local properties of the FRC by making full diffusion models, similar to the GALPROP models for the Milky Way, for galaxies across the entire range of the FRC.  A complete theory might have to include time evolution as well: \citet{Murphy08} found that synchrotron emission is better correlated spatially with star formation in regions of high star formation, possibly because the CR electrons have not yet had time to diffuse.  Spatial diffusion and time dependence would make modeling vastly more complicated, but including them may eventually be worthwhile with future improvements in radio and infrared observations.

\acknowledgments
We thank the GALPROP team for making their code and its subroutines freely available.  Also, we would like to thank Igor Moskalenko for sharing his group's estimates of the Milky Way's total luminosity.  We thank E. J. Murphy, Rainer Beck, and the anonymous referee for a careful reading of the text and comments that improved the paper.  We also thank Eli Waxman and John Beacom for stimulating discussions and the Aspen Center for Physics where a part of this work was completed.  T. A. T. is supported in part by an Alfred P. Sloan Fellowship.  E. Q. is supported in part by the David and Lucile Packard Foundation and NASA grant NNG05GO22H.  B. C. L. is supported in part by NASA grant NNX10AD01G to T. A. T.

\appendix
\section{Variants}
\label{sec:Variants}

\begin{figure}
\centerline{\includegraphics[width=8cm]{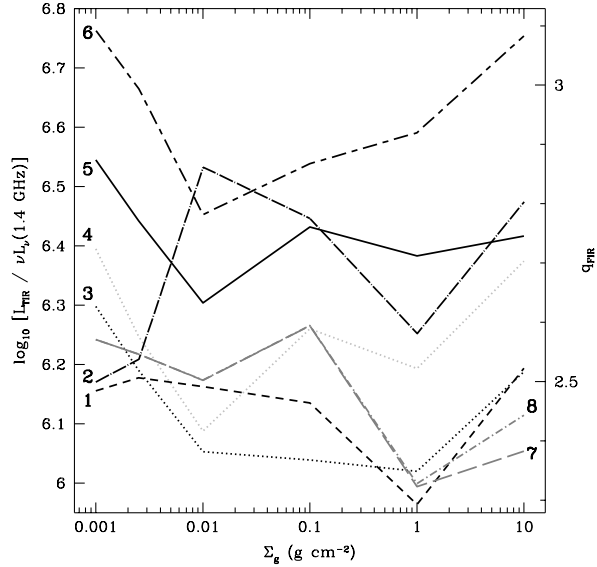}}
\figcaption[figure]{The FIR-radio correlation, as reproduced in our standard model and several variants.  We plot the values when $\xi = 0.008$ in each case.  Key: 1 (dash, black) -- $B \propto \rho^a$ with 300~$\kms$ wind in starbursts ($p = 2.2$, $f = 1.0$, $\tilde\delta = 67$, $a = 0.5$); 2 (long dash-dot, black) -- $B \propto \rho^a$ and strong winds of 175~$\kms$ in $\Sigma_g = 0.01~\gcm2$, 600~$\kms$ for starbursts ($p = 2.2$, $f = 1.5$, $\tilde\delta = 45$, $a = 0.5$); 3 (dotted, black) -- standard model with $\kappa_{\rm FIR} = 1~\cmg21$ ($p = 2.2$, $f = 1.0$, $\tilde\delta = 45$, $a = 0.7$); 4 (dotted, light gray) -- model with $\kappa_{\rm FIR} = 10~\cmg21$ ($p = 2.2$, $f = 2.0$, $\tilde\delta = 22$, $a = 0.8$); 5 (solid, black) -- standard model ($p = 2.3$, $f = 1.5$, $\tilde\delta = 48$, $a = 0.7$); 6 (long dash short dash, black) -- fast diffusive escape ($p = 2.2$, $f = 2.0$, $a = 0.6$, $\tilde\delta = 45$); 7 (long dash, gray) -- constant $D_z$, $B \propto \rho^a$, and winds of 300~$\kms$ in starbursts ($p = 2.2$, $f = 1.5$, $\tilde\delta = 34$, $a = 0.6$); 8 (dash-dot, gray) -- same as (7) but with FIR opacity of $\kappa_{\rm FIR} = 1~\cmg21$. \label{fig:LFIRRadio}}
\end{figure}

Our standard model (Sections~\ref{sec:Procedure} and \ref{sec:StandardModel}) does not include a variety of effects that may alter the CR populations in star-forming galaxies.  Our one-zone models of CR injection, cooling, and escape allow us to efficiently survey many scenarios.  These include variations in essentially all of the parameterizations of Section~\ref{sec:Procedure}.  We search for models that are successful under the observational constraints described in Section~\ref{sec:Constraints}.  We test several combinations of these effects with sparser grids of models, spanning $p = 2.0, 2.2, 2.4$ and $2.6$, and choose $f$ so that $fh = 1.0~\kpc$ and $2.0~\kpc$.

We summarize our results in Table~\ref{table:WorkingModels}.  Table~\ref{table:WorkingModels} shows the values of $p$, $f$, $a$, and $\tilde\delta$ that satisfy local-based constraints, the integrated Milky Way pion gamma-ray luminosity, and both together (Section~\ref{sec:Constraints}).  Despite the large number of scenarios tried, we find similar successful parameters in those scenarios that worked at all.  The FRC in some of these variants is shown in Figure~\ref{fig:LFIRRadio}.   Overall, we did not strongly constrain $p$ or $f$, with the range in allowed $f$ being determined mainly by the observed beryllium isotope ratios at Earth.  Specific variants occasionally imposed stricter constraints on $p$.  However, our models did place strong limits on the magnetic field energy density (in the form of $a$; eq.~\ref{eqn:StandardB}) and the CR energy density (in the form of $\tilde\delta$; eq.~\ref{eqn:DeltaTilde}).  We found $a$ to be 0.5 - 0.8, depending on whether $B$ was parametrized to vary with density or surface density.  Table~\ref{table:WorkingModels} shows the allowed $\tilde\delta$, because $\delta$ decreases by a factor of 100 as $p$ increases from 2.0 to 2.6 from the normalization issue discussed in Section~\ref{sec:xiEffects}.  In most cases, $\tilde\delta$ is within the range 35 - 100, as expected from the local proton-to-electron ratio $p/e$ of about 100.

\subsection{$B \propto \rho^a$}
\label{sec:Basn}
While our standard model assumes that $B \propto \Sigma_g^a$ (eq.~\ref{eqn:StandardB}), the magnetic field in galaxies may vary as $\rho^a$ \citep{Groves03}.  A $B \propto \rho^{0.5}$ scaling is also observed in Galactic molecular clouds \citep{Crutcher99}.  The difference matters most when comparing the FRC between starburst and normal galaxies, where the scale height changes by an order of magnitude in our models.  At the transition between the two regimes, since $h$ decreases by a factor of 10 for the starbursts, $B$ in this parametrization jumps by a factor of $10^a$, strengthening synchrotron cooling (Section~\ref{sec:BEffects}).

This variation on our standard model typically breaks the FRC at the transition between normal galaxies and starbursts.  Starbursts generally have a $L_{\rm TIR}/L_{\rm radio}$ that is several times smaller than normal galaxies, because of the dramatic increase in the magnetic field and corresponding synchrotron emission.  No model retains the correlation and also fulfills the local-based constraints on $e^+ / (e^- + e^+)$ and $p/e$, and only one model predicts the integrated Milky Way gamma-ray luminosity from pions.  Models which do satisfy local constraints on the proton normalization have $L_{\rm TIR}/L_{\rm radio}$ that vary by 2.4 at $p = 2.6$ and more at lower $p$ ($\ge 2.7$ at $p = 2.2$).  However, other variants in combination with $B \propto \rho^a$ turn out to restore the FRC (Section~\ref{sec:Winds}).

\subsection{Winds}
\label{sec:Winds}

We test the effects of including a wind of $300~\kms$ (a relatively ``weak'' wind) in all starbursts with $\Sigma_g \ge 0.05~\gcm2$, as discussed in Section~\ref{sec:Escape} (eq.~\ref{eqn:WindLife}).  We find that the FRC is broken for all values of the parameters.  In the best cases, $L_{\rm TIR}/L_{\rm radio}$ varies by $\sim 2.3$ over the range in $\Sigma_g$, while our upper limit to the allowed variation was 2.0 (eq.~\ref{eqn:FRCExists}), and this does not consider additional constraints on the proton normalization.  The difference comes from the $\Sigma_g = 0.1~\gcm2$ models, because the radio emission is halved as the $1.4~\GHz$ CR electrons escape before cooling: essentially, electron calorimetry fails for weak starbursts.  While the FRC is broken in models with winds and $B \propto \Sigma_g^a$ because synchrotron cooling is not strong enough, recall that the problem with the $B \propto \rho^a$ models (Section~\ref{sec:Basn}) was that synchrotron cooling was too strong in low-$\Sigma_g$ starbursts.  This suggests that models with both winds and $B \propto \rho^a$ might work.  

We test this conjecture by testing cases with winds and $B \propto \rho^a$. We then do find some models that satisfy our local constraints, including preserving the FRC.  Overall, the derived constraints on $p$, $f$, and $\delta$ are similar to our standard model (Section~\ref{sec:StandardModel}).  There is one noticeable difference in the allowed parameter space compared to our standard models: the models that do work now have $a = 0.5 - 0.6$ instead of $0.6 - 0.8$.  However, this still implies strong magnetic fields in starbursts, because $B$ does not just depend on $\Sigma_g$ but also $h$, where our adopted $h$ is 10 times smaller in starbursts than in the Milky Way.  If $B \propto \rho^{0.5}$, then the magnetic field strengths are $0.12~\mGauss$ in weak starbursts ($\Sigma_g = 0.1~\gcm2$) and $1.2~\mGauss$ in extreme starbursts ($\Sigma_g = 10~\gcm2$).  These magnetic fields are comparable in strength to those that would be present if $B \propto \Sigma_g^{0.7}$.  A model with $p = 2.2$, $\tilde\delta = 67$, $a = 0.5$, $f = 1.0$, and $\xi = 0.0119$ satisfies both local-based constraints and the integrated Milky Way $\gamma$-ray luminosity.  It also reproduces both the CR electron and proton fluxes at Earth above 10 GeV to within a factor of $\sim 2$.  The spectral slope $\alpha$ in starbursts is comparable to our standard models, being slightly lower because the CRs escape quicker.  

In this model the high-$\Sigma_g$ conspiracy operates in starbursts largely as it does in our fiducial model (see Figure~\ref{fig:HighSigmaGVariants}).  However, its onset is more gradual because of the winds in the $\Sigma_g = 0.1~\gcm2$ case, reducing secondary production in the weakest starbursts.  Furthermore, the low $a$ strengthens the magnetic field in the weakest starbursts, so that synchrotron can compete more effectively with bremsstrahlung and secondaries.  In dense starbursts, magnetic fields are relatively weak and a strong high-$\Sigma_g$ conspiracy sets the radio luminosity.

We also test stronger winds in combination with $B \propto \rho^a$, motivated by the inference of a wind in the inner regions of the Milky Way \citep{Everett08}, as well as the high wind speeds observed in starbursts \citep{Heckman00}.  The new wind speeds are $v = 175~\kms$ in the $\Sigma_g = 0.01~\gcm2$ model (comparable to the \citealt{Everett08} wind), and $v = 600~\kms$ in the starbursts.  This variant tends to break the FRC, mainly because CRs escaped too quickly in the $\Sigma_g = 0.01~\gcm2$ Milky Way analog model.  With a wind advecting away CRs at lower surface densities where cooling is weaker, most of these models are not sufficiently good electron calorimeters to preserve the FRC.  Further variants with strong winds are discussed in Section~\ref{sec:MultipleEffects}.

\subsection{Other Disk Scale Heights}
\label{sec:ScaleHeight}
\emph{Small normal galaxy disk scale heights.} In our standard scenario, the scale height jumps down from 1 kpc at $\Sigma_g = 0.01~\gcm2$ to 100 pc at $\Sigma_g = 0.1~\gcm2$.  The jump can cause discontinuities in the FRC if the parameters are not fine-tuned.  However, the gas disk of the Milky Way is only 100 pc thick, and there is a thin radio disk associated with it.  Thus, we try models where $h = 100~\pc$ for all $\Sigma_g$.  In this run, we consider $f$ values of 0.1 and 0.2 to continue to match the inferred ISM density that CRs propagate in the Milky Way from beryllium isotopes.  Since $h$ does not vary, $\Sigma_g$ and $\rho$ are directly proportional.

We find that models with $p = 2.2 - 2.6$, $a = 0.5 - 0.6$, $f = 0.1 - 0.2$, and $\tilde\delta = 34 - 100$ satisfy both the integrated Galactic pion luminosity, and the local $e^+ / (e^+ + e^-)$ and $p/e$ constraints.  The relatively low $a$ is preferred since $f$ is low implying weak bremsstrahlung and ionization.  Since there is no jump in bremsstrahlung and ionization losses for starbursts because they have the same scale height as normal galaxies, a strong magnetic field would make starbursts too radio bright.  However, $p = 2.4 - 2.6$ produces CR spectra that are steeper than observed.  Higher $p$ are somewhat preferred: there is a window in parameter space with $a = 0.5$, $f = 0.1$, and $\tilde\delta = 50 - 100$ where $L_{\rm TIR}/L_{\rm radio}$ varies over the range in $\Sigma_g$ by 2.26 to 2.75 for $p = 2.0$ but 1.94 to 2.21 for $p = 2.4$, so that higher $p$ passes marginally.  The reason for the relatively high variation in $L_{\rm TIR}/L_{\rm radio}$ is that $f$ is low: starbursts are still proton calorimeters, producing secondaries that contribute to the radio; bremsstrahlung and ionization losses, which compete for the energy available for radio emission, are much weaker than in the standard model (Section~\ref{sec:FRCCauses}).  At high surface densities, secondaries become important and can make starbursts too radio bright; they are diluted more at higher $p$ (Section~\ref{sec:SpectralSlopeEffects}), so that $L_{\rm TIR}/L_{\rm radio}$ varies slightly less over the range in $\Sigma_g$.  The high-$\Sigma_g$ conspiracy is present in starbursts for the models, but its onset is more gradual, with non-synchrotron losses growing from a factor of $\sim 5$ at $\Sigma_g = 0.1~\gcm2$ to $\sim 10$ at $\Sigma_g = 10~\gcm2$ (see Figure~\ref{fig:HighSigmaGVariants}).  The more gradual onset arises because we use low $f$ to match the beryllium isotope constraints at Earth, weakening bremsstrahlung and ionization and reducing secondaries in weak starbursts; and low $a$, which increases relative synchrotron strength in weak starbursts and suppresses synchrotron in dense starbursts.

Furthermore, we find that the CR proton fluxes predicted at Earth at 1, 10, and 100 GeV are about 4 - 20 times higher than observed.  The predicted electron fluxes are also about 10 times higher than observed.  This can easily be understood: if $h$ is shrunk 10 times, the same number of CRs are being injected into a smaller volume, so that their number density is higher.  However, it is possible that the Earth resides in an atypical region of the Galaxy, in which case our integrated constraints alone allow this variant.

We also consider a less extreme version of this model, where $h_{\rm norm} = 300~\pc$, about the height of the Milky Way's thin radio disk \citep{Beuermann85}.  We use $f$ of 0.3-0.6 to match the beryllium isotope measurements.  This variant is less restrictive, allowing $p \ge 2.2$ for all considered constraints.  In all of the allowed models, $a = 0.6$ where $B \propto \Sigma_g^a$, midway between the $a = 0.5$ case preferred when $h_{\rm norm} = 100~\pc$ and the $a = 0.7$ preferred when $h_{\rm norm} = 1~\kpc$.  The CR flux predicted at Earth is still about 2 - 5 times higher than observed at Earth, for both protons and electrons in these models.  

\emph{Large normal galaxy disk scale heights.}  While the gas disks of normal galaxies are thin, the scale heights of the CRs themselves are estimated to be several kpc (see the discussion in Section~\ref{sec:Height}).  We therefore considered models with $h = 2\ \kpc$ and $h = 4\ \kpc$.  When $h = 2\ \kpc$, we find that the allowed parameters are similar to our standard value.  Higher $a$ and lower $fh$ is slightly preferred, because the jump in density between normal galaxies and starbursts is even greater than when $h_{\rm norm} = 1\ \kpc$; therefore, either higher magnetic fields or lower gas densities are needed to prevent bremsstrahlung and ionization from overwhelming synchrotron losses.  CR proton fluxes at Earth are somewhat small by a factor of $\sim 1.25 - 2$ when $p = 2.2$.  In these models, the high-$\Sigma_g$ conspiracy is present at an even greater magnitude than in our standard model, with non-synchrotron losses suppressing synchrotron emission by a factor of $\sim 15$ instead of $\sim 10$ (see Figure~\ref{fig:HighSigmaGVariants}).

When $h = 4\ \kpc$, no models preserve the FRC.  Essentially, since $f$ is constant for all $\Sigma_g$ in our models, and since it must be large to match the beryllium isotope constraints in the Milky Way, bremsstrahlung and ionization are inevitably extremely strong in starbursts.  While the density increases drastically from normal galaxies to starbursts because of the decreasing scale height, magnetic fields do not suddenly increase if they go as $\Sigma_g^a$, and therefore synchrotron cannot properly balance bremsstrahlung and ionization for a linear FRC.

\subsection{Optically Thick Galaxies}
\label{sec:OpticalThickFIR}
Normally, our models assume that the CRs propagated in ISM that was optically thin to FIR light.  Then, as stated in Section~\ref{sec:Environment}, $U_{\rm ph,\star} = F_{\star} / c$.  However, in a scattering atmosphere, the photon energy density may actually be greater if the environment is embedded in an optically thick region.  In that case, $U_{\rm ph,\star} =  (1 + \tau_{\rm FIR}) F / c$, where $\tau_{\rm FIR} = \kappa_{\rm FIR} \Sigma_g / 2$ acts as a midplane scattering optical depth.  

In models with $\kappa_{\rm FIR} = 1~\cmg21$, we find that we are still able to recreate the FRC and match both local and integrated Galactic constraints.  The parameter space allowed by this scenario is similar to our standard model (Section~\ref{sec:StandardModel}).  The increased IC scattering in extreme starbursts rules out $a = 0.6$, so that the magnetic field energy density remains comparable to the increased photon energy density.  We also typically recover the CR flux at Earth to within a factor of 2 for these models when $p = 2.2$ or $2.4$.

When $\kappa_{\rm FIR} = 10~\cmg21$, the FRC does not survive in any of our models.  The minimum variation in $L_{\rm TIR} / L_{\rm radio}$ is 2.03 when $a = 0.8$, $f = 2.0$, $\tilde\delta \la 25$.  Synchrotron losses need to keep up with IC losses in extreme starbursts, which would favor high $a$.  However, $a = 0.8 - 0.9$ often caused too severe synchrotron losses compared to bremsstrahlung and ionization in dense starbursts, and too weak synchrotron losses in low surface density galaxies.  Those models that nearly preserve an FRC have low $\delta$ and high $f$, which reduces the number of secondaries and increased bremsstrahlung and ionization cooling to compensate for the increased magnetic field strength and keep $L_{\rm TIR}/L_{\rm radio}$ sufficiently high.  

\subsection{$U_B = U_{\rm ph}$} 
\label{sec:UBeqUph}
Radiation pressure may drive turbulence in the ISM, until the energy densities in radiation and kinetic motions are comparable.  The turbulence can, in turn, generate magnetic fields.  As a result, it is possible that $U_{\rm ph} \approx U_{\rm turb} \approx U_{B}$ \citep{Thompson08}.  

We test models where $U_B$ was forced to equal $U_{\rm ph}$, where the radiation energy density includes both the CMB and starlight.  In optically thin models ($\tau_{\rm FIR} = 0$), the fiducial values for $p$, $f$, and $\tilde\delta$ (given in Section~\ref{sec:StandardModel}) recreate the FRC and match both local and integrated proton constraints (see Table~\ref{table:WorkingModels}).  Some models, generally those with $p = 2.2 - 2.4$, also correctly predict to within a factor of 2 the CR proton flux at Earth at energies $E = 1$, $10$, and $100~\GeV$, as well as the CR electron flux at Earth at $E = 10~\GeV$.  

A few optically thick models with $\kappa_{\rm FIR} = 1~\cmg21$ where $U_B$ is forced to $U_{\rm ph}$ satisfy the integrated Galactic $\pi^0$ luminosity, though only one satisfies local constraints.  In these models, the effectiveness of synchrotron increased in high surface density galaxies, because $U_B$ equaled the quickly increasing $U_{\rm ph}$.  The FRC generally fails because of a tension between intermediate densities and high densities.  Bremsstrahlung and ionization are stronger at intermediate densities ($0.01~\gcm2 \la \Sigma_g \la 1~\gcm2$) than in the standard model (since $U_B$ is lower than our standard prescription), requiring a high secondary fraction to compensate.  But when $\Sigma_g = 10~\gcm2$, synchrotron cooling overwhelms bremsstrahlung and ionization, making those galaxies too radio bright.  The $\nu_C$ effect (Section~\ref{sec:FRCCauses}) becomes particularly strong as $\tau_{\rm FIR}$ becomes appreciable, since $B$ is rapidly increasing, causing the radio emission to increase further.  The model that does satisfy all constraints has low $p$, increasing the secondary abundance but weakening the $\nu_C$ effect.

Increasing the FIR opacity to $\kappa_{\rm FIR} = 10~\cmg21$ in these variants completely breaks the FRC in all attempted models.  Again, the bremsstrahlung and ionization losses are strong at intermediate densities ($0.01~\gcm2 \la \Sigma_g \la 1~\gcm2$) but small at high densities, because both the magnetic field energy density and the radiation field increase sharply at $\Sigma_g \approx 1~\gcm2$.  Therefore starbursts would appear too radio bright to maintain a linear FRC.

\subsection{Fast Diffusive Escape}
\label{sec:FastEscape}
We have used a scale height of $h = 1\ \kpc$ for normal galaxies in most variants, based on the scale heights of radio disks.  However, the diffusive escape time we use in Equation~(\ref{eqn:StandardCRLife}) applies to the entire CR scale height of the Milky Way, which is of order $2 - 4\ \kpc$ (see the discussion in Section~\ref{sec:Height}).  The escape time from the radio disk itself may be significantly shorter.  Variants in which the scale height of normal galaxies is increased (Appendix~\ref{sec:ScaleHeight}) still may not properly model the radio disk, because they use the midplane magnetic field strength, possibly overestimating the effectiveness of synchrotron losses.  

We consider the effect of faster diffusive escape of by running models in which the diffusive escape time in galaxies is shortened by a factor of $4$ from the nominal lifetime in equation~\ref{eqn:StandardCRLife}.  We find that the FRC tends is broken in almost all models, because escape reduces the radio luminosity of the lowest surface density galaxies.  The minimum variation in $L_{\rm FIR}/L_{\rm radio}$ in any model is 1.998, barely under our criterion of 2.000; this models also satisfies the integrated Galactic pion luminosity ($p = 2.0$, $f = 2.0$, $a = 0.6$, $\tilde\delta = 35$, $\xi = 0.018$).  The models with the most-preserved FRC tend to have low $a$ of 0.6, weakening the synchrotron emission in high surface density galaxies.  On the other hand, the spectral slope of normal galaxies in these models is $0.85 - 0.90$ when $p = 2.2$, which is closer to the observed values than in our standard model (see Section~\ref{sec:AlphaDiscussion}).  The CR flux at Earth in models where $L_{\rm FIR}/L_{\rm radio}$ varies by $\le 2.2$ and $p = 2.2$ ranges from $\sim 35\%$ to $140\%$ of the observed values ($68\% - 142\%$ when the local and integrated Galactic proton constraints hold).

\subsection{Varying Escape Times}
\label{sec:DVary}
\emph{Constant $D_z$.} So far, we have been simply assuming that escape time by diffusion for CRs is the same in all galaxies and starbursts.  However, in our models, the scale height of starbursts is 10 times smaller than that of normal disk galaxies.  Another simple assumption would be that the vertical diffusion constant $D_z$ at any given energy is constant across star-forming galaxies and starbursts.  Then, since $t_{\rm diff} = h^2 / D_z$, the escape time would be a hundred times smaller for starbursts.  This could break the FRC at the transition between normal galaxies and starbursts.

To see whether this variation had any effect, we modify the escape time to $t_{\rm diff}(E) = t_{\rm diff,MW}(E) (h / h_{\rm MW})^2$.  We ran our grid for two cases that had worked previously: the standard model where $B \propto \Sigma_g^a$ and with no winds, and the case with (weak) winds and $B \propto \rho^a$ (see Section~\ref{sec:Winds}).  In standard models with constant $D_z$, the vastly more efficient escape of CRs in weak starbursts ($\Sigma_g = 0.1~\gcm2$) broke the FRC.  Models with $B \propto \rho^a$ and winds did reproduce the FRC and were able to satisfy local and integrated constraints.  The allowed values for $p$, $f$, $a$, and $\tilde\delta$ are similar to those for our models with $B \propto \rho^a$, winds, and constant $t_{\rm diff}$ (c.f. Section~\ref{sec:Winds} and Table~\ref{table:WorkingModels}). The increased magnetic field strength in starbursts (from $B \propto \rho^a$, where $\rho$ jumps up for starbursts) compensates for the decreased diffusive escape time to restore the FRC in these models.  The models generally predict low CR flux at Earth when $p = 2.2$, with a proton flux of about 50\% - 120\% of its observed Earth value.

\begin{figure}
\centerline{\includegraphics[width=8cm]{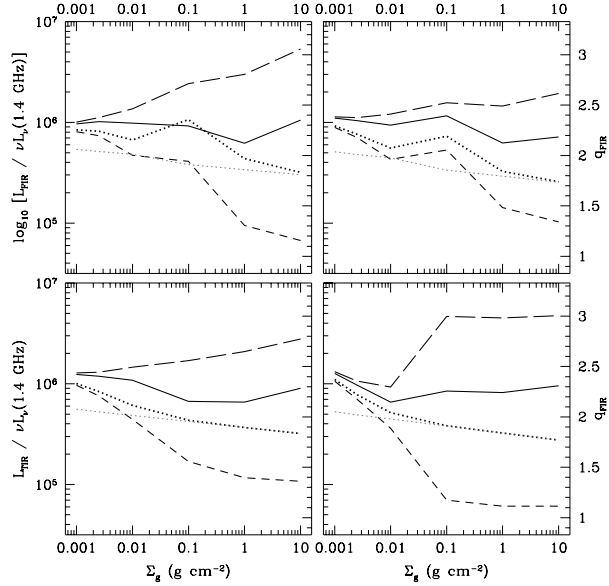}}
\figcaption[figure]{The high-$\Sigma_g$ conspiracy in several of our variants.  The line styles are the same as in Figure~\ref{fig:HighSGConspiracy}.  At upper left is a model with $B \propto \rho^a$ and moderate winds ($p = 2.2$, $f = 1.0$, $\tilde\delta = 67$, $a = 0.5$; Section~\ref{sec:Winds}); at upper right, a model with $B \propto \rho^a$, moderate winds, and constant $D_z$ ($p = 2.2$, $f = 1.5$, $\tilde\delta = 34$, $a = 0.6$; Section~\ref{sec:DVary}); at lower left, a model with $h = 100\ \pc$ in normal galaxies ($p = 2.2$, $f = 0.2$, $\tilde\delta = 34$, $a = 0.6$; Section~\ref{sec:ScaleHeight}); and at lower right, we show a model with $h = 2\ \kpc$ in normal galaxies ($p = 2.2$, $f = 2.0$, $\tilde\delta = 67$, $a = 0.7$; Section~\ref{sec:ScaleHeight}).  Although the strength and the onset of the high-$\Sigma_g$ conspiracy varies in these scenarios, it is always present in some form in the densest starbursts.  \label{fig:HighSigmaGVariants}}
\end{figure}

The models with constant $D_z$ weaken the high-$\Sigma_g$ conspiracy somewhat, but it remains present (Figure~\ref{fig:HighSigmaGVariants}).  The very strong diffusive losses and the winds mean that starbursts with $\Sigma_g \la 1~\gcm2$ are not proton calorimeters.  Since $B$ scales with $\rho^a$, synchrotron losses can compete more effectively with bremsstrahlung and ionization in weak starbursts.  Bremsstrahlung and ionization mainly balance the $\nu_C$ effect in these galaxies, which only lowers $L_{\rm TIR}/L_{\rm radio}$ by a factor of $\sim 50\%$ from $\Sigma_g = 0.001~\gcm2$ normal galaxies.  In denser starbursts, bremsstrahlung and ionization grow in importance, but are balanced by secondaries.  Overall the conspiracy presented in Figure~\ref{fig:HighSigmaGVariants} is weaker than in the fiducial model, with non-synchrotron losses suppressing synchrotron losses by only $\sim 2 - 8$ for starbursts, because $a$ is relatively high.

Note that constant $D_z$ from normal galaxies to starbursts is not necessarily correct.  The speed CRs can stream out of galaxies is expected to be limited to the Alfv\'en speed, $v_A = B / \sqrt{4 \pi \rho}$ (Section~\ref{sec:Calorimeter}; \citealt{Kulsrud69}).  For $B \propto \rho^{0.5}$, this comes out to a diffusive escape time of $3.7 h_{100} \times 10^6 \yr$, where $h_{100} = h / (100\ \pc)$.  While this is shorter than the Milky Way CR escape time (eq.~\ref{eqn:StandardCRLife}), it is also 10 times longer than the constant $D_z$ escape time (see also Figure~\ref{fig:tCool}).  These considerations suggest that diffusion is weaker in starbursts than in constant $D_z$ models, which would help preserve the FRC.  

\emph{$D_z \propto \rho^{-1/3}$.} \citet{Helou93} proposed that the diffusion constant scales as $\rho^{-1/3}$, which would make escape less efficient in high-density galaxies.  We try models where the escape time equaled its local value (eq.~\ref{eqn:StandardCRLife}) for the local surface density ($\Sigma_g = 0.0025~\gcm2$) and increasing as $\rho^{1/3}$ to account for this effect.  Testing this assumption against both our standard assumptions (no winds and $B \propto \Sigma_g^a$) and $B \propto \rho^a$ with winds, we find similar results to the constant $D_z$ case, although $a$ and $f$ are more severely constrained.  The faster escape time still breaks the FRC in otherwise-standard models.  When $B \propto \rho^a$, though, the models restore the FRC and satisfy local and integrated constraints.  Again, the allowed parameters are similar to the case with $B \propto \rho^a$, winds, and constant $t_{\rm diff}$ (cf. Section~\ref{sec:Winds} and Table~\ref{table:WorkingModels}).  The CR flux at Earth is correct, within about a factor of 2 of the observed values for the energies we considered.

\emph{$D_z \propto \rho^{-1}$.} Finally, we try a rapidly scaling diffusion constant, motivated by the parametrization in \citet{Murphy08}.  As before, the escape time is normalized to its local value (eq.~\ref{eqn:StandardCRLife}) at the local surface density ($\Sigma_g = 0.0025~\gcm2$), but now increasing as $\rho$.  We once again test it against $B \propto \Sigma_g^a$ with no winds and $B \propto \rho^a$ with winds.  This variant fails in both cases to create the FRC, because escape is too efficient in the lowest density galaxies ($\Sigma_g = 0.001~\gcm2$).

\subsection{Multiple Effects and Other Variants}
\label{sec:MultipleEffects}
\emph{Combinations with weak winds.} We finally consider scenarios that combine most of the previous variants.  We include starburst winds, the FIR optical depth, $B \propto \rho^a$, and constant $D_z$, $D_z \propto \rho^{-1/3}$, or $D_z \propto \rho^{-1}$ simultaneously.  Our results are essentially the same as the models with constant or varying $D_z$ considered in Section~\ref{sec:DVary} with $B \propto \rho^a$ and winds.  For constant $D_z$, the models reproduce the FRC and satisfy both local and integrated constraints.  As before, the models predict low CR proton flux at Earth.  Some models with $D_z \propto \rho^{-1/3}$ also are consistent with the FRC, but only only fulfills both local and integrated Galactic proton normalization constraints.  The increased photon energy density suppresses radio in $\Sigma_g = 10~\gcm2$ starbursts, thus making it hard to maintain the FRC over the entire range in $\Sigma_g$.  As before models with $D_z \propto \rho^{-1}$ fail to reproduce the FRC.

\emph{Combinations with strong winds.} The main problem with the strong wind scenario (Section~\ref{sec:Winds}) is the rapid escape of the CR electrons when $\Sigma_g = 0.01~\gcm2$.  More synchrotron is emitted before escape if the CRs have to travel a larger distance.  We therefore consider models with strong winds and $h_{\rm norm} = 2~\kpc$, with $f$ scaled to 2.0, 3.0, and 4.0 to match local isotope measurements.  We use a constant diffusion rate $D_z$, and try models with and without FIR opacity.  In both cases, we are able to satisfy both local and integrated constraints.

\begin{deluxetable}{ccll}
\tablecaption{List of Symbols Used}
\tablehead{\colhead{Symbol} & \colhead{Section} & \colhead{Standard} & \colhead{Definition} \\ & & \colhead{Value} & }
\startdata
\cutinhead{Derived parameters}
$p$ 		& \ref{sec:PrimaryInjection}; \ref{sec:SpectralSlopeEffects} & $2.3$ & Power law index of the injected spectrum of primary cosmic rays\\
$\xi$ 		& \ref{sec:PrimaryInjection}; \ref{sec:xiEffects} & $0.023$ & Fraction of supernova kinetic energy injected into primary CR electrons\\
$\eta$ 		& \ref{sec:PrimaryInjection} & $0.12$ & Fraction of supernova kinetic energy injected into primary CR protons\\
$\delta$	& \ref{sec:PrimaryInjection}; \ref{sec:xiEffects} & $5$ & $\eta / \xi$, the ratio of energy in injected protons to injected electrons\\
$\tilde\delta$  & \ref{sec:xiEffects} & $48$ & Proton-to-electron ratio at relativistic energy at injection; $\delta$ renormalized to remove $p$ dependence\\
$a$		& \ref{sec:MagneticField}; \ref{sec:BEffects} & $0.7$ & Power law scaling of galactic magnetic fields with surface or volume density\\
$f$		& \ref{sec:Density}; \ref{sec:DensityEffects} & $1.5$ & Ratio of density through which CRs propagate to average ISM density\\
\cutinhead{Other input parameters}
$\Sigma_g$		& \ref{sec:Procedure} & \nodata & Average gas column density\\
$\Sigma_{\rm SFR}$	& \ref{sec:Procedure} & \nodata & Star formation rate per unit area\\
$h$			& \ref{sec:Procedure}; \ref{sec:Height} & $1~\kpc - 100~\pc$ & Scale height of CR disk\\
$E$ 			& \ref{sec:Procedure} & \nodata & Total energy of cosmic ray\\
$t_{\rm life}(E)$       & \ref{sec:Procedure}; \ref{sec:Escape} & \nodata & Escape time for a particle of energy $E$ from the galaxy, includes both advection and diffusion\\
$Q(E)$			& \ref{sec:Procedure}; \ref{sec:PrimaryInjection} & \nodata & Energy spectrum of primary CRs injected into the ISM per unit volume\\
$b(E)$			& \ref{sec:Procedure} & \nodata & Energy loss rate per particle (positive for energy loss)\\
$C$			& \ref{sec:PrimaryInjection} & \nodata & Normalization of the injected energy spectrum of primary CRs\\
$\gamma$ 		& \ref{sec:PrimaryInjection} & \nodata & Lorentz factor of cosmic ray, $E / (mc^2)$\\
$\gamma_{\rm max}$	& \ref{sec:PrimaryInjection} & $10^6$ & Maximum Lorentz factor of a cosmic ray at injection\\
$K$ 			& \ref{sec:PrimaryInjection} & \nodata & Kinetic energy of cosmic ray\\
$\varepsilon$		& \ref{sec:PrimaryInjection} & $3.8 \times 10^{-4}$ & Radiative efficiency of stellar population\\
$E_{51}$		& \ref{sec:PrimaryInjection} & $1$ & Mechanical energy per supernova, in units of $10^{51}$ ergs\\
$\psi_{17}$    		& \ref{sec:PrimaryInjection} & $1$ & Conversion rate between the supernova rate per unit mass $\Gamma_{\rm SN}$ and starlight emissivity $\epsilon_{\rm ph}$\\
$t_{\rm diff}(E)$	& \ref{sec:Escape} & \nodata & Escape time for particle of energy $E$ from the galaxy by diffusion\\
$t_{\rm adv}$		& \ref{sec:Escape} & $\infty$ & Escape time for particle from the galaxy by advection in a wind\\
$\mean{n}$              & \ref{sec:Density} & \nodata & Average number density of hydrogen, $\Sigma_g / (2h)$\\
$n_{\rm eff}$		& \ref{sec:Density} & \nodata & Average hydrogen number density the CRs encounter, $f \mean{n}$\\
$U_{\rm ph,\star}$	& \ref{sec:RadiationField}; \ref{sec:UphEffects} & \nodata & Energy density in starlight (UV or reprocessed FIR)\\
$F_{\star}$		& \ref{sec:RadiationField} & \nodata & Starlight energy flux\\
$U_{\rm ph,CMB}$	& \ref{sec:RadiationField}; \ref{sec:UphEffects} & \nodata & Energy density of CMB\\
$\kappa_{\rm FIR}$ 	& \ref{sec:RadiationField}; \ref{sec:UphEffects} & $0$ & Effective ISM opacity to far-infrared light (FIR) \\
$B$			& \ref{sec:MagneticField}; \ref{sec:BEffects} & \nodata & ISM magnetic field strength\\
$\kappa_{\rm UV}$ 	& \ref{sec:Constraints} & $500~\cmg21$ & Effective ISM opacity to ultraviolet (UV) light\\
$\nu_C$			& \ref{sec:BEffects} & \nodata & Critical frequency of synchrotron radiation.\\
\cutinhead{Output}
$N(E)$		& \ref{sec:Procedure} & \nodata & Final steady-state spectrum of CRs, calculated per unit volume\\
$\epsilon$      & \ref{sec:Constraints} & \nodata & Emissivity (here, power per volume), for radiation or a CR loss process\\
$\epsilon_{\nu}$& \ref{sec:Constraints} & \nodata & Specific emissivity (emissivity per unit frequency), generally for synchrotron radio emission\\
$L_{\rm TIR}$	& \ref{sec:Constraints} & \nodata & Total infrared emission from young stars\\
$L_{\rm radio}$	& \ref{sec:Constraints} & \nodata & Nonthermal synchrotron radio emission\\
$q_{\rm FIR}$             & \ref{sec:Constraints} & \nodata & Rescaled, observed logarithm of ratio $L_{\rm FIR} / L_{\rm radio}$\\
$e^+ / (e^+ + e^-)$           & \ref{sec:Constraints} & \nodata & Fraction of CR positrons in CR electrons and positrons at Earth, usually at 1~\GeV\\
$p/e$           & \ref{sec:Constraints} & \nodata & Observed ratio of protons to electrons at Earth, usually at 1~\GeV\\
$L_{\pi^0}$	& \ref{sec:Constraints} & \nodata & Gamma-ray emission from $\pi^0$ production\\
$dI_e(E) / dE$	& \ref{sec:Constraints} & \nodata & Spectrum of CR electrons at Earth from Milky Way sources\\
$dI_p(E) / dE$	& \ref{sec:Constraints} & \nodata & Spectrum of CR protons at Earth from Milky Way sources\\
$\alpha$	& \ref{sec:Constraints} & \nodata & Power law slope of the observed radio flux, $d\log F_{\nu} / d\log \nu$, usually measured between 1.4 and 4.8 GHz\\
$\cal{P}(E)$	& \ref{sec:SpectralSlopeEffects} & \nodata & Spectral slope of the final steady-state CR spectrum, $d\log N(E) / d\log E$\\
$e^-_{\rm sec} / e^-$         & \ref{sec:xiEffects} & \nodata & Fraction of CR electrons that are secondaries\\
$F_{\rm cal}$ & \ref{sec:Calorimeter} & \nodata & Fraction of CR proton luminosity going into pion losses\\
\enddata
\label{table:SymbolList}
\end{deluxetable}

\begin{deluxetable}{lcccccccccc}
\tablecaption{$\pi^0$ $\gamma$-Ray (and $\pi^{\pm}$ Neutrino\tablenotemark{a}) Fluxes}
\tablehead{\colhead{Galaxy} & \colhead{$\log_{10} \Sigma_g$} & \colhead{R} & \colhead{D} & \colhead{$\Gamma_{\rm SN}$} & \multicolumn{6}{c}{Integrated $\pi^0$ Photon Flux $>E$} \\ & & & & & \colhead{100 MeV (Total\tablenotemark{b})} & \colhead{1 GeV} & \colhead{10 GeV} & \colhead{100 GeV} & \colhead{300 GeV} & \colhead{1 TeV} \\ & ($\gcm2$) & (\kpc) & (\Mpc) & ($\yr^{-1}$) & \colhead{($\phFluxUnitsWOph$)}  & \colhead{($\phFluxUnitsWOph$)} & \colhead{($\phFluxUnitsWOph$)} & \colhead{($\phFluxUnitsWOph$)} & \colhead{($\phFluxUnitsWOph$)} & \colhead{($\phFluxUnitsWOph$)}}
\startdata
\cutinhead{Standard}
Milky Way\tablenotemark{c} & -2.0 & 4 & 0.008 & 0.018 & 2.7E-4 (5.1E-4) & 3.4E-5 & 8.5E-7 & 1.5E-8 & 2.3E-9 & 2.8E-10\\
M31 & -3.0 & 20.9 & 0.9 & 0.019 & 3.7E-9 (1.2E-8) & 3.9E-10 & 8.1E-12 & 1.4E-13 & 2.0E-14 & 2.4E-15\\
NGC 253 & -0.33 & 0.21 & 3.5 & 0.011 & 2.6E-9 (4.7E-9) & 5.1E-10 & 3.2E-11 & 1.6E-12 & 3.8E-13 & 7.5E-14\\
M82 & -0.16 & 0.23 & 3.6 & 0.022 & 5.1E-9 (9.1E-9) & 1.0E-9 & 6.3E-11 & 3.2E-12 & 7.8E-12 & 1.6E-13\\
Arp 220 (east) & 0.78 & 0.12 & 76.6 & 0.12 & 6.4E-11 (1.1E-10) & 1.3E-11 & 8.0E-13 & 4.3E-14 & 1.1E-14 & 2.3E-15\\
Arp 220 (west) & 0.94 & 0.07 & 76.6 & 0.071 & 3.6E-11 (6.0E-11) & 7.2E-12 & 4.6E-13 & 2.4E-14 & 6.1E-15 & 1.3E-15\\
Arp 220 (disk) & 1.08 & 0.37 & 76.6 & 3.1 & 1.6E-9 (2.6E-9) & 3.1E-10 & 2.0E-11 & 1.1E-12 & 2.7E-13 & 5.8E-14\\
\cutinhead{$B \propto \rho^a$ and winds}
Milky Way\tablenotemark{c} & -2.0 & 4 & 0.008 & 0.018 & 2.5E-4 (4.4E-4) & 3.3E-5 & 9.6E-7 & 2.1E-8 & 3.5E-9 & 4.8E-10\\
M31 & -3.0 & 20.9 & 0.9 & 0.019 & 3.0E-9 (7.7E-9) & 3.5E-10 & 8.8E-12 & 1.9E-13 & 3.0E-14 & 4.2E-15\\
NGC 253 & -0.33 & 0.21 & 3.5 & 0.011 & 2.4E-9 (4.1E-9) & 5.2E-10 & 4.0E-11 & 2.5E-12 & 6.5E-13 & 1.4E-13\\
M82 & -0.16 & 0.23 & 3.6 & 0.022 & 5.2E-9 (8.9E-9) & 1.1E-9 & 8.7E-11 & 5.5E-12 & 1.5E-12 & 3.3E-13\\
Arp 220 (east) & 0.78 & 0.12 & 76.6 & 0.12 & 7.9E-11 (1.4E-10) & 1.7E-11 & 1.4E-12 & 9.2E-14 & 2.6E-14 & 6.1E-15\\
Arp 220 (west) & 0.94 & 0.07 & 76.6 & 0.071 & 4.6E-11 (7.9E-11) & 1.0E-11 & 7.9E-13 & 5.3E-14 & 1.5E-14 & 3.6E-15\\
Arp 220 (disk) & 1.08 & 0.37 & 76.6 & 3.1 & 2.0E-9 (3.5E-9) & 4.4E-10 & 3.5E-11 & 2.3E-12 & 6.6E-13 & 1.6E-13
\enddata
\tablenotetext{a}{Although we calculate the $\pi^0$ $\gamma$-ray spectrum explicitly, we do not perform a similar calculation for neutrinos.  However, the neutrino flux from $\pi^{\pm}$ decay at energies much higher than $m_{\pi} c^2 \approx 140~\MeV$ is approximately equal to the $\gamma$-ray flux, if antineutrinos and all flavors are included \citep{Stecker79,Loeb06}.} 
\tablenotetext{b}{The flux in parentheses includes bremsstrahlung and IC $\gamma$-rays as well as pionic emission.}
\tablenotetext{c}{For simplicity, we treat the Milky Way as a point source at the Galactic Center, and consider only its inner regions.}
\label{table:GammaRayFluxes}
\end{deluxetable}

\clearpage

\LongTables
\begin{deluxetable}{lcccccccccc}
\tabletypesize{\scriptsize}
\tablecaption{Successful Models}
\tablehead{\colhead{\S} & \colhead{$B$} & \colhead{$v_{\rm wind}~(\Sigma_{\rm g,min})$\tablenotemark{a}} & \colhead{$\kappa_{\rm FIR}$} & \colhead{$h_{\rm norm}$} & \colhead{$t_{\rm diff}$} & \colhead{Constraints\tablenotemark{b}} & \multicolumn{4}{c}{Allowed Values\tablenotemark{c}} \\ & & $\kms~(\gcm2)$ & ($\cmg21$) & ($\pc$) & & & \colhead{$p$} & \colhead{$f$} & \colhead{$a$} & \colhead{$\tilde\delta$}}
\startdata
\ref{sec:StandardModel} & $B \propto \Sigma_g^a$ & \nodata & $0$ & $1000$ & Const. $t_{\rm diff}$ & L & $2.0 - 2.6$ & $1.0 - 2.0$ & $0.6 - 0.7$ & $34 - 100$\\
 & & & & & & G & $2.0 - 2.6$ & $1.0 - 2.0$ & $0.6 - 0.8$ & $10 - 152$\\
 & & & & & & C & $2.0 - 2.6$ & $1.0 - 2.0$ & $0.6 - 0.7$ & $34 - 100$\\
\tableline
\ref{sec:Basn} & $B \propto \rho^a$ & \nodata & $0$ & $1000$ & Const. $t_{\rm diff}$ & L & $\emptyset$ & $\emptyset$ & $\emptyset$ & $\emptyset$\\
 & & & & & & G & $\emptyset$ & $\emptyset$ & $\emptyset$ & $\emptyset$\\
 & & & & & & C & $\emptyset$ & $\emptyset$ & $\emptyset$ & $\emptyset$\\
\tableline
\ref{sec:Winds} & $B \propto \Sigma_g^a$ & $300~(0.05)$ & $0$ & $1000$ & Const. $t_{\rm diff}$ & L & $\emptyset$ & $\emptyset$ & $\emptyset$ & $\emptyset$\\
 & & & & & & G & $\emptyset$ & $\emptyset$ & $\emptyset$ & $\emptyset$\\
 & & & & & & C & $\emptyset$ & $\emptyset$ & $\emptyset$ & $\emptyset$\\
\tableline
\ref{sec:Winds} & $B \propto \rho^a$ & $300~(0.05)$ & $0$ & $1000$ & Const. $t_{\rm diff}$ & L & $2.0 - 2.6$ & $1.0 - 2.0$ & $0.5 - 0.6$ & $34 - 100$\\
 & & & & & & G & $2.0 - 2.6$ & $1.0 - 2.0$ & $0.5 - 0.6$ & $15 - 100$\\
 & & & & & & C & $2.0 - 2.6$ & $1.0 - 2.0$ & $0.5 - 0.6$ & $34 - 100$\\
\tableline
\ref{sec:Winds} & $B \propto \rho^a$ & $175~(0.01)$ & $0$ & $1000$ & Const. $t_{\rm diff}$ & L & $\emptyset$ & $\emptyset$ & $\emptyset$ & $\emptyset$\\
 & & $600~(0.05)$ & & & & G & $\emptyset$ & $\emptyset$ & $\emptyset$ & $\emptyset$\\
 & & & & & & C & $\emptyset$ & $\emptyset$ & $\emptyset$ & $\emptyset$\\
\tableline
\ref{sec:ScaleHeight}\tablenotemark{d} & $B \propto \Sigma_g^a$ & \nodata & $0$ & $100$ & Const. $t_{\rm diff}$ & L & $2.2 - 2.6$ & $0.1 - 0.2$ & $0.5 - 0.6$ & $34 - 100$ \\
 & & & & & & G & $2.0 - 2.6$ & $0.1 - 0.2$ & $0.5 - 0.6$ & $15 - 100$\\
 & & & & & & C & $2.2 - 2.6$ & $0.1 - 0.2$ & $0.5 - 0.6$ & $34 - 100$\\
\tableline
\ref{sec:ScaleHeight} & $B \propto \Sigma_g^a$ & \nodata & $0$ & $300$ & Const. $t_{\rm diff}$ & L & $2.0 - 2.6$ & $0.3 - 0.6$ & $0.6$ & $34 - 100$\\
 & & & & & & G & $2.0 - 2.6$ & $0.3 - 0.6$ & $0.6 - 0.7$ & $15 - 100$\\
 & & & & & & C & $2.0 - 2.6$ & $0.3 - 0.6$ & $0.6$ & $34 - 100$\\
\tableline
\ref{sec:ScaleHeight} & $B \propto \Sigma_g^a$ & \nodata & $0$ & $2000$ & Const. $t_{\rm diff}$ & L & $2.0 - 2.4$ & $2.0$ & $0.7$ & $35 - 67$\\
 & & & & & & G & $2.0 - 2.4$ & $2.0$ & $0.7$ & $35 - 67$\\
 & & & & & & C & $2.0 - 2.4$ & $2.0$ & $0.7$ & $35 - 67$\\
\tableline
\ref{sec:ScaleHeight} & $B \propto \Sigma_g^a$ & \nodata & $0$ & $4000$ & Const. $t_{\rm diff}$ & L & $\emptyset$ & $\emptyset$ & $\emptyset$ & $\emptyset$\\
 & & & & & & G & $\emptyset$ & $\emptyset$ & $\emptyset$ & $\emptyset$\\
 & & & & & & C & $\emptyset$ & $\emptyset$ & $\emptyset$ & $\emptyset$\\
\tableline
\ref{sec:OpticalThickFIR} & $B \propto \Sigma_g^a$ & \nodata & $1$ & $1000$ & Const. $t_{\rm diff}$ & L & $2.0 - 2.6$ & $1.0 - 2.0$ & $0.7$ & $34 - 91$ \\
 & & & & & & G & $2.0 - 2.6$ & $1.0 - 2.0$ & $0.7 - 0.8$ & $10 - 91$\\
 & & & & & & C & $2.0 - 2.6$ & $1.0 - 2.0$ & $0.7$ & $34 - 91$\\
\tableline
\ref{sec:OpticalThickFIR} & $B \propto \Sigma_g^a$ & \nodata & $10$ & $1000$ & Const. $t_{\rm diff}$ & L & $\emptyset$ & $\emptyset$ & $\emptyset$ & $\emptyset$\\
 & & & & & & G & $\emptyset$ & $\emptyset$ & $\emptyset$ & $\emptyset$\\
 & & & & & & C & $\emptyset$ & $\emptyset$ & $\emptyset$ & $\emptyset$\\
\tableline
\ref{sec:UBeqUph} & $U_B = U_{\rm ph}$ & \nodata & $0$ & $1000$ & Const. $t_{\rm diff}$ & L & $2.0 - 2.6$ & $1.0 - 2.0$ & \nodata & $34 - 100$\\
 & & & & & & G & $2.0 - 2.6$ & $1.0 - 1.5$ & \nodata & $34 - 100$\\
 & & & & & & C & $2.0 - 2.6$ & $1.0 - 1.5$ & \nodata & $34 - 100$\\
\tableline
\ref{sec:UBeqUph} & $U_B = U_{\rm ph}$ & \nodata & $1$ & $1000$ & Const. $t_{\rm diff}$ & L & $2.0$ & $1.5$ & \nodata & $50$\\
 & & & & & & G & $2.0$ & $1.0 - 1.5$ & \nodata & $20 - 50$\\
 & & & & & & C & $2.0$ & $1.5$ & \nodata & $50$\\
\tableline
\ref{sec:UBeqUph} & $U_B = U_{\rm ph}$ & \nodata & $10$ & $1000$ & Const. $t_{\rm diff}$ & L & $\emptyset$ & $\emptyset$ & \nodata & $\emptyset$\\
 & & & & & & G & $\emptyset$ & $\emptyset$ & \nodata & $\emptyset$\\
 & & & & & & C & $\emptyset$ & $\emptyset$ & \nodata & $\emptyset$\\
\tableline
\ref{sec:FastEscape} & $B \propto \Sigma_g^a$ & \nodata & $0$ & $1000$ & Const. $t_{\rm diff}$ & L & $\emptyset$ & $\emptyset$ & $\emptyset$ & $\emptyset$\\
 & & & & & ($1/4$ nominal) & G & $2.0$ & $2.0$ & $0.6$ & $35.0$\\
 & & & & & & C & $\emptyset$ & $\emptyset$ & $\emptyset$ & $\emptyset$\\
\tableline
\ref{sec:DVary} & $B \propto \Sigma_g^a$ & \nodata & $0$ & $1000$ & Const. $D_z$ & L & $\emptyset$ & $\emptyset$ & $\emptyset$ & $\emptyset$\\
 & & & & & & G & $\emptyset$ & $\emptyset$ & $\emptyset$ & $\emptyset$\\
 & & & & & & C & $\emptyset$ & $\emptyset$ & $\emptyset$ & $\emptyset$\\
\tableline
\ref{sec:DVary} & $B \propto \Sigma_g^a$ & \nodata & $0$ & $1000$ & $D_z \propto \rho^{-1/3}$ & L & $\emptyset$ & $\emptyset$ & $\emptyset$ & $\emptyset$\\
 & & & & & & G & $\emptyset$ & $\emptyset$ & $\emptyset$ & $\emptyset$\\
 & & & & & & C & $\emptyset$ & $\emptyset$ & $\emptyset$ & $\emptyset$\\
\tableline
\ref{sec:DVary} & $B \propto \Sigma_g^a$ & \nodata & $0$ & $1000$ & $D_z \propto \rho^{-1}$ & L & $\emptyset$ & $\emptyset$ & $\emptyset$ & $\emptyset$\\
 & & & & & & G & $\emptyset$ & $\emptyset$ & $\emptyset$ & $\emptyset$\\
 & & & & & & C & $\emptyset$ & $\emptyset$ & $\emptyset$ & $\emptyset$\\
\tableline
\ref{sec:DVary} & $B \propto \rho^a$ & $300~(0.05)$ & $0$ & $1000$ & Const. $D_z$ & L & $2.2 - 2.6$ & $1.0 - 2.0$ & $0.5 - 0.6$ & $34 - 100$\\
 & & & & & & G & $2.0 - 2.6$ & $1.0 - 2.0$ & $0.5 - 0.6$ & $10 - 91$\\
 & & & & & & C & $2.2 - 2.6$ & $1.0 - 2.0$ & $0.5 - 0.6$ & $34 - 91$\\
\tableline
\ref{sec:DVary} & $B \propto \rho^a$ & $300~(0.05)$ & $0$ & $1000$ & $D_z \propto \rho^{-1/3}$ & L & $2.0 - 2.6$ & $1.0 - 2.0$ & $0.5$ & $34 - 100$\\
 & & & & & & G & $2.0 - 2.4$ & $1.0 - 2.0$ & $0.5 - 0.6$ & $10 - 67$\\
 & & & & & & C & $2.0 - 2.4$ & $1.0 - 1.5$ & $0.5$ & $34 - 67$\\
\tableline
\ref{sec:DVary} & $B \propto \rho^a$ & $300~(0.05)$ & $0$ & $1000$ & $D_z \propto \rho^{-1}$ & L & $\emptyset$ & $\emptyset$ & $\emptyset$ & $\emptyset$\\
 & & & & & & G & $\emptyset$ & $\emptyset$ & $\emptyset$ & $\emptyset$\\
 & & & & & & C & $\emptyset$ & $\emptyset$ & $\emptyset$ & $\emptyset$\\
\tableline
\ref{sec:MultipleEffects} & $B \propto \rho^a$ & $300~(0.05)$ & $1$ & $1000$ & Const. $D_z$ & L & $2.0 - 2.6$ & $1.0 - 2.0$ & $0.5 - 0.6$ & $34 - 100$\\
 & & & & & & G & $2.0 - 2.6$ & $1.0 - 2.0$ & $0.5 - 0.6$ & $10 - 91$\\
 & & & & & & C & $2.0 - 2.6$ & $1.0 - 2.0$ & $0.5 - 0.6$ & $34 - 91$\\
\tableline
\ref{sec:MultipleEffects} & $B \propto \rho^a$ & $300~(0.05)$ & $1$ & $1000$ & $D_z \propto \rho^{-1/3}$ & L & $2.6$ & $1.0 - 1.5$ & $0.5$ & $91$\\
 & & & & & & G & $2.0 - 2.6$ & $1.0 - 2.0$ & $0.5 - 0.6$ & $10 - 91$\\
 & & & & & & C & $2.6$ & $1.0$ & $0.5$ & $91$\\
\tableline
\ref{sec:MultipleEffects} & $B \propto \rho^a$ & $300~(0.05)$ & $1$ & $1000$ & $D_z \propto \rho^{-1}$ & L & $\emptyset$ & $\emptyset$ & $\emptyset$ & $\emptyset$\\
 & & & & & & G & $\emptyset$ & $\emptyset$ & $\emptyset$ & $\emptyset$\\
 & & & & & & C & $\emptyset$ & $\emptyset$ & $\emptyset$ & $\emptyset$\\
\tableline
\ref{sec:MultipleEffects} & $B \propto \rho^a$ & $175~(0.01)$ & $0$ & $2000$ & Const. $D_z$ & L & $2.2$ & $3.0 - 4.0$ & $0.6$ & $34 - 45$\\
 & & $600~(0.05)$ & & & & G & $2.0 - 2.2$ & $3.0 - 4.0$ & $0.6$ & $20 - 45$\\
 & & & & & & C & $2.2$ & $3.0 - 4.0$ & $0.6$ & $34 - 45$\\
\tableline
\ref{sec:MultipleEffects} & $B \propto \rho^a$ & $175~(0.01)$ & $1$ & $2000$ & Const. $D_z$ & L & $2.2, 2.6$ & $2.0 - 4.0$ & $0.6, 0.5$ & $34 - 91$\\
 & & $600~(0.05)$ & & & & G & $2.0 - 2.6$ & $2.0 - 4.0$ & $0.5 - 0.6$ & $20 - 91$\\
 & & & & & & C & $2.2, 2.6$ & $2.0 - 4.0$ & $0.6, 0.5$ & $34 - 91$
\enddata
\tablenotetext{a}{Models without a wind have a \nodata entry.  Otherwise, there is a wind of speed $v_{\rm wind}$ in models with a surface density $\Sigma_g$ of at least $\Sigma_{\rm g,min}$.  In the strong wind variants with more than one $\Sigma_{\rm g,min}$ listed, the wind speed is that with the greatest $\Sigma_{\rm g,min}$ less than $\Sigma_g$ for each model.}
\tablenotetext{b}{The FIR-radio correlation must always hold.  Additional constraints on the proton normalization: L -- local constraints ($e^+ / (e^- + e^+)$ and $p/e$).  G -- integrated constraint (Milky Way gamma-ray luminosity from $\pi^0$ production).  C -- both local and Galactic constraints.}
\tablenotetext{c}{Variants with $\emptyset$ entries could not satisfy the given constraints.  Models where the magnetic field has no free parameters have a \nodata entry in the $a$ column.}
\tablenotetext{d}{Since the scale height does not vary, $D_z$ is constant and $B \propto \rho^a$ for these models.}
\label{table:WorkingModels}
\end{deluxetable}


\begin{thebibliography}{}

\bibitem[Abdo et al.(2010)]{Abdo10} Abdo, A.~A., et al.\ 2010, \apjl, 709, L152 

\bibitem[Acciari et al.(2009)]{Acciari09} Acciari, V.~A., et al.\ 2009, \nat, 462, 770 

\bibitem[Acero et al.(2009)]{Acero09} Acero, F., et al.\ 2009, Science, 326, 1080 

\bibitem[Adriani et al.(2009)]{Adriani08} Adriani, O., et al. 2009, \nat, 458, 607 

\bibitem[Aharonian et al.(2005)]{Aharonian05} Aharonian, F. et al. 2005, \aap~442, 177.

\bibitem[Albert et al.(2007)]{Albert07} Albert, J. et al. 2007, \apj~658, 245.

\bibitem[AMS Collaboration et al.(2002)]{AMS02} AMS Collaboration, et al.\ 2002, \physrep~366, 331.

\bibitem[Andrew(1966)]{Andrew66} Andrew, B. H. 1966, \mnras~132, 79.

\bibitem[Appleton et al.(2004)]{Appleton04} Appleton, P.~N., et al.\ 2004, \apjs~154, 147.

\bibitem[Arnold et al.(1961)]{Arnold61} Arnold, J. R., Honda, M., \& Lal, D. 1961, \jgr~66, 3519.

\bibitem[Beatty et al.(2004)]{Beatty04} Beatty, J. J. et al. 2004, \prl~93, 241102.

\bibitem[Beck \& Golla(1988)]{Beck88} Beck, R., \& Golla, G.\ 1988, \aap~191, L9.

\bibitem[Beck(2001)]{Beck01} Beck, R. 2001, \ssr, 99, 243.

\bibitem[Beck(2009)]{Beck09} Beck, R.\ 2009, \apss, 320, 77 

\bibitem[Bell(2003)]{Bell03} Bell, E. F. 2003, \apj~586, 794.

\bibitem[Beswick et al.(2008)]{Beswick08} Beswick, R. J., Muxlow, T. W. B, Thrall, H., Richards, A. M. S., Garrington, S. T. 2008, \mnras~385, 1143.

\bibitem[Beuermann et al.(1985)]{Beuermann85} Beuermann, K., Kanbach, G., Berkhuijsen, E. M. 1985, \aap~153, 17.

\bibitem[Bicay \& Helou(1990)]{Bicay90} Bicay, M.~D., \& Helou, G.\ 1990, \apj~362, 59.

\bibitem[Biggs \& Ivison(2008)]{Biggs08} Biggs, A.~D., \& Ivison, R.~J.\ 2008, \mnras~385, 893.

\bibitem[Bouch{\'e} et al.(2007)]{Bouche07} Bouch{\'e}, N., et al.\ 2007, \apj~671, 303.

\bibitem[Boulares \& Cox(1990)]{Boulares90} Boulares, A. \& Cox, D. P. 1990, \apj~365, 544.

\bibitem[Buat et al.(2005)]{Buat05} Buat, V., et al.\ 2005, \apjl~619, L51.

\bibitem[Calzetti et al.(2000)]{Calzetti00} Calzetti, D., Armus, L., Bohlin, R.~C., Kinney, A.~L., Koornneef, J., \& Storchi-Bergmann, T.\ 2000, \apj~533, 682.

\bibitem[Chapman et al.(2004)]{Chapman04} Chapman, S.~C., Smail, I., Windhorst, R., Muxlow, T., \& Ivison, R.~J.\ 2004, \apj~611, 732.

\bibitem[Chevalier \& Fransson(1984)]{Chevalier84} Chevalier, R.~A., \& Fransson, C.\ 1984, \apjl~279, L43.

\bibitem[Chi \& Wolfendale(1990)]{Chi90} Chi, X., \& Wolfendale, A.~W.\ 1990, \mnras~245, 101.

\bibitem[Clemens et al.(2008)]{Clemens08} Clemens, M. S. et al. 2008, \aap~477, 95.

\bibitem[Condon et al.(1991)]{Condon91} Condon, J. J., Huang, Z.-P., Yin, Q. F., \& Thuan, T. X. 1991, \apj~378, 65.

\bibitem[Condon(1992)]{Condon92} Condon, J. J. 1992, \araa~30, 575.

\bibitem[Connell(1998)]{Connell98} Connell, J. J. 1998, \apjl~501, 59.

\bibitem[Cox et al.(1988)]{Cox88} Cox, M.~J., Eales, S.~A.~E., Alexander, P., \& Fitt, A.~J.\ 1988, \mnras~235, 1227.

\bibitem[Crutcher(1999)]{Crutcher99} Crutcher, R.~M.\ 1999, \apj, 520, 706 

\bibitem[Dahlem et al.(1995)]{Dahlem95} Dahlem, M., Lisenfeld, U., \& Golla, G.\ 1995, \apj 444, 119.

\bibitem[de Cea del Pozo et al.(2009)]{deCeaDelPozo09} de Cea del Pozo, E., Torres, D. F., \& Rodriguez Marrero, A. Y. 2009, \apj, 698, 1054 

\bibitem[de Jong et al.(1985)]{deJong85} de Jong, T., Klein, U., Wielebinski, R., \& Wunderlich, E. 1985, \aap~147, L6.

\bibitem[Delahaye et al.(2009)]{Delahaye08} Delahaye, T., Lineros, R., Donato, F., Fornengo, N., Lavalle, J., Salati, P., \& Taillet, R. 2009, \aap, 501, 821 

\bibitem[Domingo-Santamar\'ia \& Torres(2005)]{Domingo05} Domingo-Santamar\'ia, E. \& Torres, D. F. 2005, \aap 444, 403.

\bibitem[Downes \& Solomon(1998)]{Downes98} Downes, D., \& Solomon, P. M. 1998, \apj~507, 615.

\bibitem[Dumke et al.(1995)]{Dumke95} Dumke, M., Krause, M., Wielebinski, R., \& Klein, U.\ 1995, \aap, 302, 691 

\bibitem[Dumke \& Krause(1998)]{Dumke98} Dumke, M., \& Krause, M.\ 1998, IAU Colloq.~166: The Local Bubble and Beyond, 506, 555 

\bibitem[Dumke et al.(2000)]{Dumke00} Dumke, M., Krause, M., \& Wielebinski, R.\ 2000, \aap, 355, 512 

\bibitem[Engelmann et al.(1990)]{Engelmann90} Engelmann, J. J., Ferrando, P., Soutoul, A., Goret, P., Juliusson, E., Koch-Miramond, L., Lund, N., Masse, P., Peters, B., Petrou, N., \& Rasmussen, I. L. 1990, \aap~233, 96.

\bibitem[Everett et al.(2008)]{Everett08} Everett, J. E. et al. 2008, \apj~674, 258.

\bibitem[Ferri\`ere (2001)]{Ferriere01} Ferri\`ere, K. M. 2001, Rev. Mod. Phys.~73, 1031.

\bibitem[Fitt et al.(1988)]{Fitt88} Fitt, A.~J., Alexander, P., \& Cox, M.~J.\ 1988, \mnras, 233, 907.

\bibitem[Freudenreich(1998)]{Freudenreich98} Freudenreich, H.~T.\ 1998, \apj~492, 495.

\bibitem[Garcia-Munoz et al.(1977)]{GarciaMunoz77} Garcia-Munoz, M., Mason, G. M., Simpson, J. A. 1977, \apj~217, 859.

\bibitem[Ginzburg \& Ptuskin(1976)]{Ginzburg76} Ginzburg, V.~L., \& Ptuskin, V.~S.\ 1976, Reviews of Modern Physics~48, 161. 

\bibitem[Greve et al.(2009)]{Greve09} Greve, T.~R., Papadopoulos, P.~P., Gao, Y., \& Radford, S.~J.~E.\ 2009, \apj~692, 1432.

\bibitem[Groves et al.(2003)]{Groves03} Groves, B. A., Cho, J., Dopita, M., \& Lazarian, A. 2003, Publ. Astron. Soc. Austaralia 20, 252.

\bibitem[Han \& Qiao(1994)]{Han94} Han, J.~L., \& Qiao, G.~J.\ 1994, \aap~288, 759.

\bibitem[Heckman et al.(2000)]{Heckman00} Heckman, T. M., Lehnert, M. D., Strickland, D. K., Armus, L. 2000, \apjs~129, 493.

\bibitem[Heckman(2003)]{Heckman03} Heckman, T. M. 2003, in Rev. Mex. AA Ser. Conf., 17, 47

\bibitem[Heesen et al.(2009)]{Heesen09} Heesen, V., Beck, R., Krause, M., \& Dettmar, R.-J.\ 2009, \aap, 494, 563 

\bibitem[Helou(1986)]{Helou86} Helou, G.\ 1986, \apjl~311, L33.

\bibitem[Helou et al.(1985)]{Helou85} Helou, G., Soifer, B. T., \& Rowan-Robinson, M. 1985, \apjl~298, 7.

\bibitem[Helou \& Bicay(1993)]{Helou93} Helou, G. \& Bicay, M. D. 1993, \apj~415, 93.

\bibitem[Hughes et al.(2006)]{Hughes06} Hughes, A., Wong, T., Ekers, R., Staveley-Smith, L., Filipovic, M., Maddison, S., Fukui, Y., \& Mizuno, N.\ 2006, \mnras~370, 363.

\bibitem[Jubelgas et al.(2008)]{Jubelgas08} Jubelgas, M., Springel, V., En{\ss}lin, T., \& Pfrommer, C.\ 2008, \aap~481, 33.

\bibitem[Kennicutt(1998)]{Kennicutt98} Kennicutt, R. C. 1998, \apj~498, 541.

\bibitem[Kogut et al.(2009)]{Kogut09} Kogut, A. et al. 2009, arXiv:0901.0562.

\bibitem[Kov{\'a}cs et al.(2006)]{Kovacs06} Kov{\'a}cs, A., Chapman, S.~C., Dowell, C.~D., Blain, A.~W., Ivison, R.~J., Smail, I., \& Phillips, T.~G.\ 2006, \apj, 650, 592 

\bibitem[Krause et al.(2006)]{Krause06} Krause, M., Wielebinski, R., \& Dumke, M.\ 2006, \aap, 448, 133 

\bibitem[Kulsrud \& Pearce(1969)]{Kulsrud69} Kulsrud, R., Pearce, W. P. 1969, \apj~156, 445.

\bibitem[Lacki \& Thompson(2009)]{Lacki09} Lacki, B.~C., \& Thompson, T.~A.\ 2009, arXiv:0910.0478 

\bibitem[Li \& Draine(2001)]{Li01} Li, A., \& Draine, B.~T.\ 2001, \apj~554, 778.

\bibitem[Lisenfeld \& V\"olk(2000)]{Lisenfeld00} Lisenfeld, U. \& V\"olk, H. J. 2000, \aap~354, 423.

\bibitem[Lisenfeld et al.(1996a)]{Lisenfeld96a} Lisenfeld, U., V\"olk, H. J., \& Xu, C. 1996, \aap~306, 677.

\bibitem[Lisenfeld et al.(1996b)]{Lisenfeld96b} Lisenfeld, U., V\"olk, H. J., \& Xu, C. 1996, \aap~314, 745.

\bibitem[Loeb \& Waxman(2006)]{Loeb06} Loeb, A. \& Waxman, E. 2006, Journal of Cosmology and Astroparticle Physics 5,3.

\bibitem[Longair(1994)]{Longair04} Longair, M. S. 1994, High Energy Astrophysics, 2nd ed. (Cambridge: Cambridge Univ. Press)

\bibitem[Lukasiak et al.(1994)]{Lukasiak94} Lukasiak, A., Ferrando, P., McDonald, F.~B., \& Webber, W.~R.\ 1994, \apj~423, 426.

\bibitem[Mannheim \& Schlickeiser(1994)]{Mannheim94} Mannheim, K., \& Schlickeiser, R.\ 1994, \aap, 286, 983 

\bibitem[Martin et al.(2005)]{Martin05} Martin, D.~C., et al.\ 2005, \apjl~619, L59.

\bibitem[Melo et al.(2002)]{Melo02} Melo, V.~P., P{\'e}rez Garc{\'{\i}}a, A.~M., Acosta-Pulido, J.~A., Mu{\~n}oz-Tu{\~n}{\'o}n, C., \& Rodr{\'{\i}}guez Espinosa, J.~M.\ 2002, \apj, 574, 709 

\bibitem[Menn et al.(2000)]{Menn00} Menn, W., et al.\ 2000, \apj~533, 281.

\bibitem[Micha{\l}owski et al.(2010)]{Michalowski09} Micha{\l}owski, M. J., Hjorth, J., \& Watson, D. 2010, \aap, in press (arXiv:0905.4499)

\bibitem[Mills(1959)]{Mills59} Mills, B. Y. 1959, in Paris Symposium of Radio Astronomy, ed. R. N. Bracewell, (Stanford, California: Stanford University Press), 431

\bibitem[Mori(1997)]{Mori97} Mori, M.\ 1997, \apj 478, 225.

\bibitem[Moskalenko \& Strong(1998)]{Moskalenko98} Moskalenko, I. V. \& Strong, A. W. 1998, \apj~493, 694.

\bibitem[Moskalenko et al.(2002)]{Moskalenko02} Moskalenko, I. V., Strong, A. W., Ormes, J. R., \& Potgieter, M. S. 2002, \apj~565, 280.

\bibitem[Murgia et al.(2005)]{Murgia05} Murgia, M., Helfer, T. T., Ekers, R., Blitz, L., Moscadelli, L., Wong, T., \& Paladino, R. 2005, \aap~437, 389.

\bibitem[Murphy et al.(2006a)]{Murphy06a} Murphy, E.~J., et al.\ 2006, \apj, 638, 157.

\bibitem[Murphy et al.(2006b)]{Murphy06b} Murphy, E. J. et al. 2006, \apjl~651, 111.

\bibitem[Murphy et al.(2008)]{Murphy08} Murphy, E. J. et al. 2008, \apj~678, 828.

\bibitem[Murphy(2009)]{Murphy09} Murphy, E.~J.\ 2009, \apj, 706, 482 

\bibitem[Niklas(1997)]{Niklas97b} Niklas, S.\ 1997, \aap~322, 29.

\bibitem[Niklas \& Beck(1997)]{Niklas97a} Niklas, S. \& Beck, R. 1997, \aap~320, 54.

\bibitem[Paladino et al.(2006)]{Paladino06} Paladino, R., Murgia, M., Helfer, T.~T., Wong, T., Ekers, R., Blitz, L., Gregorini, L., \& Moscadelli, L.\ 2006, \aap~456, 847.

\bibitem[Parker(1966)]{Parker66} Parker, E.~N.\ 1966, \apj, 145, 811 

\bibitem[Persic et al.(2008)]{Persic08} Persic, M., Rephaeli, Y., \& Arieli, Y. 2008, \aap~486, 143.

\bibitem[Platania et al.(1998)]{Platania98} Platania, P., Bensadoun, M., Bersanelli, M., de Amici, G., Kogut, A., Levin, S., Maino, D., \& Smoot, G. F. 1998, \apj~505, 473.

\bibitem[Platania et al.(2003)]{Platania03} Platania, P., Burigana, C., Maino, D., Caserini, E., Bersanelli, M., Cappellini, B., \& Mennella, A. 2003, \apj~410, 847.

\bibitem[Popescu et al.(2005)]{Popescu05} Popescu, C.~C., et al.\ 2005, \apjl~619, L75.

\bibitem[Ptuskin \& Soutoul(1998)]{Ptuskin98} Ptuskin, V.~S., \& Soutoul, A.\ 1998, \aap, 337, 859 

\bibitem[Reich \& Reich(1988)]{Reich88} Reich, P., \& Reich W. 1988, \aap Suppl. 74, 7.

\bibitem[Rengarajan(2005)]{Rengarajan05} Rengarajan, T. N. 2005, arXiv:astro-ph/0511156

\bibitem[Rephaeli et al.(2010)]{Rephaeli09} Rephaeli, Y., Arieli, Y., \& Persic, M. 2010, \mnras, 401, 473

\bibitem[Robinshaw et al.(2008)]{Robinshaw08} Robinshaw, T., Quataert, E., Heiles, C. 2008, \apj~680, 981.

\bibitem[Rogers \& Bowman(2008)]{Rogers08} Rogers, A. E. E., \& Bowman, J. D. 2008, \aj~136, 641.

\bibitem[Roussel et al.(2003)]{Roussel03} Roussel, H., Helou, G., Beck, R., Condon, J.~J., Bosma, A., Matthews, K., \& Jarrett, T.~H.\ 2003, \apj, 593, 733.

\bibitem[Rybicki \& Lightman(1979)]{Rybicki79} Rybicki, G. B. \& Lightman, A. P. 1979, \emph{Radiative Processes in Astrophysics}, (New York: Wiley-VCH).

\bibitem[Sanders et al.(2003)]{Sanders03} Sanders, D.~B., Mazzarella, J.~M., Kim, D.-C., Surace, J.~A., \& Soifer, B.~T.\ 2003, \aj, 126, 1607 

\bibitem[Schaeffer(1975)]{Schaeffer75} Schaeffer, O. A. 1975, in Proc. 14th Int. Cosmic Ray Conf., Munich, 11, 3508

\bibitem[Schlickeiser(2002)]{Schlickeiser02} Schlickeiser, R. 2002, Cosmic Ray Astrophysics, (Berlin: Springer)

\bibitem[Schmidt(1959)]{Schmidt59} Schmidt, M. 1959, \apj~129, 243.

\bibitem[Seaquist \& Odegard(1991)]{Seaquist91} Seaquist, E.~R., \& Odegard, N.\ 1991, \apj~369, 320.

\bibitem[Semenov et al.(2003)]{Semenov03} Semenov, D., Henning, Th., Helling, Ch., Ilgner, M., Sedlmayr, E. 2003, \aap~410, 611.

\bibitem[Socrates et al.(2008)]{Socrates08} Socrates, A., Davis, S. W., \& Ramirez-Ruiz, E. 2008, \apj~687, 202.

\bibitem[Sopp \& Alexander(1989)]{Sopp89} Sopp, H.~M., \& Alexander, P.\ 1989, \apss, 157, 287.

\bibitem[Stecker(1979)]{Stecker79} Stecker, F.~W.\ 1979, \apj~228, 919.

\bibitem[Strong \& Moskalenko(1998)]{Strong98} Strong, A. W., \& Moskalenko, I. V. 1998, \apj~509, 212.

\bibitem[Strong et al.(2000)]{Strong00} Strong, A. W., Moskalenko, I. V., Reimer, O. 2000, \apj~537, 763.

\bibitem[Thompson et al.(2006)]{Thompson06} Thompson, T. A. et al. 2006, \apj~645, 186.

\bibitem[Thompson, Quataert, \& Waxman(2007)]{Thompson07} Thompson, T. A., Quataert, E., Waxman, E. 2007, \apj~654, 219.

\bibitem[Thompson(2008)]{Thompson08} Thompson, T.~A.\ 2008, \apj~684, 212.

\bibitem[Torres(2004)]{Torres04} Torres, D. F. 2004, \apj~617, 966.

\bibitem[van der Kruit(1971)]{vanDerKruit71} van der Kruit, P. C. 1971, \aap~15, 110.

\bibitem[van der Kruit(1973)]{vanDerKruit73} van der Kruit, P. C. 1973, \aap~29, 263.

\bibitem[Vlahakis et al.(2007)]{Vlahakis07} Vlahakis, C., Eales, S., \& Dunne, L. 2007, \mnras~379, 1042.

\bibitem[V\"olk(1989)]{Volk89} V\"olk, H. J. 1989, \aap~218, 67.

\bibitem[Walter et al.(2009)]{Walter09} Walter, F., Riechers, D., Cox, P., Neri, R., Carilli, C., Bertoldi, F., Weiss, A., Maiolino, R. 2009, Nature 457, 699.

\bibitem[Warren et al.(2005)]{Warren05} Warren, J. S. et al. 2005, \apj~634:376.

\bibitem[Webber \& Soutoul(1998)]{Webber98} Webber, W.~R., \& Soutoul, A.\ 1998, \apj, 506, 335 

\bibitem[Webber et al.(2003)]{Webber03} Webber, W. R., McDonald, F. B., \& Lukasiak, A. 2003, \apj~599, 582.

\bibitem[Webber \& Higbie(2008)]{Webber08} Webber, W.~R., \& Higbie, P.~R.\ 2008, Journal of Geophysical Research (Space Physics), 113, 11106 

\bibitem[Webster(1974)]{Webster74} Webster, A. S. 1974, \mnras~166, 355.

\bibitem[Williams \& Bower(2010)]{Williams10} Williams, P.~K.~G., \& Bower, G.~C.\ 2010, \apj, 710, 1462 

\bibitem[Wunderlich et al.(1987)]{Wunderlich87} Wunderlich, E., Wielebinski, R., \& Klein, U.\ 1987, \aaps~69, 487.

\bibitem[Xu \& Buat(1995)]{Xu95} Xu, C., \& Buat, V.\ 1995, \aap~293, L65.

\bibitem[Younger et al.(2008)]{Younger08} Younger, J.~D., et al.\ 2008, \apj~688, 59.

\bibitem[Yun et al.(2001)]{Yun01} Yun, M. S., Reddy, N. A., \& Condon, J. J. 2001, \apj~554, 803.

\end{thebibliography}
\end{document}